\pdfoutput=1
\documentclass[12pt]{article}
\usepackage[utf8]{inputenc}
\usepackage{pdflscape,url}
\usepackage{bbm,amsmath,color,amsfonts,rotating,amsthm,amssymb,mathtools,setspace,fullpage, mathabx,animate}
\usepackage[authoryear, sort]{natbib}
\setcitestyle{authoryear, open={(},close={)}}
\usepackage{float} 
\usepackage[left=1in, right=1in, top=1in, bottom=1in]{geometry}

\title{A Bayesian framework for incorporating exposure uncertainty into health analyses with application to air pollution and stillbirth}
\author{Saskia Comess$^1$, Howard H.\ Chang$^2$, Joshua L.\ Warren$^{3*}$}
\date{\small $^1$Emmett Interdisciplinary Program in Environment and Resources, Stanford University, Stanford, CA 94305, USA\\ \small $^2$Department of Biostatistics and Bioinformatics, Emory University, Atlanta, GA 30329, USA\\ \small $^3$Department of Biostatistics, Yale University, New Haven, CT 06510, USA\\
\small ($^*$E-mail: joshua.warren@yale.edu)} 

\begin{document}
\maketitle
\begin{abstract}
\noindent 
Studies of the relationships between environmental exposures and adverse health outcomes often rely on a two-stage statistical modeling approach, where exposure is modeled/predicted in the first stage and used as input to a separately fit health outcome analysis in the second stage.  Uncertainty in these predictions is frequently ignored, or accounted for in an overly simplistic manner, when estimating the associations of interest.  Working in the Bayesian setting, we propose a flexible kernel density estimation (KDE) approach for fully utilizing posterior output from the first stage modeling/prediction to make accurate inference on the association between exposure and health in the second stage, derive the full conditional distributions needed for efficient model fitting, detail its connections with existing approaches, and compare its performance through simulation.  Our KDE approach is shown to generally have improved performance across several settings and model comparison metrics.  Using competing approaches, we investigate the association between lagged daily ambient fine particulate matter levels and stillbirth counts in New Jersey (2011-2015), observing an increase in risk with elevated exposure three days prior to delivery.  The newly developed methods are available in the R package \textbf{KDExp}.

\end{abstract}


\doublespacing

\clearpage
\section{Introduction}
Environmental health studies require the linking of environmental exposure information for each observation in the analysis (e.g., individual or time point) in order to estimate the association with adverse health outcomes.  Because exposure data are typically not available at every spatial location and time period covered by the study, researchers often rely on predictions from a first stage statistical model to fill in the spatiotemporal gaps.  For example, several advanced statistical methods have been developed to interpolate ambient air pollution concentrations using monitoring data combined with estimates from deterministic models and other data sources (e.g., \cite{fuentes2005model, berrocal2010spatio, berrocal2010bivariate, mcmillan2010combining, berrocal2012space, reich2014spectral, guan2019multivariate, Warren2021_downscaling}). 

These methods yield point predictions of exposure at the relevant locations and times of interest, but also provide measures of uncertainty.  Because many of the methods are fitted within a Bayesian framework using Markov chain Monte Carlo (MCMC) techniques, samples from posterior predictive distributions (ppd) are also available.  Incorporating this exposure uncertainty into the subsequent health analysis is important for correctly characterizing uncertainty in the association between exposure and health.  Specifically, health observations linked with exposure estimates with higher uncertainties should contribute less to the overall health effect estimate.  However, previous studies often ignore this uncertainty entirely which may impact inference for these associations, although the full implications of this approach are not currently clear.  Several methods for propagating exposure uncertainty have been developed and used previously (e.g., \cite{gryparis2009measurement,lee2010spatial, peng2010spatial, chang2011estimating, Szpiro2011, warren2012spatial, szpiro2013measurement, Blangiardo2016, lee2017rigorous, huang2018multivariate}), and we provide full details on many of them in Section 2.  Although prior work in this area has recommended an investigation and comparison of their performances \citep{lee2017rigorous}, to our knowledge such an analysis has yet to be conducted. 

In this work, we present a flexible framework for exposure uncertainty propagation, carry out a simulation study to compare its performance to existing methods, and apply several of the methods to better understand the relationship between acute, population-level fine particulate matter (PM$_{2.5}$) exposure and daily stillbirth counts using data from three counties in New Jersey (NJ), 2011-2015.  The proposed framework uses kernel density estimation (KDE) with a Gaussian kernel function to specify prior distributions for the exposures within the health model.  The resulting model fitting derivations suggest that this represents a hybrid between two existing approaches; allowing for more flexibility while also avoiding some of the limiting assumptions of those approaches.  It is also shown to maintain computational efficiency for several common health outcome analysis types.  Through simulation, we show that the new approach is flexible enough to accurately characterize uncertainty in the predictions, leading to improved estimation of the association of interest in the health analysis compared to existing approaches.  Differences between the methods are also observed in the NJ stillbirth case study results, indicating the importance of selecting the optimal method in future applications.     

\section{Background}
We specify that the primary epidemiological health outcome analysis of interest consists of $n$ data points (e.g., individuals, time periods) where an exposure level (e.g., ambient air pollution) is assigned to each data point in order to examine its association with the outcome.  For presentation purposes, we introduce the framework using a single exposure while noting that it is straightforward to extend the following results/derivations to accommodate multiple additive exposures.  Statistical modeling of the exposures and health outcome are assumed to take place in the Bayesian setting, as is common in the environmental health statistical methodology literature (e.g., \cite{berrocal2010spatio, chang2011estimating}).  

We assume that $m$ samples from the exposure ppd have been obtained based on the modeling and prediction of observed exposures in a first stage Bayesian framework.  While many modeling options are available, the end result is the same across all approaches; an $n$ by $m$ matrix of ppd samples (i.e., $\textbf{Z}^*$) is obtained such that \begin{equation*} \textbf{Z}^* =
\begin{bmatrix}
    \text{z}_{11}^* & \text{z}_{12}^* & \text{z}_{13}^* & \dots  & \text{z}_{1m}^* \\
    \text{z}_{21}^* & \text{z}_{22}^* & \text{z}_{23}^* & \dots  & \text{z}_{2m}^* \\
    \vdots          & \vdots          & \vdots          & \ddots & \vdots          \\
    \text{z}_{n1}^* & \text{z}_{n2}^* & \text{z}_{n3}^* & \dots  & \text{z}_{nm}^*
\end{bmatrix}. 
\end{equation*}  In this matrix, $\text{z}^*_{ij}$ represents the $j^{th}$ ppd sample of exposure for health data point $i$.  In terms of notation, it is helpful to define the exposure matrix in terms of row and column vectors such that \begin{equation*} \textbf{Z}^* = \begin{bmatrix}
    \textbf{z}_{1.}^{*\text{T}} \\
    \vdots \\
    \textbf{z}_{n.}^{*\text{T}}
\end{bmatrix} =
\begin{bmatrix}
    \textbf{z}_{.1}^* & \hdots & \textbf{z}_{.m}^*
\end{bmatrix}   
\end{equation*} where $\textbf{z}^*_{i.}$ represents the complete set of exposure ppd samples for data point $i$ and $\textbf{z}^*_{.j}$ is the vector of exposures for all $n$ data points from the $j^{\text{th}}$ ppd sample.  

Within a column of $\textbf{Z}^*$ the exposures for the different data points could be collected independently (i.e., from the marginal ppds) or jointly (i.e., from the joint ppd).  Sampling from the marginal ppds, which may be necessary due to computational considerations when working with a large spatial/temporal domain, results in independence across the rows of the $\textbf{Z}^*$, whereas sampling from the joint ppd retains the correlation across the rows.  In either case, trend in exposures across the rows may be present depending on the structure of the data points (e.g., air pollution concentrations across time).   We note that independence across the columns of $\textbf{Z}^*$ in practice is achieved through Monte Carlo sampling or MCMC sampling and thinning of the collected ppd samples.    

Once the exposure ppd samples are obtained from the first stage modeling, they are used in a subsequent epidemiological health outcome analysis to determine their association with the outcome.  This second stage health model typically follows a regression framework of the form \begin{align}\begin{split}
    &Y_i|\mu_i, \boldsymbol{\zeta} \stackrel{\text{ind}}{\sim} f\left(y|\mu_i,\boldsymbol{\zeta}\right),\ i=1,\hdots,n,\\
    &g\left(\mu_i\right) = \text{O}_i + \textbf{x}_i^{\text{T}}\boldsymbol{\beta} + \text{z}_i \theta
\end{split}\end{align} where $n$ is the previously defined number of data points; $Y_i$ represents the health outcome for data point $i$; $f(.|.)$ is the probability density function (pdf) of the outcome; $\mu_i$ is the mean of this distribution with $\boldsymbol{\zeta}$ representing additional parameters that often define variance/dispersion or auxiliary variables used to improved the efficiency of posterior sampling; $g\left(.\right)$ is a link function to connect the mean with a set of covariates; $\text{O}_i$ is an offset term sometimes used in the modeling of count data but will be zero otherwise; and $\textbf{x}_i$ is a vector of covariates unrelated to the primary exposure of interest, including an intercept term, with $\boldsymbol{\beta}$ the vector of corresponding regression parameters.  

The true but unobserved exposure for data point $i$ is denoted by $\text{z}_i$, where $\theta$ describes the association between exposure and outcome.  In this work, we assume that the set of true exposures, $\textbf{z}^{\text{T}} = \left(\text{z}_1, \hdots, \text{z}_n\right)$, follows the ppd derived in the first stage analysis.  However, we avoid the likely unrealistic assumptions made in some previous work that $\textbf{z}$ is included as one of the columns of $\textbf{Z}^*$ (i.e., $\textbf{z} \neq \textbf{z}^*_{.j}$ for any $j=1,\hdots,m$) (e.g., \cite{peng2010spatial, chang2011estimating}).  Instead, we treat entries of $\textbf{z}$ as unknown parameters in (1) with the first stage ppd representing our current state of knowledge about their values.  Therefore, $\textbf{z}$ is thought to arise from the same process that produced the columns of $\textbf{Z}^*$, but is not actually observed in the finite set of $m$ samples collected in the first stage.  To complete the model specification, we assign weakly informative prior distributions for each of the introduced parameters in (1) with specific settings based on the likelihood choice.

\subsection{Existing approaches}
Given $\textbf{Z}^*$ and the fully specified health model from (1), the question becomes how to efficiently utilize the information contained in the full set of ppd samples to accurately quantify uncertainty in the exposures when making inference on $\theta$.  A number of approaches, ranging in conceptual and computational complexity, have been proposed and we detail several of them below.  In Figure S1 of the Supplement, we present an overview of the different approaches.

\subsubsection*{Plug-in exposures}
The simplest approach used in previous work replaces $\text{z}_i$ from (1) with $\widehat{\text{z}}_i = T\left(\textbf{z}_{i.}^*\right)$ where $T\left(.\right)$ is a function of the input ppd samples (e.g., median, mean) (e.g., \cite{warren2021stillbirth}).  We refer to this as the \textit{Plug-in} approach.  Using only a summary measure of the ppd samples entirely ignores uncertainty in the exposures which may affect uncertainty estimation for $\theta$.  

\subsubsection*{Multiple imputation}
The multiple imputation (\textit{MI}) approach incorporates uncertainty in the exposures by fitting the health model in (1) separately for each of the $m$ columns of $\textbf{Z}^*$, replacing $\textbf{z}$ with $\textbf{z}^*_{.j}$ \citep{Blangiardo2016}.  During model fit $j$, $s$ posterior samples (post-convergence and possibly thinned) from $\theta$ are collected and denoted as $\boldsymbol{\theta}^{(j)\text{T}} = \left(\theta_1^{(j)}, \hdots, \theta_s^{(j)}\right)$.  After fitting the model to all columns of $\textbf{Z}^*$, posterior inference is conducted based on the combined samples across all $m$ model fits; $\boldsymbol{\theta}^{\text{T}} = \left(\boldsymbol{\theta}^{(1)\text{T}}, \hdots, \boldsymbol{\theta}^{(m)\text{T}}\right)$ \citep{zhou2010note}.  Depending on how long it takes to fit the health model in (1), which is impacted by the likelihood choice and sample size, \textit{MI} may be computationally demanding as $m$ increases.  This method also assumes that the columns of $\textbf{Z}^*$ resemble $\textbf{z}$, which is not necessarily true and depends on the amount of variability in the first stage ppd.  

\subsubsection*{Multiple imputation approximation}
The multiple imputation approximation (\textit{MIA}) approach approximates the results from \textit{MI} while only requiring a single fit of the health model in (1); representing a major computational improvement.  Specifically, during each iteration of the MCMC algorithm developed for the health model in (1), \textit{MIA} randomly selects a new column of exposures for the $n$ data points with replacement from $\textbf{Z}^*$ (i.e., $\textbf{z}^*_{.j}$) and completes a full sweep of the algorithm (i.e., collecting samples from all of the introduced model parameters) \citep{lee2017rigorous}.  Posterior inference for $\theta$ is made based on the $s$ MCMC samples collected from this algorithm.

\subsubsection*{Discrete uniform prior distribution}
Similar to \textit{MIA}, the discrete uniform (\textit{DU}) approach requires only a single fit of the health model in (1) and uses columns directly from $\textbf{Z}^*$ during model fitting.  However, instead of randomly selecting a column during each MCMC iteration, \textit{DU} incorporates the health data in the decision making, resulting in the selection of more probable columns during posterior sampling.  It does so by assigning a prior distribution to $\textbf{z}$ using the collected ppd samples in $\textbf{Z}^*$ and carrying out full Bayesian inference for the health model in (1).  Specifically, \textit{DU} makes the likely unrealistic assumption that the true vector of exposures is contained in $\textbf{Z}^*$ (i.e., $\textbf{z} = \textbf{z}^*_{.j}$ for some $j=1,\hdots,m$).  Based on this assumption, a prior distribution for $\textbf{z}$ is specified such that $$\text{P}\left(\textbf{z} = \textbf{z}_{.j}^*\right) = \frac{1}{m},\ j=1,\hdots,m$$ \citep{peng2010spatial, chang2011estimating}.  Use of \textit{DU} yields a semi-conjugate full conditional distribution for $\textbf{z}$ regardless of the choice for $f\left(.|.\right)$ in (1), allowing for convenient updating during MCMC sampling.  When $m$ becomes large, a more computationally efficient Metropolis algorithm can be used to propose/evaluate columns from $\textbf{Z}^*$ using a likelihood ratio calculation. 

\subsubsection*{Multivariate normal prior distribution}
Similar to \textit{DU}, the multivariate normal (\textit{MVN}) approach assigns a prior distribution to $\textbf{z}$ but avoids the assumption that $\textbf{z} = \textbf{z}^*_{.j}$ for some $j=1,\hdots,m$.  Specifically, a MVN prior distribution for $\textbf{z}$ is specified such that $$\textbf{z} \sim \text{MVN}\left(\widehat{\textbf{z}} = \begin{bmatrix}
    T\left(\textbf{z}_{1.}^*\right) \\
    \vdots \\
    T\left(\textbf{z}_{n.}^*\right)
\end{bmatrix},\ \widehat{\Sigma} = \frac{1}{m-1} \sum_{j=1}^m \left(\textbf{z}_{.j}^* - \widebar{\textbf{z}}^*\right)\left(\textbf{z}_{.j}^* - \widebar{\textbf{z}}^*\right)^{\text{T}}\right)$$ where $\widehat{\textbf{z}}$ and $T\left(.\right)$ have been previously described, and $\widebar{\textbf{z}}^* = \frac{1}{m} \sum_{j=1}^m \textbf{z}_{.j}^*$ is the length $n$ vector of average exposures across all $m$ ppd samples \citep{warren2012spatial, lee2017rigorous}.  

Updating $\textbf{z}$ within an MCMC algorithm is straightforward (i.e., the vector has a standard, closed-form full conditional distribution) for multiple likelihood choices that cover a number of relevant health outcome data types, including Gaussian with identity link function (continuous outcome), Bernoulli with logit link function (binary outcome), and negative binomial with logit link function (count data).  The latter two likelihood/link function results are made possible by the work of \cite{polson2013bayesian}.  Details for deriving this distribution are provided in Section S1 of the Supplement.  However, posterior sampling will be increasingly time consuming as $n$ increases given the large dimension of the full conditional distribution covariance matrix.  Additionally, \textit{MVN} may struggle when the shape of the exposure ppd deviates substantially from normality (e.g., skewness, multiple modes).  

\section{Kernel density estimation prior distributions}
We propose using univariate and multivariate KDE with a Gaussian kernel function to fully leverage the information contained in $\textbf{Z}^*$ when estimating the health model in (1), detail its intuitive connections with existing approaches, and consider its computational requirements.  We show that assigning prior distributions for $\textbf{z}$ based on KDE results in a more flexible hybrid between \textit{DU} and \textit{MVN}, allowing us to avoid potentially problematic assumptions made by existing methods without significantly increasing the computational burden.

For the univariate version of KDE with a Gaussian kernel function (\textit{UKDE}), the prior distributions for $\text{z}_i,\ i=1,\hdots,n,$ are specified independently as \begin{equation*}f\left(\text{z}_i\right) = \frac{1}{m} \sum_{j=1}^m \frac{1}{\sqrt{2 \pi h_i^2}} \exp\left\{-\frac{1}{2h_i^2} \left(\text{z}_i - \text{z}_{ij}^*\right)^2\right\}\end{equation*} where $h_i$ is the bandwidth variable that controls the level of smoothness of the density function and is estimated using standard approaches (e.g., \cite{sheather1991reliable}) based on the the ppd samples in $\textbf{z}_{i.}^*$.  This prior distribution represents a mixture of $m$ equally-weighted normal distributions centered at the observed samples from the ppd (i.e., $\text{z}_{ij}^*$) with standard deviation $h_i$.  This allows information from each individual ppd sample to be utilized when fitting the health model in (1) and thereby avoids relying on overly simplistic summaries of the samples used by existing methods (e.g., \textit{Plug-in}, \textit{MVN}).        

Because we select a Gaussian kernel function, the full conditional distribution for each $\text{z}_{i}$ has a closed form for the likelihood/link function combinations mentioned previously in Section 2, allowing for convenient updates within an MCMC algorithm.  Specifically, based on the health model in (1) the full conditional distribution for $\text{z}_i$ is a mixture of univariate normal distributions such that 
\begin{align}\begin{split}
&f\left(\text{z}_{i} | \boldsymbol{Y}, \boldsymbol{\beta}, \theta, \textbf{z}_{-i}, \boldsymbol{\zeta}\right) =\\ &\sum_{j=1}^m \left(\frac{c_{ij}}{\sum_{k=1}^m c_{ik}}\right) \frac{\sqrt{\theta^2 h_i^2 \Omega_{ii} + 1}}{h_i} \phi\left(\frac{\sqrt{\theta^2 h_i^2 \Omega_{ii} + 1}}{h_i} \left[\text{z}_i - \left\{\frac{\left(\widetilde{Y}_i - \text{O}_i - \textbf{x}_i^{\text{T}}\boldsymbol{\beta}\right) \theta h_i^2 \Omega_{ii} + \text{z}_{ij}^*}{\theta^2 h_i^2 \Omega_{ii} + 1}\right\}\right]\right)
\end{split}\end{align} 
where $\boldsymbol{Y}$ is the complete vector of $n$ data points; $\textbf{z}_{-i}$ is the complete vector of true exposures with $\text{z}_i$ removed; $\phi\left(.\right)$ is the pdf of the standard univariate normal distribution; $\Omega$ is an $n$ by $n$ diagonal matrix with \[\Omega_{ii} = \left\{\begin{array}{cl}
1/\sigma^2_{\epsilon}, &\text{Gaussian/identity}\\
\stackrel{\text{ind}}{\sim} \text{P\'olya-Gamma}\left(1, \text{O}_i + \textbf{x}_i^{\text{T}}\boldsymbol{\beta} + \text{z}_i \theta\right), &\text{Bernoulli/logit}\\
\stackrel{\text{ind}}{\sim} \text{P\'olya-Gamma}\left(r + Y_i, \text{O}_i + \textbf{x}_i^{\text{T}}\boldsymbol{\beta} + \text{z}_i\theta\right), &\text{Negative binomial/logit;}\end{array}\right.\] and \[\widetilde{Y}_{i} = \left\{\begin{array}{cl}
Y_i, &\text{Gaussian/identity}\\
\left(Y_i - 0.50\right)/\Omega_{ii}, &\text{Bernoulli/logit}\\
0.50\left(Y_i - r\right)/\Omega_{ii}, &\text{Negative binomial/logit.}\end{array}\right.\]  In the case of Gaussian distributed health outcome data, $\boldsymbol{\zeta}$ includes the error variance parameter, $\sigma^2_{\epsilon}$, and in the negative binomial case it includes the dispersion parameter, $r$.  The mixture weights in (2) are defined as \begin{equation}c_{ij} = \exp\left\{-\frac{1}{2}\left[\frac{\left(\widetilde{Y}_i - \text{O}_i - \textbf{x}_i^{\text{T}}\boldsymbol{\beta}\right)^2 h_i^2 \Omega_{ii} + \text{z}_{ij}^{*2}}{h_i^2} - \frac{\left\{\left(\widetilde{Y}_i - \text{O}_i - \textbf{x}_i^{\text{T}}\boldsymbol{\beta}\right) \theta h_i^2 \Omega_{ii} + \text{z}_{ij}^* \right\}^2}{h_i^2 \left(\theta^2 h_i^2 \Omega_{ii} + 1\right)}\right]\right\}\end{equation} where all remaining terms have been previously described.  Further details for these derivations are provided in Section S1 of the Supplement. 

The form of the full conditional distribution in (2-3) suggests that the use of \textit{UKDE} for specifying exposure prior distributions results in a hybrid approach, combining features of \textit{DU} with those of \textit{MVN}.  Specifically for updating $\text{z}_i$, we sample from this mixture distribution in two steps. \begin{enumerate}
    \item Choosing a probable observed exposure:  We compute $c_{ij}$ in (3), which depends on the health data and ppd exposure value $\text{z}^*_{ij}$, for every $j=1,\hdots,m$.  Each $c_{ij}/\sum_{k=1}^m c_{ik}$ is computed and a random $j$ index corresponding to $\text{z}^*_{ij}$ is selected based on these probabilities.
    \item Updating exposure based on selected value:  Based on the selected $\text{z}^*_{ij}$ ppd sample, the true exposure $\text{z}_i$ is drawn from the normal distribution in (2) whose mean depends on $\text{z}^*_{ij}$ and none of the other collected ppd samples.  
\end{enumerate}  The first step resembles \textit{DU} in that the health data are used to inform a probable ppd sample.  However, once this choice is made, \textit{DU} assigns it as the true exposure unlike \textit{UKDE}.  The second step resembles \textit{MVN} since the true exposure is simulated from a distribution and is therefore, not required to be one of the observed columns of $\textbf{Z}^*$.  However, with \textit{MVN} the prior mean of the exposures remains the same across all MCMC iterations for the corresponding full conditional distribution update (i.e., $T\left(\textbf{z}^*_{i.}\right)$).  For \textit{UKDE}, the mean in (2) changes each time a new $\text{z}^*_{ij}$ is selected.  

Therefore, \textit{UKDE} represents a more flexible alternative to \textit{DU} and \textit{MVN} as it avoids the assumption that the true exposures are observed in $\textbf{Z}^*$ and allows the prior mean of the exposures to effectively vary in the full conditional distribution update across MCMC iterations.  \textit{UKDE} should also be better able to handle non-symmetric ppd behavior than \textit{MVN}; although it does not directly account for correlation between the exposures corresponding to different data points.  In Section S2 of the Supplement, we extend the \textit{UKDE} framework to the multivariate setting (i.e., \textit{MKDE}) by specifying a prior distribution on the entire vector $\textbf{z}$, detail similar connections with existing approaches, and note some limitations of \textit{MKDE} due to the large dimension of the exposure data in most environmental health applications.

\section{Simulation study}
We design a simulation study to compare each of the methods detailed in Sections 2 and 3 with respect to estimating the association between exposure and health while accounting for exposure uncertainty.  

\subsection{Data generation}
We begin by defining the health model based on (1) and using a Gaussian likelihood with identity link function, no offset term, and no intercept/covariates such that \begin{equation}\label{eq:health_model_simstudy}
Y_i = \theta \text{z}_i + \epsilon_i,\ \epsilon_i \stackrel{\text{iid}}{\sim} \text{N}\left(0,1\right)\end{equation} for $i=1,\hdots,n = 250$, where all terms have been previously described in Section 2.  Two-dimensional spatial locations for the $250$ \textit{individuals} in the study are simulated randomly within the unit square.  We consider two different settings for the primary risk parameter, $\theta = 0$ and $\theta = 1$.  These settings allows us to investigate the type I error rate and power of the competing approaches, respectively.  

Next, we simulate exposure data for analysis.  In each setting we define a unique ppd and use it to create $\textbf{Z}^*$, which in this case represents a $250 \times 1,000$ matrix of exposures (i.e., $m=1,000$).  We define the true exposures by simulating an additional vector from the same ppd and assigning it to $\textbf{z}$.  The true exposures are used to simulate the health data from (4) and are not included in $\textbf{Z}^*$. 

When defining the ppds, we vary several factors to test the performances of the competing methods across different settings.  Specifically, we consider correlation between exposures (uncorrelated, correlated), skewness of the exposure distribution (symmetric, skewed), and variability of the means of the marginal exposure distributions (low, high).  Each column of $\textbf{Z}^*$ is simulated based on these settings such that \begin{align}\begin{split}\label{eq:sim_study_model} &\textbf{z}_{.j}^* = h\left(\textbf{w}_{.j}^*\right),\ j=1,\hdots,m=1000,\\ &\textbf{w}_{.j}^* \stackrel{\text{iid}}{\sim} \text{MVN}\left(\boldsymbol{\delta}, \Sigma\right), \text{ and }\\ &\delta_i|\tau^2 \stackrel{\text{iid}}{\sim}\text{N}\left(0,\tau^2\right), \ i=1,\hdots,n=250. \end{split}\end{align}  We standardize $\textbf{Z}^*$ by subtracting the mean of the entire matrix from each element of $\textbf{Z}^*$ and dividing these centered values by the standard deviation of the entire matrix. 

\begin{table}[h!]
\centering
\footnotesize
\caption{Simulation study settings for exposure posterior predictive distributions based on (5).}
\begin{tabular}{lll}
\hline
Factor                      & Setting      & Form                                                       \\
\hline
Associated  &  No  & $\theta = 0.00$ \\
            &  Yes & $\theta = 1.00$ \\
\hline
Correlated  & No        & $\Sigma_{ij} = 1\left(i = j\right)$                                      \\
                            & Yes   & $\Sigma_{ij} = \exp\left\{2\ln\left(0.05\right) ||\textbf{s}_i - \textbf{s}_j||\right\}$ \\
\hline
Skewed                      & No        & $h\left(x\right) = x$                                      \\
                            & Yes     & $h\left(x\right) = \exp\left\{x\right\}$                   \\
\hline
$\delta_i$ Variance           & Low     & $\tau^2 = 0.10$                                            \\
                            & High     & $\tau^2 = 1.00$                                            \\
\hline
\end{tabular}
\end{table}

For uncorrelated exposures, we define $\Sigma = I_{250}$ (i.e., the $250$ by $250$ identity matrix) while for correlated data we define $\Sigma_{ij} = \exp\left\{-\phi ||\textbf{s}_i - \textbf{s}_j||\right\}$, where $||.||$ represents the Euclidean distance between individuals $i$ and $j$, and $\phi = -2\ln\left(0.05\right)$ which allows for the correlation between two individuals to equal $0.05$ at a distance of $0.50$ (recall that individuals are simulated within the unit square where the maximum possible distance between two individuals is $\sqrt{2}$).  For symmetric data, $h\left(.\right)$ is defined as the identity function while $h\left(.\right) = \exp\left\{.\right\}$ for skewed data.  Finally, $\tau^2 = 0.10$ represents low variability in the means of the marginal ppds while $\tau^2 = 1.00$ represents high variability.  As $\tau^2$ increases, these means become further separated and we expect improved performance across all methods since there will be less relative uncertainty in the exposure distributions.  Full details on all considered settings are given in Table 1 while randomly selected simulated datasets from each setting are presented in Figure 1 and Figure S2 of the Supplement.  

\begin{figure}[h!]
\centering
\includegraphics[scale = 0.24]{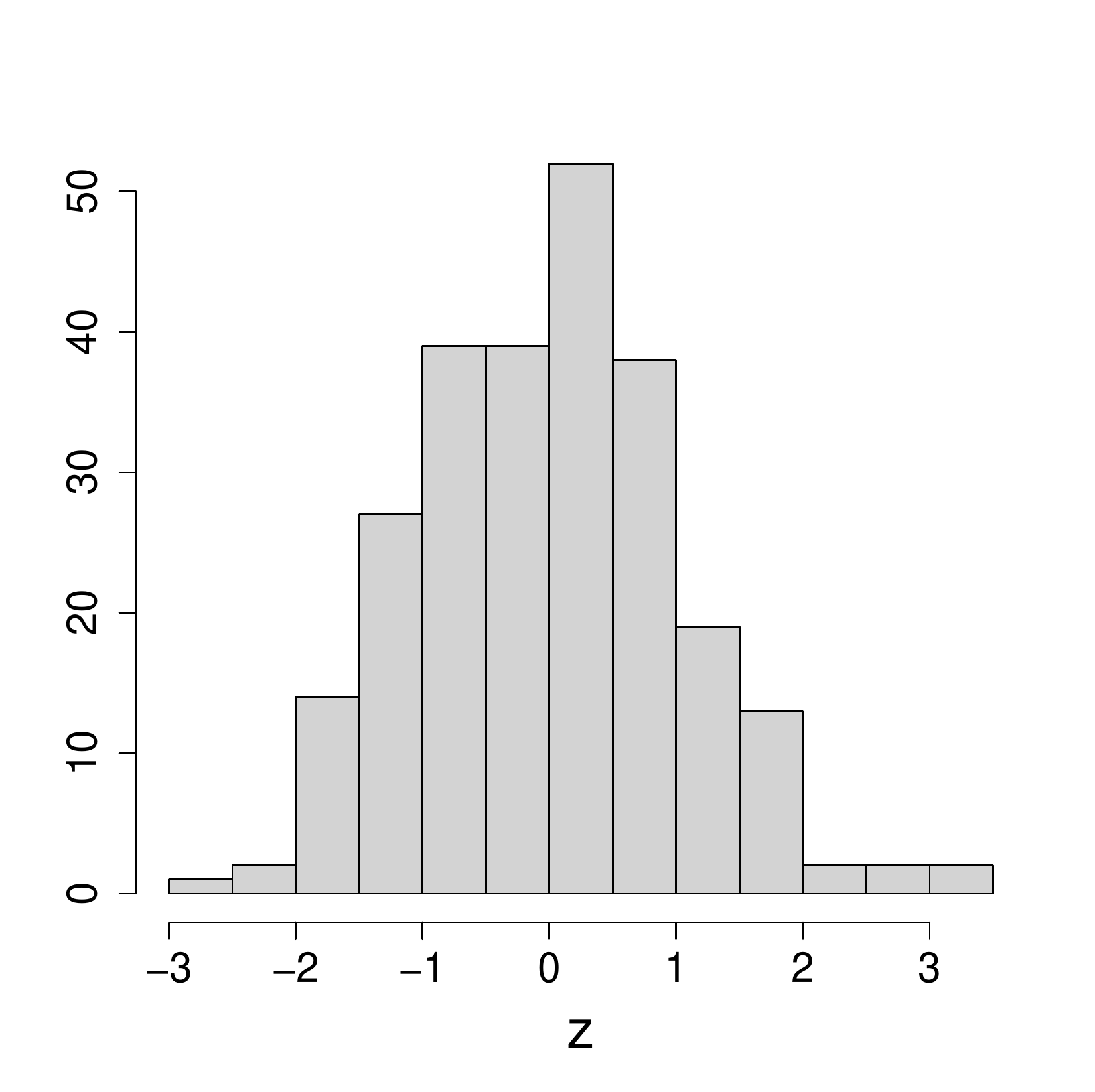}
\includegraphics[scale = 0.24]{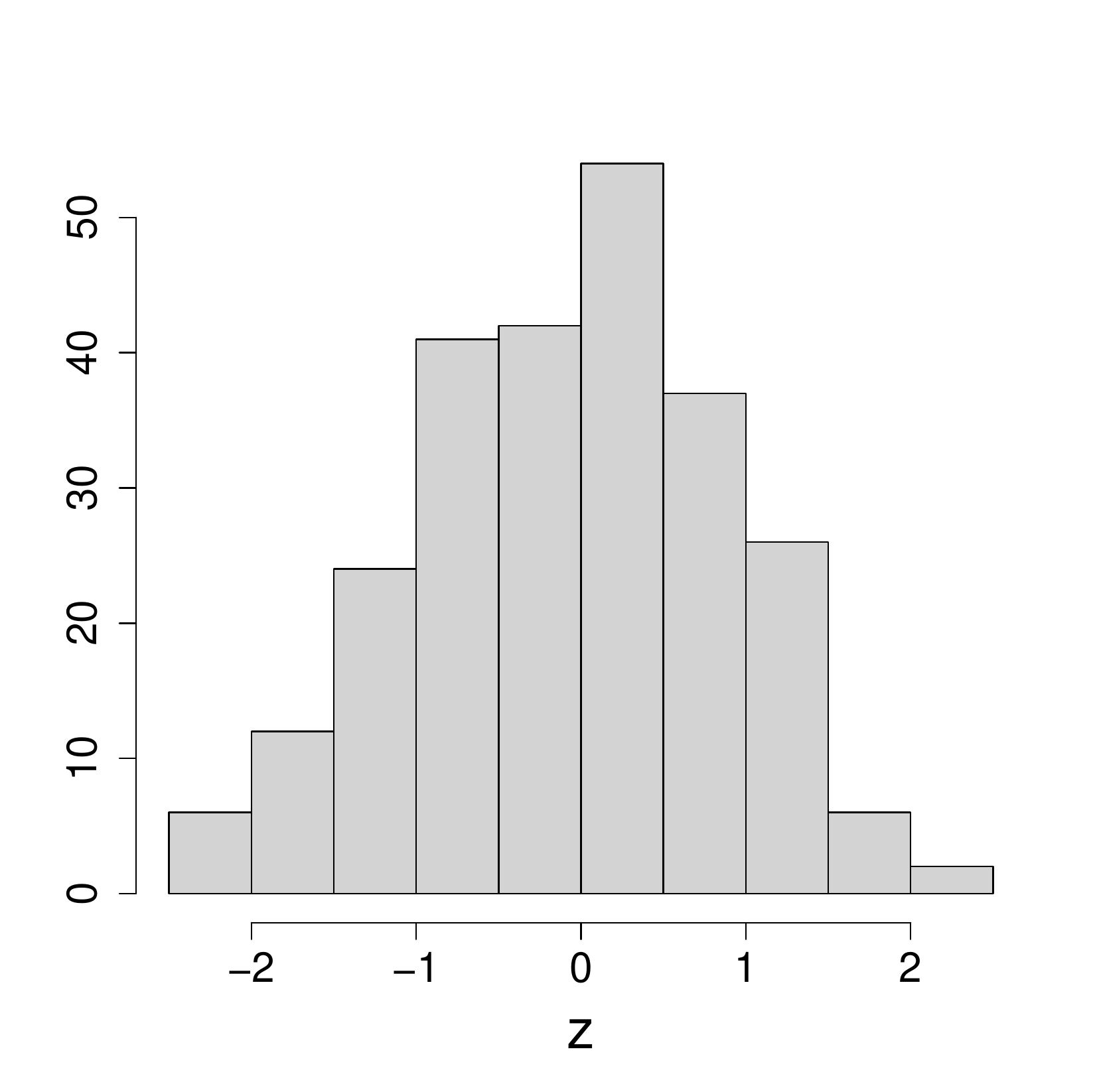}
\includegraphics[scale = 0.24]{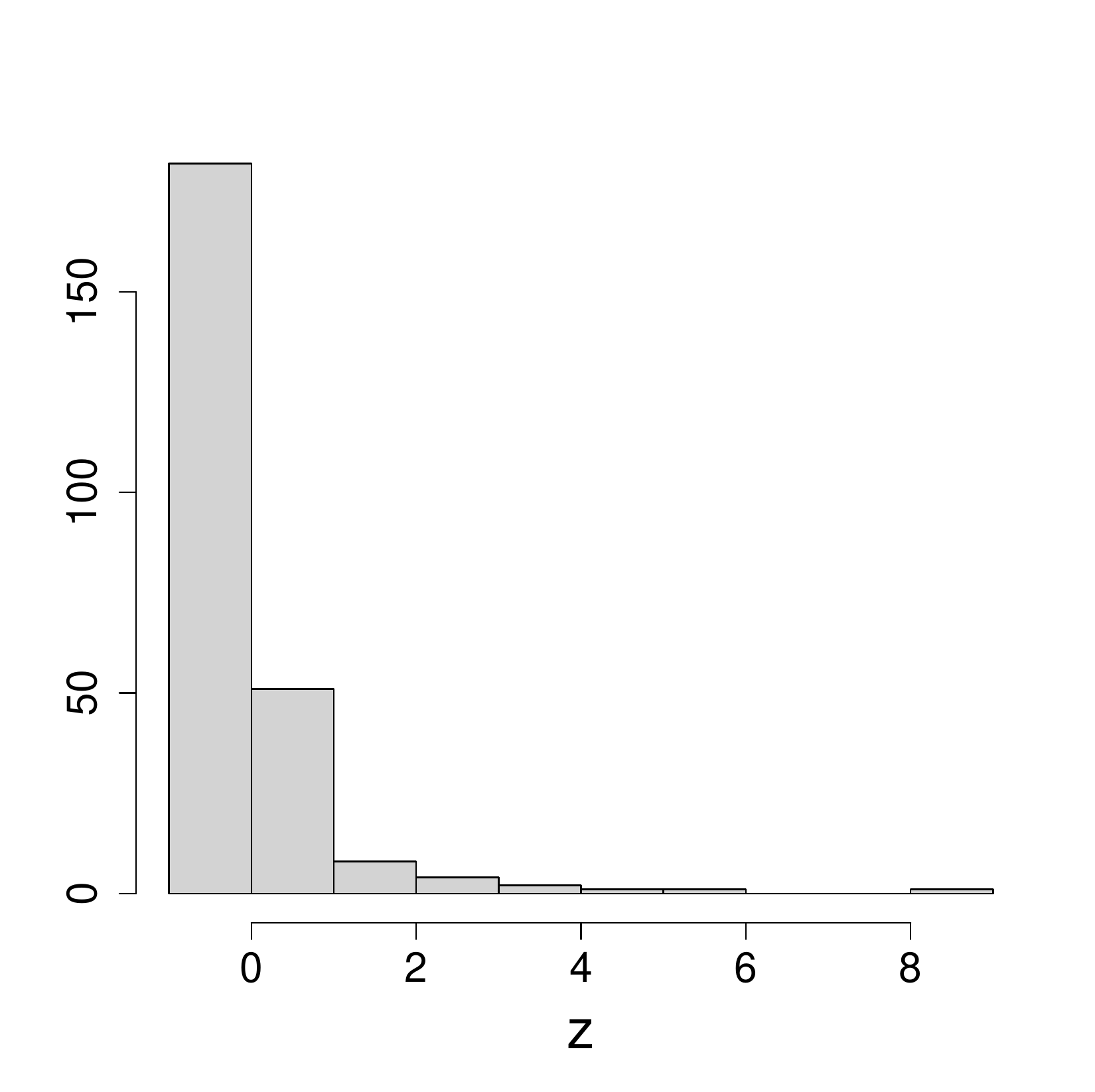}
\includegraphics[scale = 0.24]{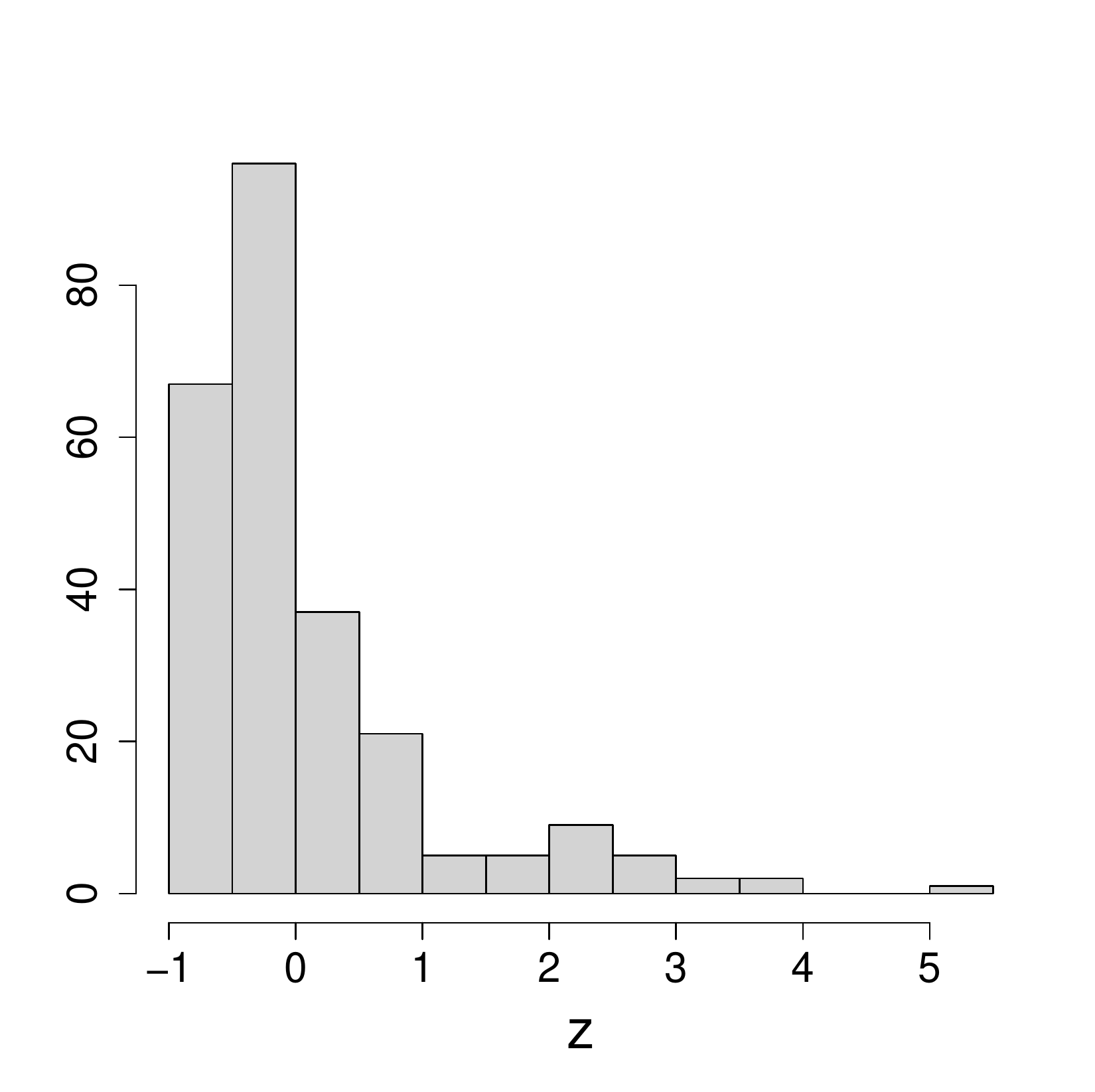}
\includegraphics[scale = 0.18]{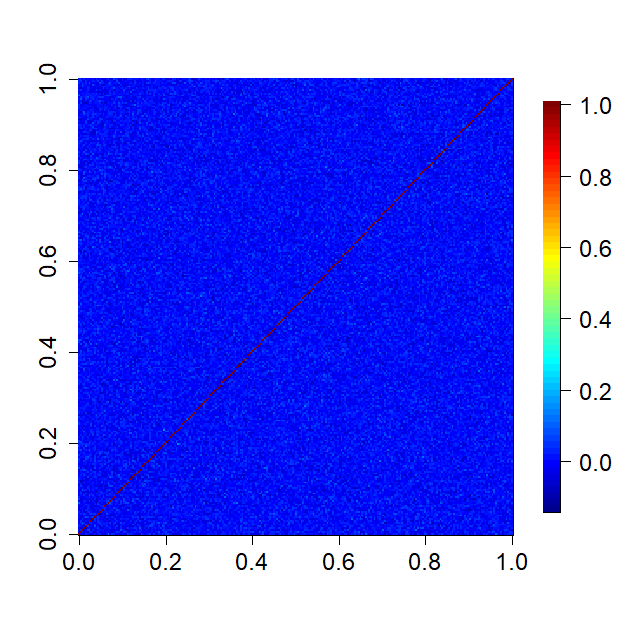}
\includegraphics[scale = 0.18]{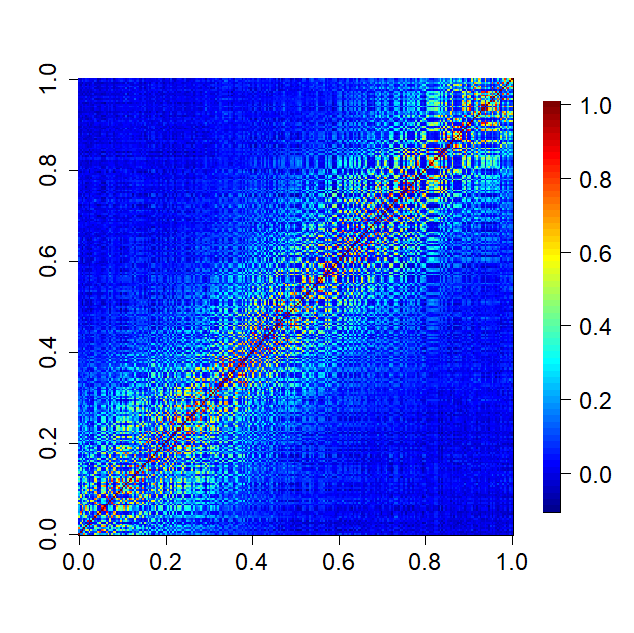}
\includegraphics[scale = 0.18]{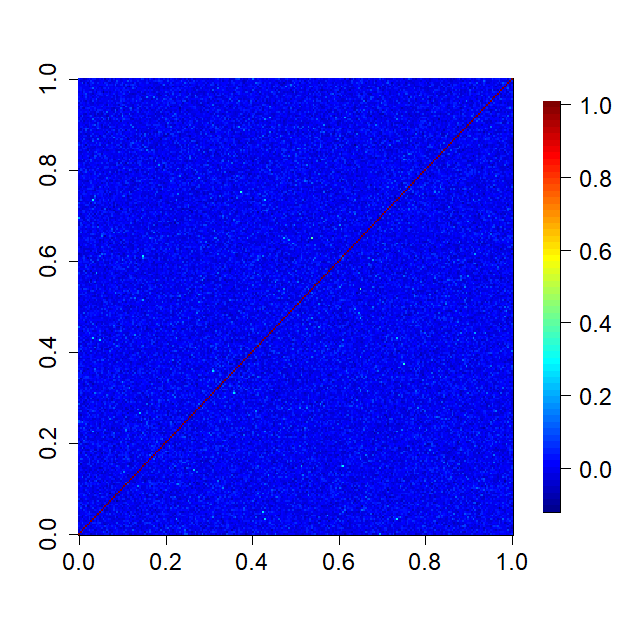}
\includegraphics[scale = 0.18]{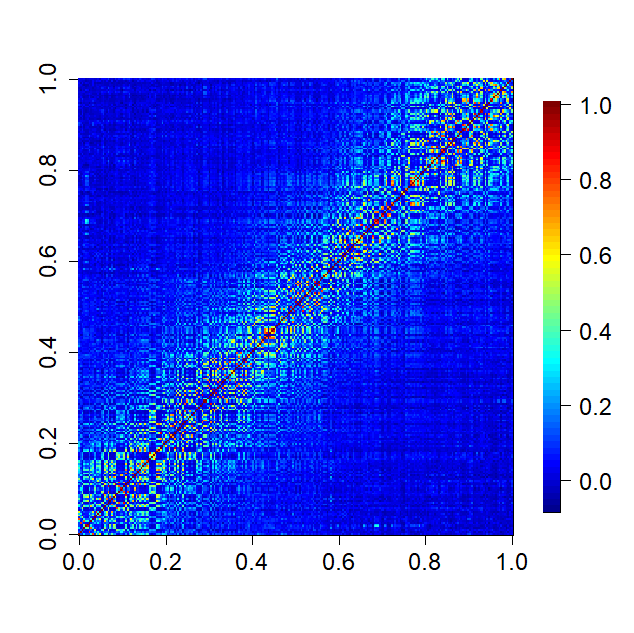}
\includegraphics[scale = 0.24]{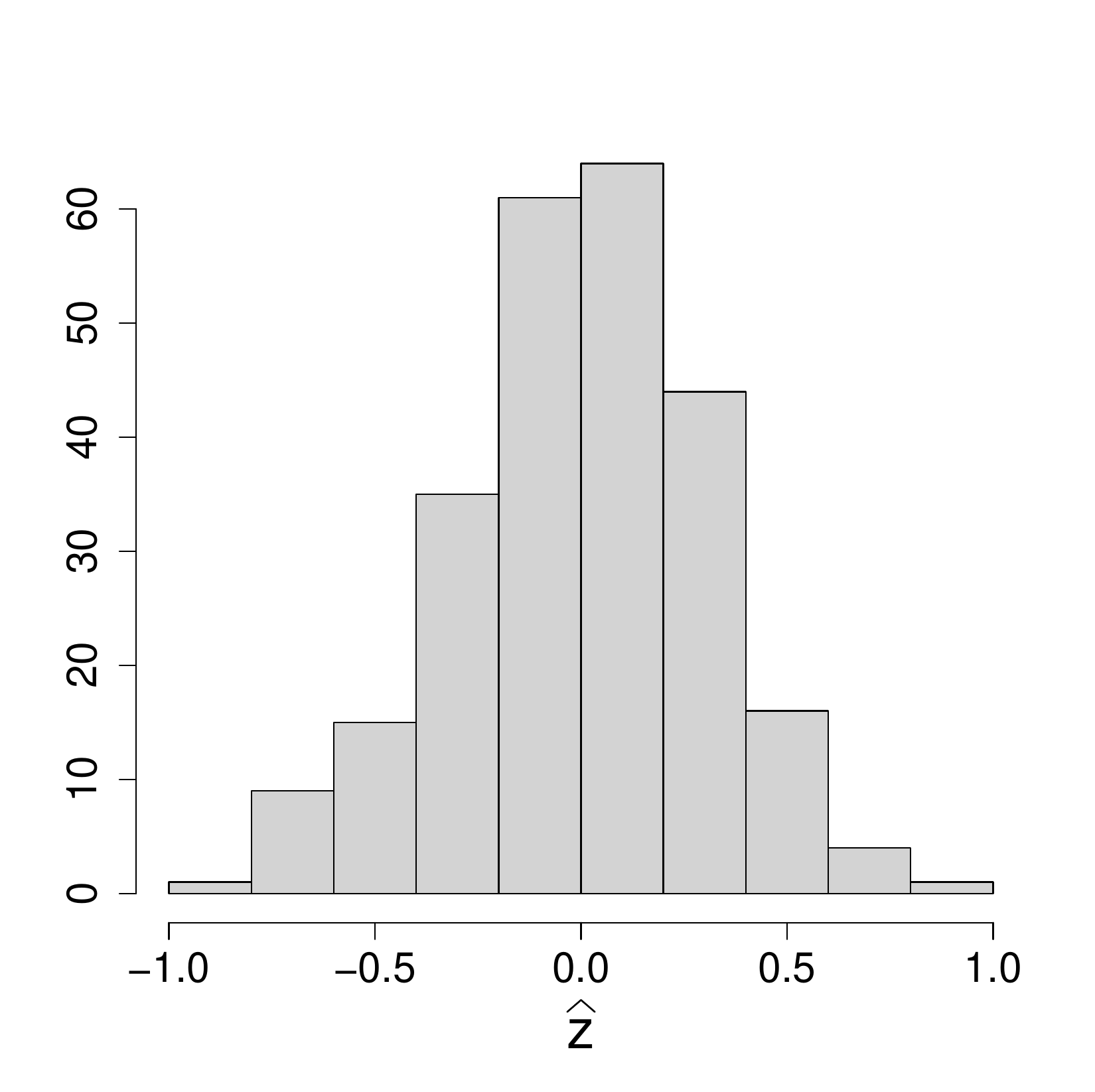}
\includegraphics[scale = 0.24]{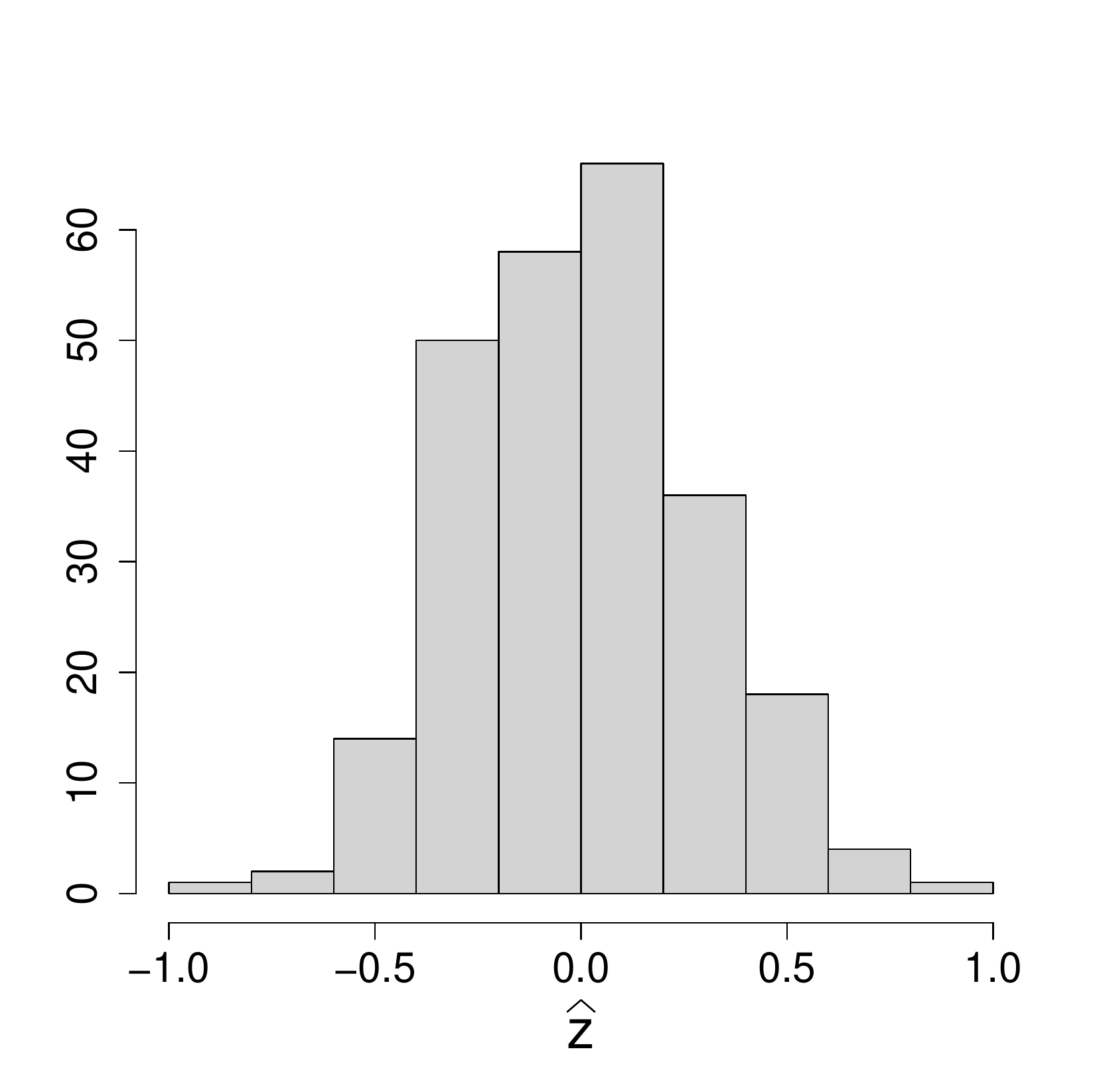}
\includegraphics[scale = 0.24]{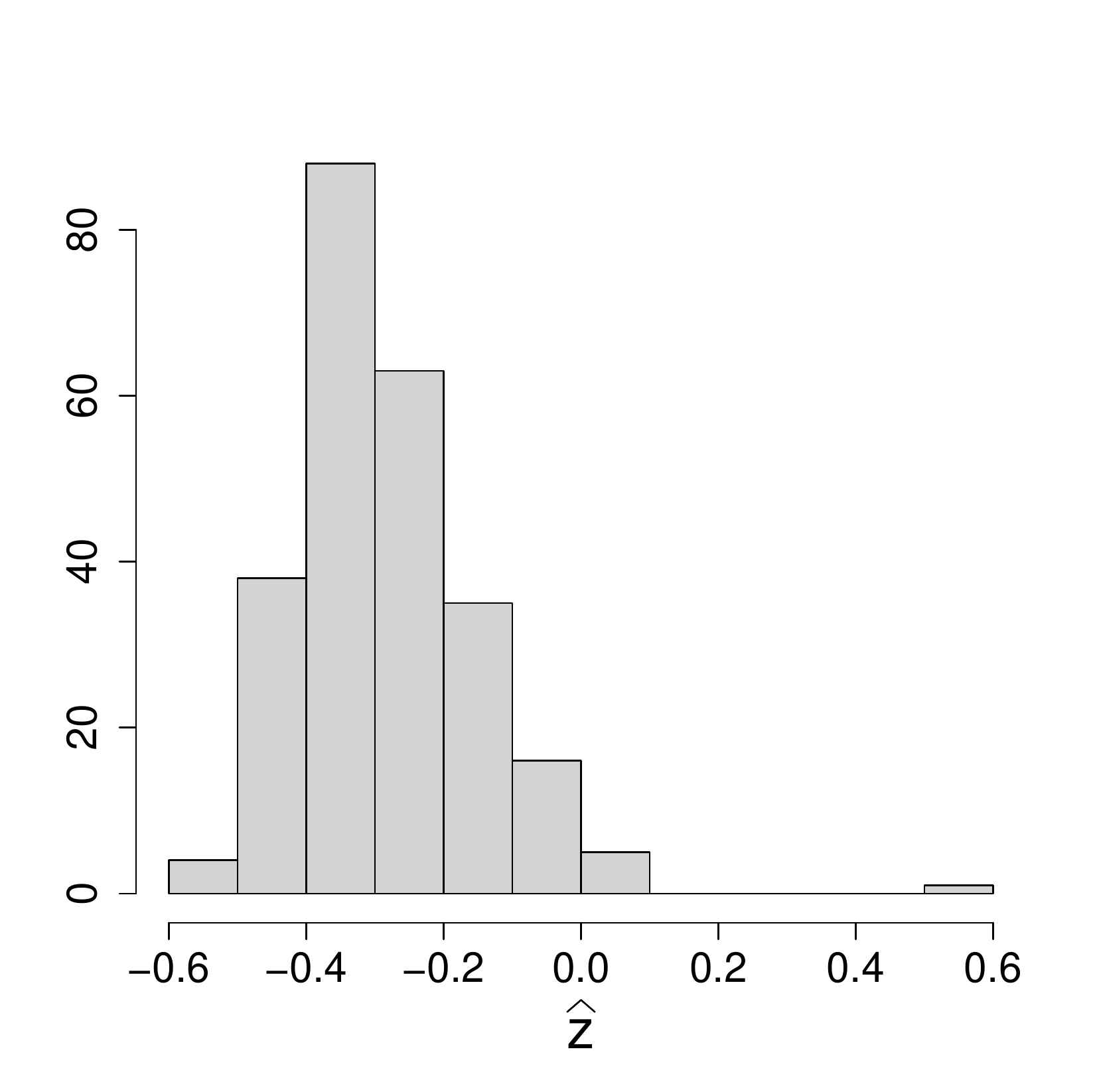}
\includegraphics[scale = 0.24]{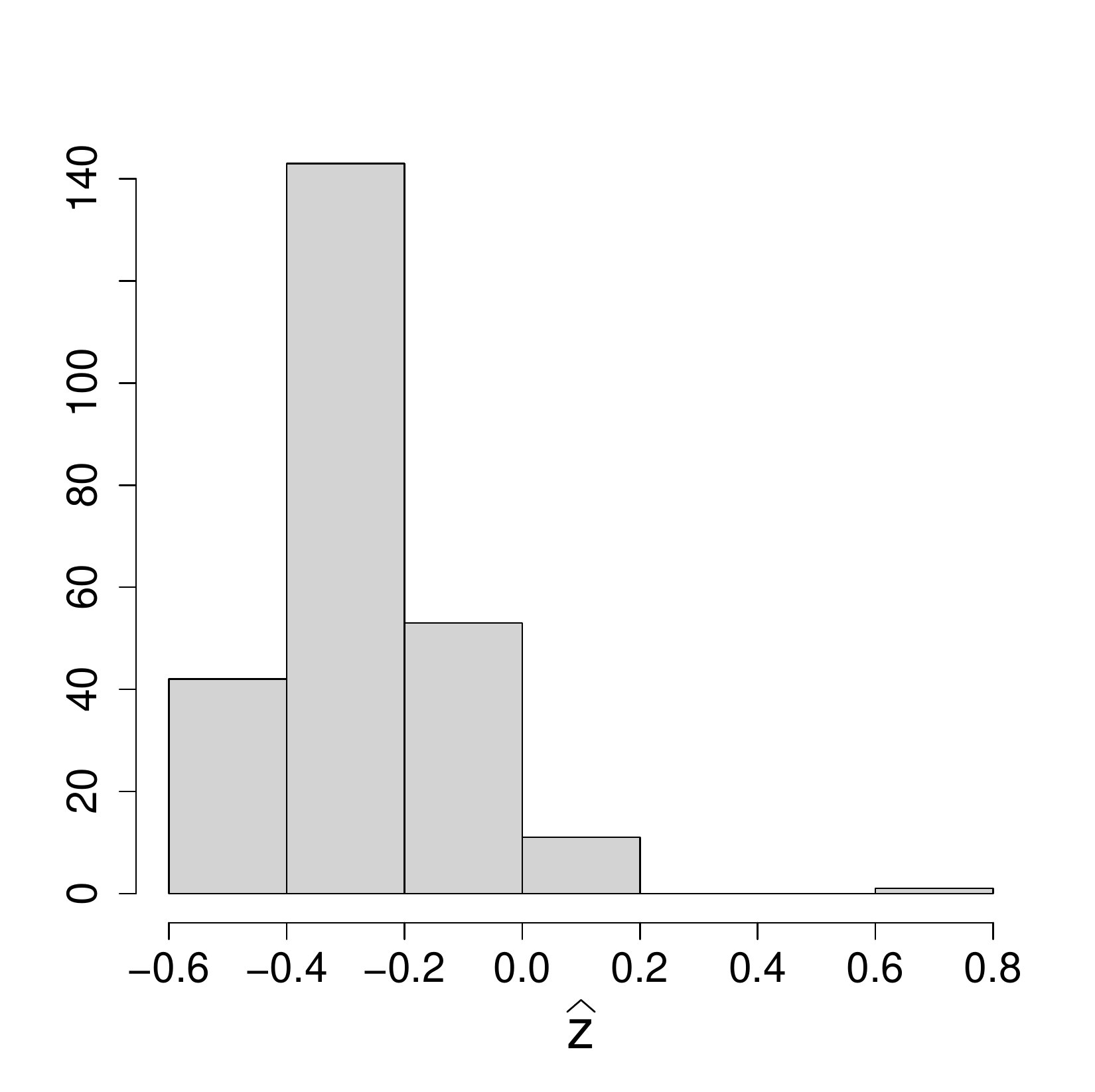}
\caption{Simulated datasets from each setting in Table 1 where $\tau^2 = 0.10$ (first column: uncorrelated, not skewed; second column: correlated, not skewed; third column: uncorrelated, skewed; fourth column: correlated, skewed).  The first row shows the histogram of true exposures (i.e., $\textbf{z}$), the second row shows the sample covariance matrix of the ppd samples (i.e., $\textbf{Z}^*$), and the third row shows the histogram of the median of each row of $\textbf{Z}^*$ (i.e., $\widehat{\text{z}}_i$).  Examples for $\tau^2 = 1.00$ are shown in Figure S2 of the Supplement.}
\end{figure}

For each simulated dataset, we generate a new set of spatial locations for the $250$ individuals and unique set of exposures (i.e., $\textbf{Z}^*$ and $\textbf{z}$).  For every combination of factors in Table 1, we simulate 500 datasets and analyze each one using the approaches described in Sections 2 and 3.  

\subsection{Data analysis}
The health model, which is shared across all methods, is given as \begin{equation}
Y_i = \beta_0 + \theta \text{z}_i + \epsilon_i,\ \epsilon_i|\sigma^2_{\epsilon} \stackrel{\text{iid}}{\sim}\text{N}\left(0, \sigma^2_{\epsilon}\right)\end{equation} with prior distributions $\sigma^2_{\epsilon} \sim \text{Inverse Gamma}\left(0.01, 0.01\right)$ (all methods), $\beta_0, \theta \sim \text{N}\left(0, 100^2\right)$ (\textit{MIA}, \textit{MVN}, \textit{DU}, \textit{UKDE}, \textit{MKDE}), and $\beta_0, \theta \sim \text{Flat}$ (\textit{Plug-in}, \textit{MI}).  The decision to use different prior distributions for the regression parameters across the methods is made for computational reasons only, as the flat priors allow for more efficient Monte Carlo posterior sampling for \textit{Plug-in} and \textit{MI}, saving considerable computing time.  We do not anticipate major changes in the results due to these differences given that neither distribution contains informative prior information about the parameters.  For \textit{UKDE} we use the method from \cite{sheather1991reliable} to estimate the bandwidth variables $h_i$, while for \textit{MKDE} we rely on Scott's rule \citep{scott2015multivariate} and the sample covariance of $\textbf{Z}^*$ to estimate the full bandwidth matrix variable $H$ (i.e., $H_{ij} = m^{-2/\left(n + 4\right)} \widehat{\Sigma}_{ij}$).  

From each method we collect $1,000$ posterior samples with which to make inference.  For the approaches where MCMC sampling is required, we run the algorithms for $11,000$ iterations, removing the first $1,000$ as a burn-in period, and thinning the remaining samples by a factor of $10$.  For the Monte Carlo sampling approaches, we directly obtain $1,000$ independent samples from the joint posterior distributions.  We estimate $\theta$ using the posterior mean and quantify uncertainty using the 95\% quantile-based, equal-tailed credible interval.   

We compare the ability of each method to estimate $\theta$ under the different simulation settings by estimating and comparing the bias and mean squared error (MSE) of the posterior mean, empirical coverage (EC) of the 95\% credible interval, and power/type I error rate.  For reference, we also calculate these quantities for the model in (6) where the true exposures are used - an analysis not possible in practice.

\subsection{Results}
In Figure 2 we display boxplots of the $\theta$ estimates across all 500 analyses for each method and correlation/skewness setting for the $\theta = 1.00$ and $\tau^2 = 0.10$ scenario.  Similar boxplots for the other $\theta$ and $\tau^2$ combinations are shown in Figures S3-S5 of the Supplement.  In Table 2 we present the simulation study results, also from the $\theta = 1.00$ and $\tau^2 = 0.10$ scenario.  Similar results are displayed in Tables S1-S3 for the other $\theta$ and $\tau^2$ combinations.  The $\theta = 1.00$, $\tau^2 = 0.10$ results suggest that \textit{MI} and \textit{MIA} perform very similarly overall as expected, and that both struggle to estimate $\theta$ well across all settings.  While \textit{DU} outperforms \textit{MI} and \textit{MIA}, it tends to have larger bias and MSE, and lower EC than some of the remaining approaches.  These methods may only be appropriate when the centers of the ppds are well separated with low uncertainty such that each column of $\textbf{Z}^*$ begins to resemble the true exposures $\textbf{z}$.   

\begin{figure}[ht]
\centering
\includegraphics[trim={0.5cm 0cm 1.5cm 0cm}, clip, scale = 0.27]{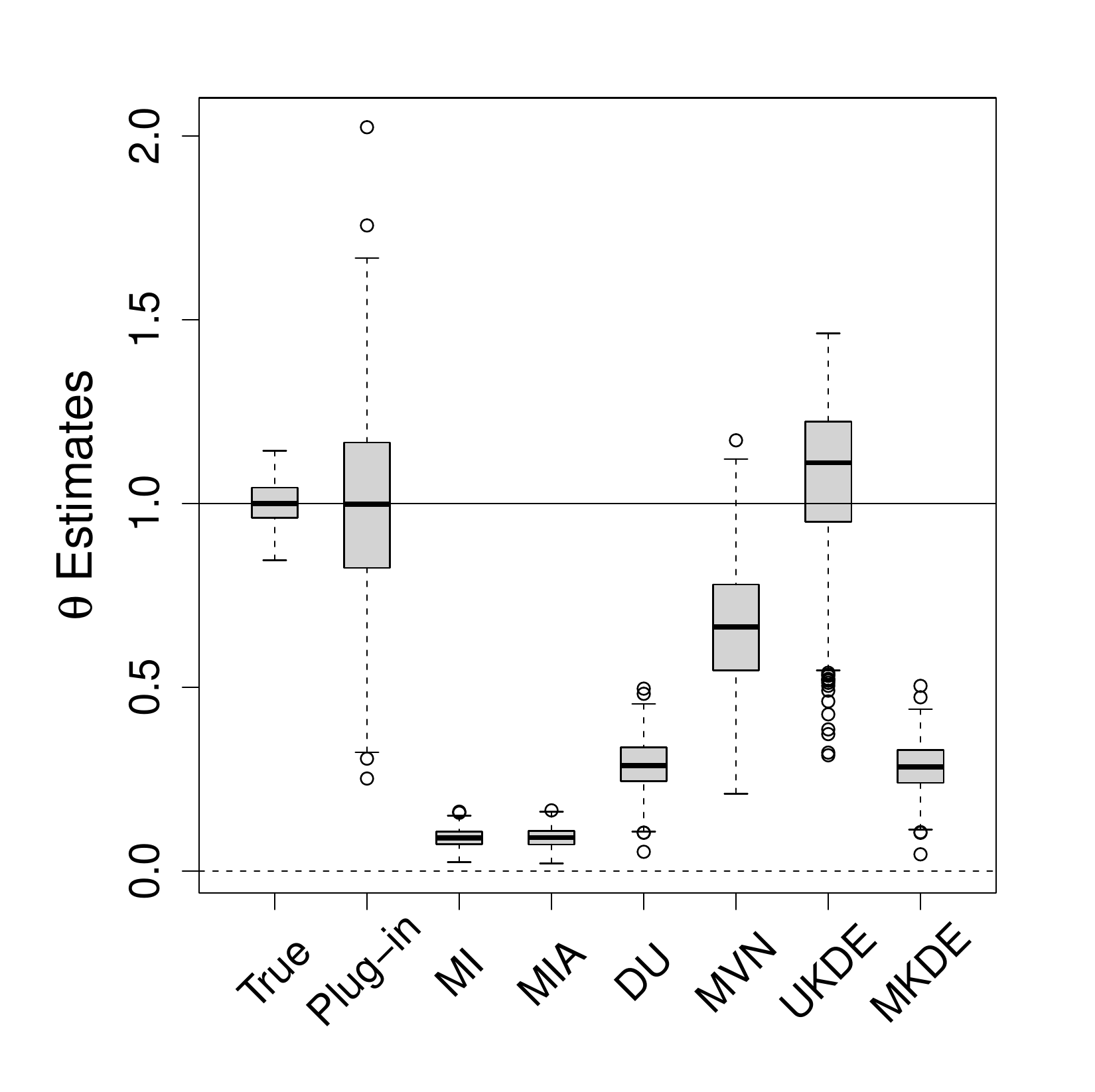}
\includegraphics[trim={0.5cm 0cm 1.5cm 0cm}, clip, scale = 0.27]{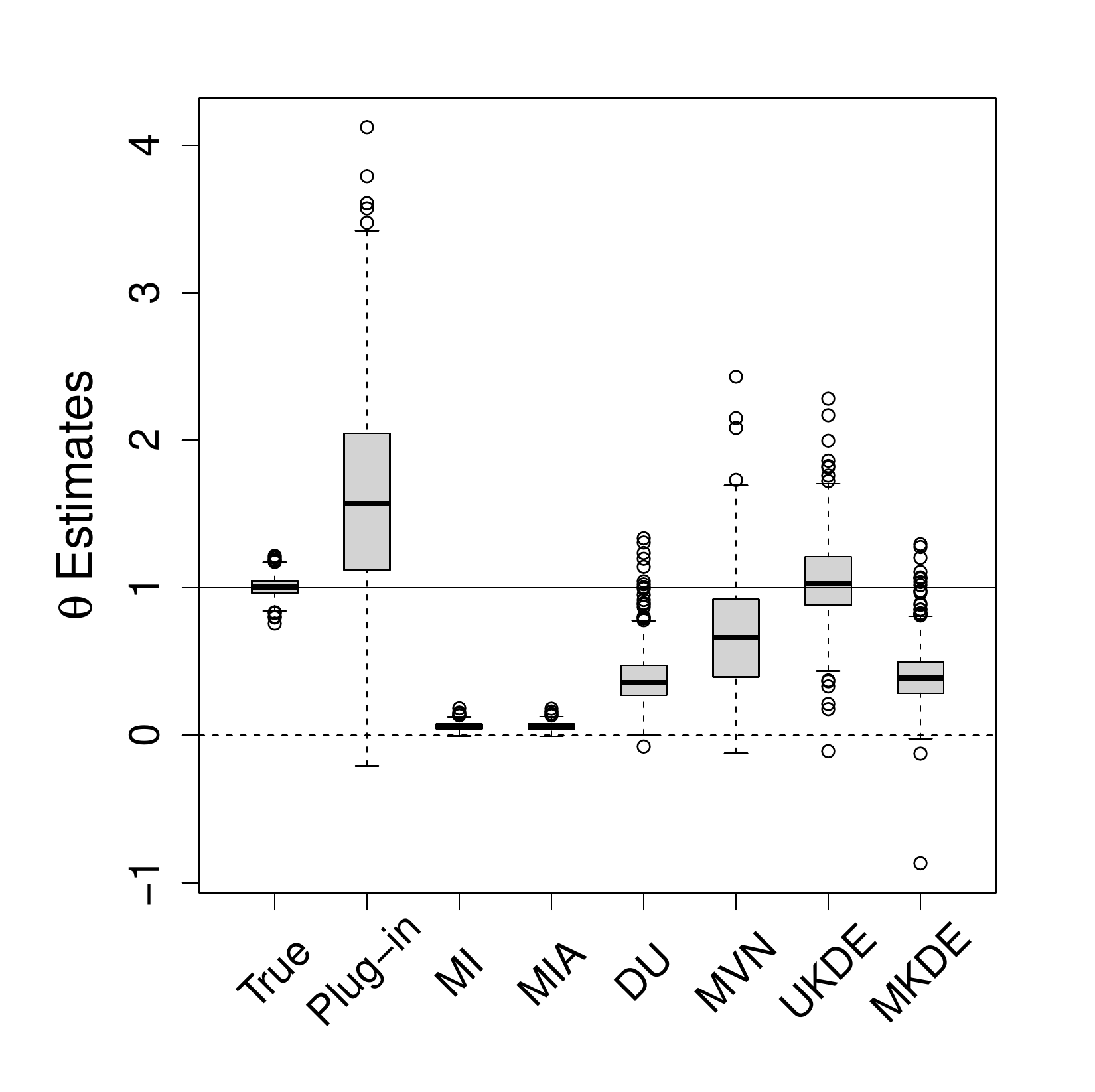}
\includegraphics[trim={0.5cm 0cm 1.5cm 0cm}, clip, scale = 0.27]{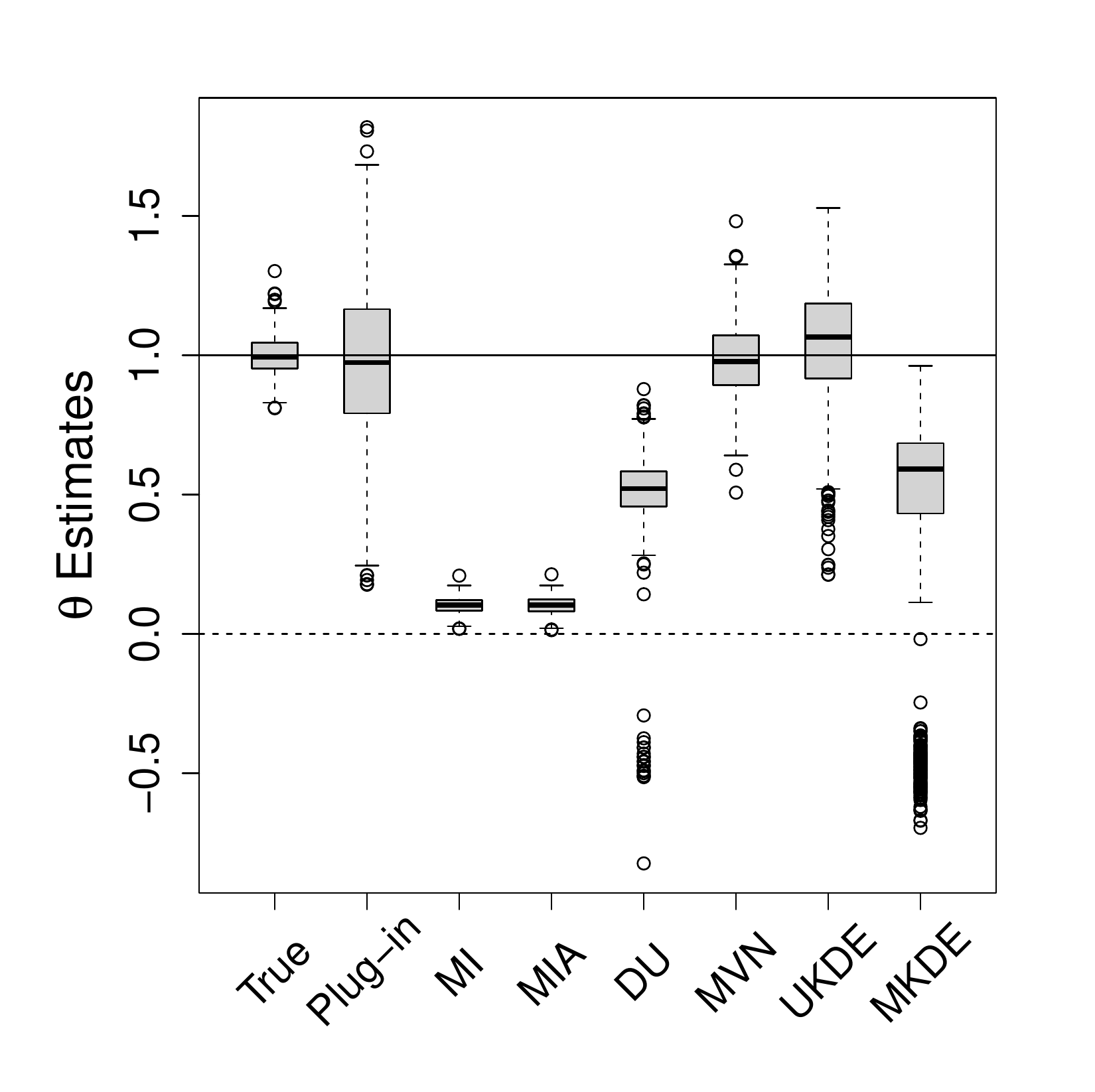}
\includegraphics[trim={0.5cm 0cm 1.5cm 0cm}, clip, scale = 0.27]{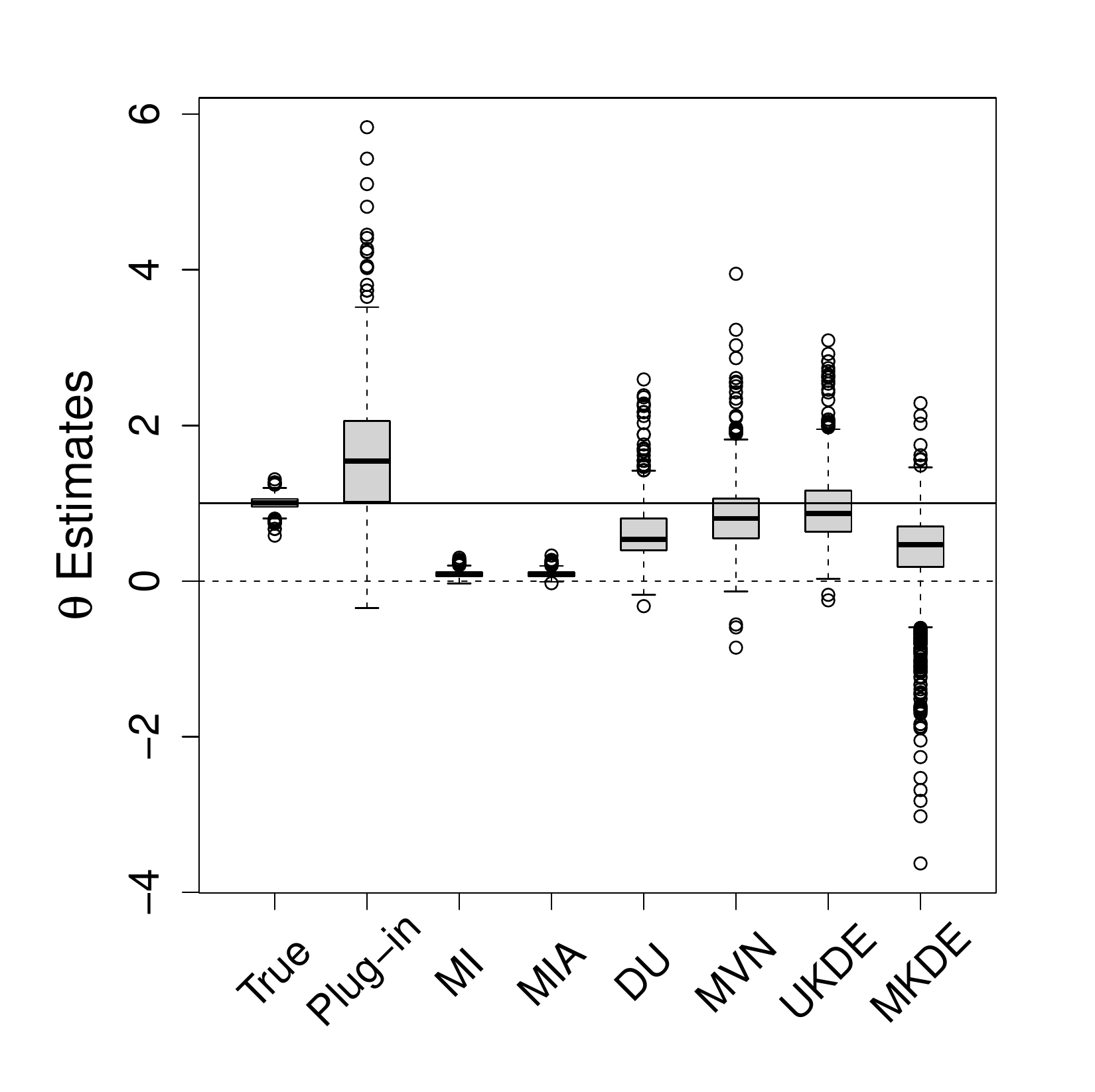}
\caption{Posterior mean estimates of $\theta$ for each method and correlation/skewness setting across all 500 analyses for the $\theta=1.00$ and $\tau^2 = 0.01$ scenario (first panel: uncorrelated, not skewed; second panel: uncorrelated, skewed; third panel: correlated, not skewed; fourth panel: correlated, skewed).  The solid horizontal line represents the true value of $\theta$.}
\end{figure}

When the ppds are not skewed, \textit{Plug-in} generally performs well overall which is an encouraging sign for past studies that have implemented this approach.  However, \textit{Plug-in} struggles greatly when the ppds become skewed, leading to elevated bias and MSE.  \textit{MVN} generally handles skewness better than \textit{Plug-in} and has improved performance when the ppds are correlated.  However, it tends to have lower EC, particularly when the ppds are independent and/or skewed.  

\textit{MKDE} performs similarly to \textit{DU} across most metrics.  It likely struggles due to the difficulty in selecting the bandwidth matrix variable $H$ when the dimension of the data is large.  Overall, \textit{UKDE} has the best balance of performance compared to the other methods, especially when the ppds are skewed.  For $\tau^2 = 1.00$, several of the methods perform more similarly to each other and we anticipate that this trend will continue as the ppds become further separated and the relative error is effectively reduced.  

When $\theta=0$, we see that all methods perform as expected while \textit{MI}, \textit{MIA}, \textit{DU}, and \textit{MDKE} appear to be preferred in terms of bias and MSE primarily because they produce estimates of $\theta$ near zero regardless of its true value.  Therefore, none of the considered methods are falsely identifying associations at a higher rate than expected; an important conclusion for previous studies that found evidence of significant associations between health and exposure using these techniques.  

\begin{landscape}
\begin{table}[h!]
\centering
\small
\caption{Simulation study results for $\theta = 1.00$ and $\tau^2 = 0.10$.  Estimates are presented with the range of standard errors for a group of estimates provided in parentheses.  Bold entries represent the optimal estimate across an entire row (i.e., closest to zero for bias, smallest for mean squared error (MSE), closest to 95 for empirical coverage (EC), and largest for power).  All results are multiplied by 100 for presentation purposes.}
\begin{tabular}{llllrrrrrrrr}
\hline
 & \multicolumn{2}{c}{Settings} & & \multicolumn{8}{c}{Methods} \\
\cline{2-3} \cline{5-12}
Metric & Corr       & Skew   & & True        & Plug-in        & MI          & MIA         & DU          & MVN             & UKDE           & MKDE         \\
\hline
Bias   & No         & No     & &   0.15      & \textbf{-1.03} & -90.95      & -90.96      & -71.26      & -33.68          &  5.71          & -71.65       \\
       & No         & Yes    & &   0.30      &  60.77         & -94.07      & -94.07      & -61.23      & -30.73          & \textbf{4.59}  & -59.06       \\
       & Yes        & No     & &  -0.05      &  -2.58         & -89.66      & -89.64      & -50.49      & \textbf{-2.25}  &  2.69          & -61.78       \\
       & Yes        & Yes    & &   0.00      &  62.14         & -90.83      & -90.83      & -35.10      & -15.02          & \textbf{-4.46} & -73.23       \\
       &            &        & & (0.27-0.40) & (1.21-3.24)    & (0.11-0.21) & (0.12-0.21) & (0.31-1.80) & (0.61-2.33)     & (0.99-2.25)    & (0.31-3.57)  \\
\hline
MSE    & No         & No     & &   0.36      &   7.29         &  82.79      &  82.80      &  51.25      &  14.17          & \textbf{5.22}  &  51.80       \\
       & No         & Yes    & &   0.46      &  89.38         &  88.56      &  88.57      &  40.90      &  24.01          & \textbf{7.89}  &  38.99       \\
       & Yes        & No     & &   0.47      &   7.78         &  80.48      &  80.45      &  30.14      & \textbf{1.89}   &  5.29          &  60.53       \\
       & Yes        & Yes    & &   0.82      & 116.24         &  82.72      &  82.72      &  28.44      &  29.36          & \textbf{25.41} & 117.26       \\
       &            &        & & (0.02-0.07) & (0.47-10.62)   & (0.21-0.37) & (0.21-0.37) & (0.44-1.29) & (0.55-2.83)     & (0.29-2.28)    & (0.44-10.38) \\

\hline
EC     & No         & No     & &  96.00      & \textbf{97.00} &   0.00      &   0.00      &   0.00      &  58.80          & \textbf{93.00} &   0.00       \\
       & No         & Yes    & &  94.80      & 80.40          &   0.00      &   0.00      &   2.40      &  49.40          & \textbf{86.00} &   4.00       \\
       & Yes        & No     & &  94.40      & \textbf{95.00} &   0.00      &   0.00      &   1.00      &  92.80          & 93.80          &   3.80       \\
       & Yes        & Yes    & &  95.00      & \textbf{81.60} &   8.00      &   8.20      &  30.60      &  46.00          & 64.80          &  19.80       \\
       &            &        & & (0.88-1.03) & (0.76-1.78)    & (0.00-1.21) & (0.00-1.23) & (0.00-2.06) & (1.16-2.24)     & (1.08-2.14)    & (0.00-1.76)  \\
\hline
Power  & No         & No     & & 100.00      & 92.60          &   0.00      &   0.00      &  54.80      &  88.40          & \textbf{94.80} &  52.40       \\
       & No         & Yes    & & 100.00      & 70.80          &   0.00      &   0.00      &  67.20      &  54.60          & \textbf{98.20} &  70.00       \\
       & Yes        & No     & & 100.00      & 93.60          &   0.00      &   0.00      &  93.00      & \textbf{99.80}  & 94.60          &  75.20       \\
       & Yes        & Yes    & & 100.00      & 70.20          &   0.00      &   0.00      &  85.60      &  78.80          & \textbf{89.00} &  63.60       \\
       &            &        & & (0.00-0.00) & (1.09-2.05)    & (0.00-0.00) & (0.00-0.00) & (1.14-2.23) & (0.20-2.23)     & (0.59-1.40)    & (1.93-2.23)  \\
\hline
\end{tabular}
\end{table}
\end{landscape}

\section{PM$_{2.5}$ and stillbirth in New Jersey}
Stillbirth is generally defined as the loss of a fetus or baby before or during a delivery that occurs on or after 20 completed weeks of gestation \citep{cdc}.  Recent literature reviews suggest that exposure to ambient air pollution during pregnancy may be associated with increased risk of stillbirth \citep{Bekkar2020, zhang2021ambient}, though further studies are needed to better understand this relationship.  In this application, we examine the relationship between acute exposure to PM$_{2.5}$ in the days prior to delivery and risk of stillbirth using a population-level time series approach in NJ, 2011-2015.  We also investigate the impact of the different methods for accounting for exposure uncertainty in the health analysis, detailed in Sections 2 and 3, on the findings.     

\subsection{Data description}
We obtain daily counts of fetal deaths and live birth across the three NJ counties located in the New York (NY)-White Plains-Wayne, NY-NJ Metropolitan Division (i.e., Bergen, Hudson, and Passaic) between 2011 and 2015 from the Division of Family Health Services in the NJ Department of Health.  Similar to \cite{warren2021stillbirth}, we only include singleton deaths/births and those with a clinically estimated gestational age of $\geq$ 20 weeks.  A fetal death occurring on or after 20 weeks of gestation is defined as a stillbirth.  In Figure 3, we display the study area as well as the proportion of stillbirths across time for this area.      

\begin{figure}[h!]
\centering
\includegraphics[scale = 0.48]{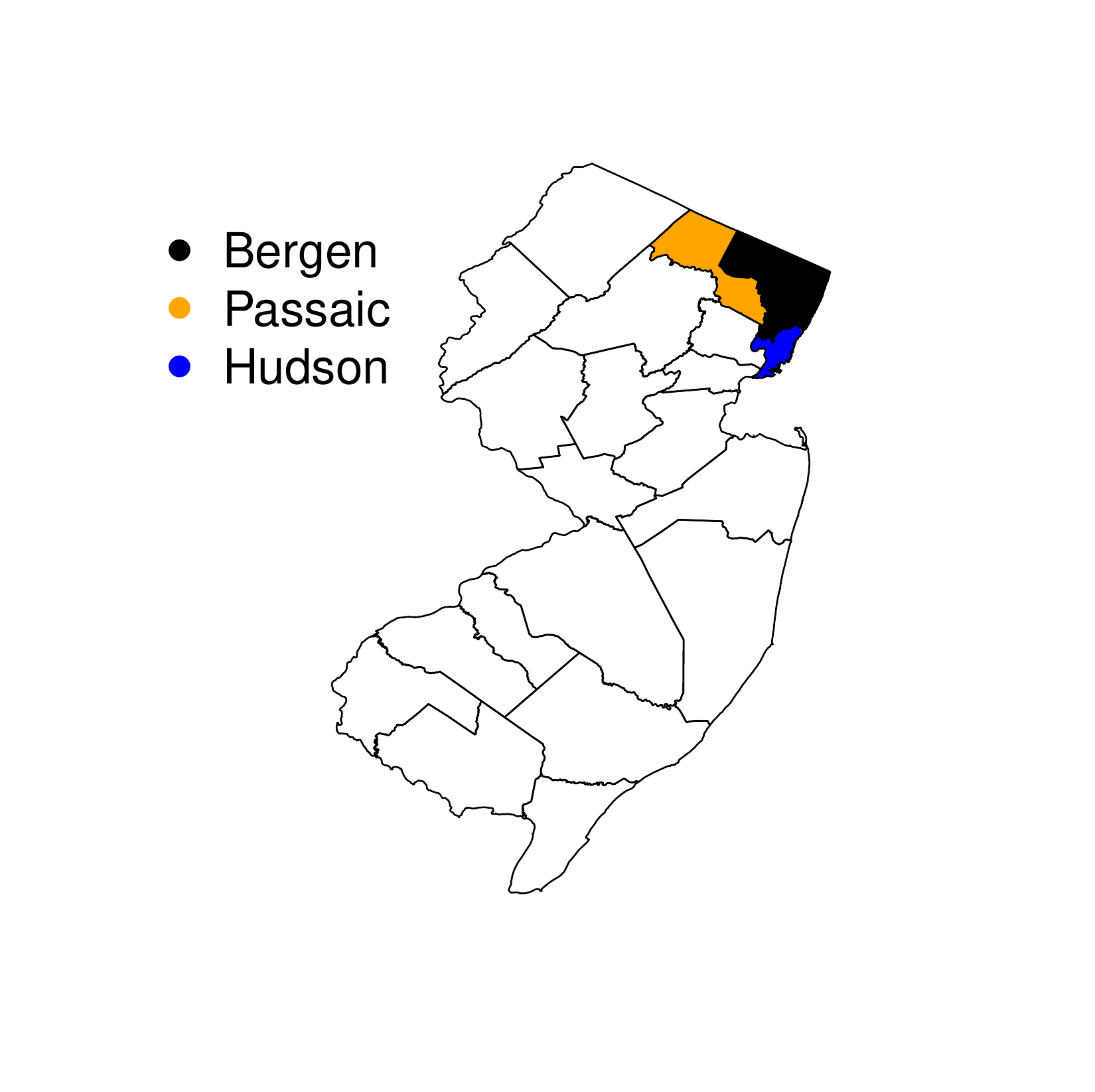}
\includegraphics[scale = 0.48]{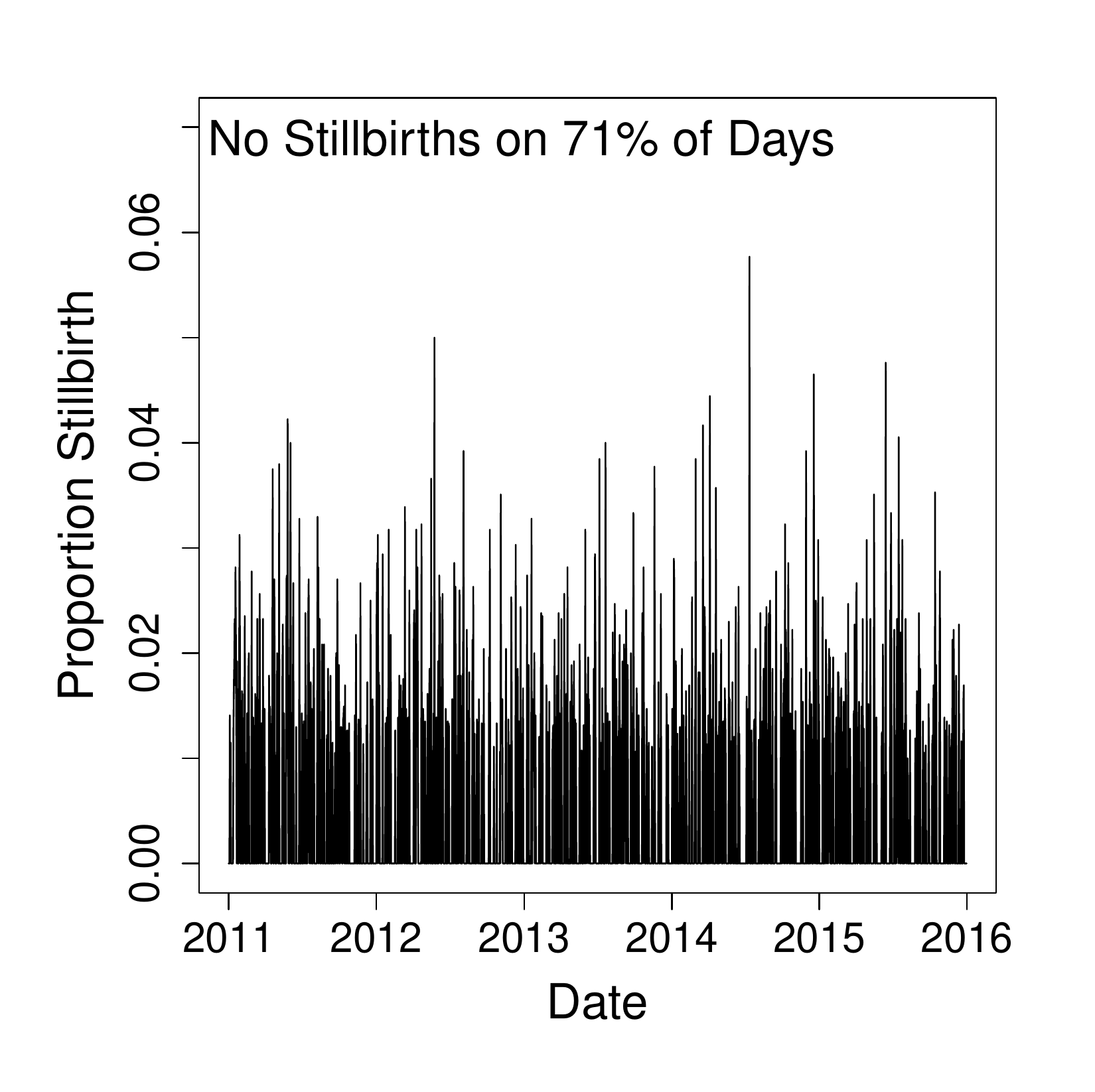}
\caption{The three New Jersey counties included in the study (left) and the daily proportion of births that resulted in stillbirth in those counties in 2011-2015 (right).}
\end{figure}

Observed air pollution data and model-derived estimates are both obtained from the United States Environmental Protection Agency (US EPA).  Specifically, for each day in 2002-2015 we access 24-hour average PM$_{2.5}$ concentrations (micrograms per cubic meter ($\mu$g/m$^3$)) measured from all active monitors located in NJ, NY, Delaware, and Pennsylvania from the US EPA's Air Quality System (AQS) \citep{AQS}.  Model-derived daily estimates of 24-hour average PM$_{2.5}$ concentrations ($\mu$g/m$^3$) from the Community Multiscale Air Quality (CMAQ) model, a deterministic numerical air quality model, are obtained on a 12 by 12 kilometer grid across the same study area and time period \citep{CMAQ}.  All data and estimates were downloaded from the US EPA's Remote Sensing Information Gateway website \citep{Fused}.  Figure 4 displays the locations of the AQS and CMAQ data from 2002-2015.

We also obtain daily estimates of the minimum and maximum temperatures at a one kilometer resolution across the three NJ counties between 2011-2015 from Daymet \citep{Daymet}.  The Daymet framework employs statistical modeling techniques to produce spatially-temporally interpolated temperature estimates using observed ground-based data as input.  
On each study day, we average the estimates within the three counties to obtain a single daily average minimum/maximum temperature estimate for the region. 

\begin{figure}[h!]
\centering
\includegraphics[scale = 0.32]{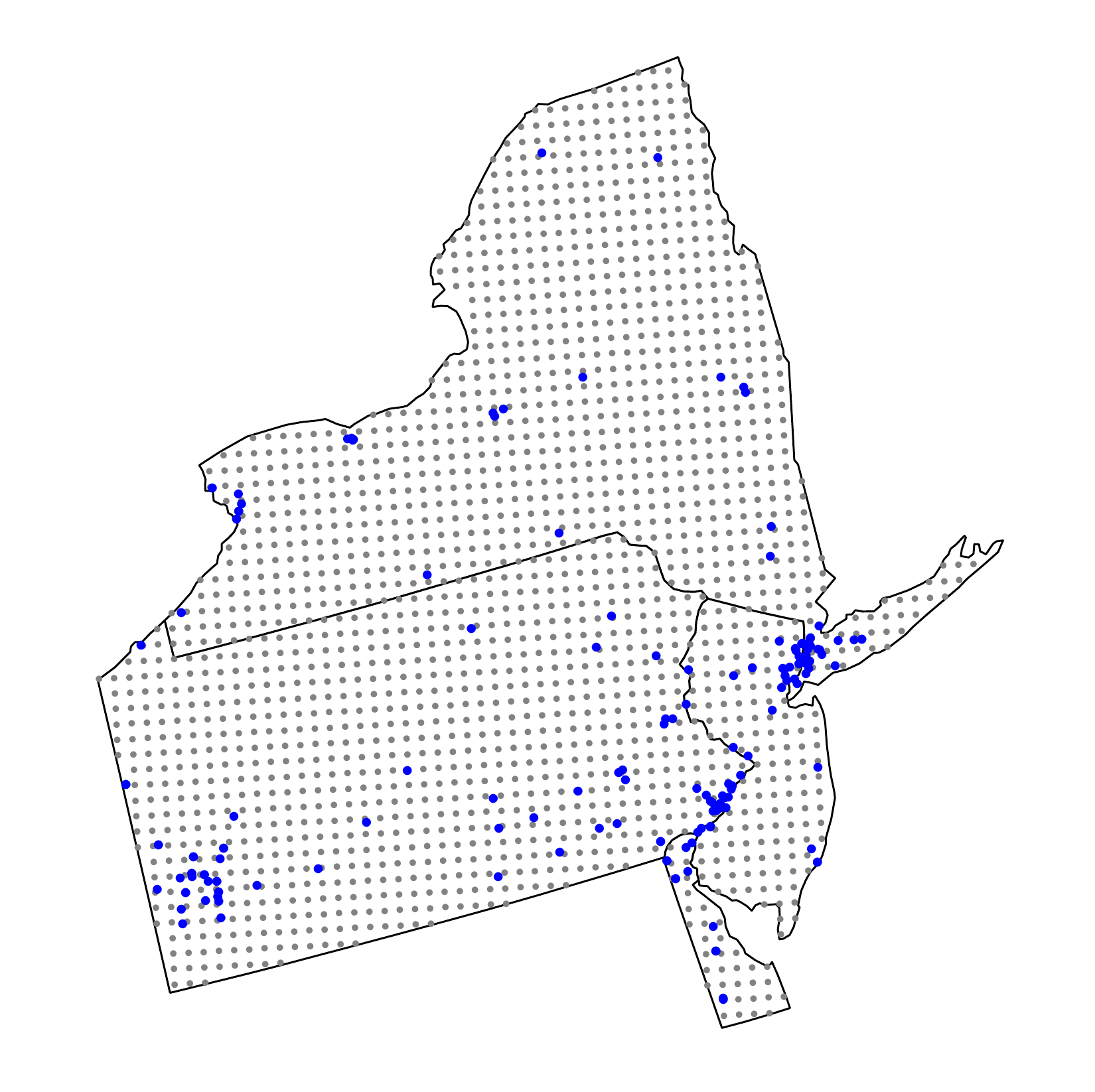}
\includegraphics[scale = 0.32]{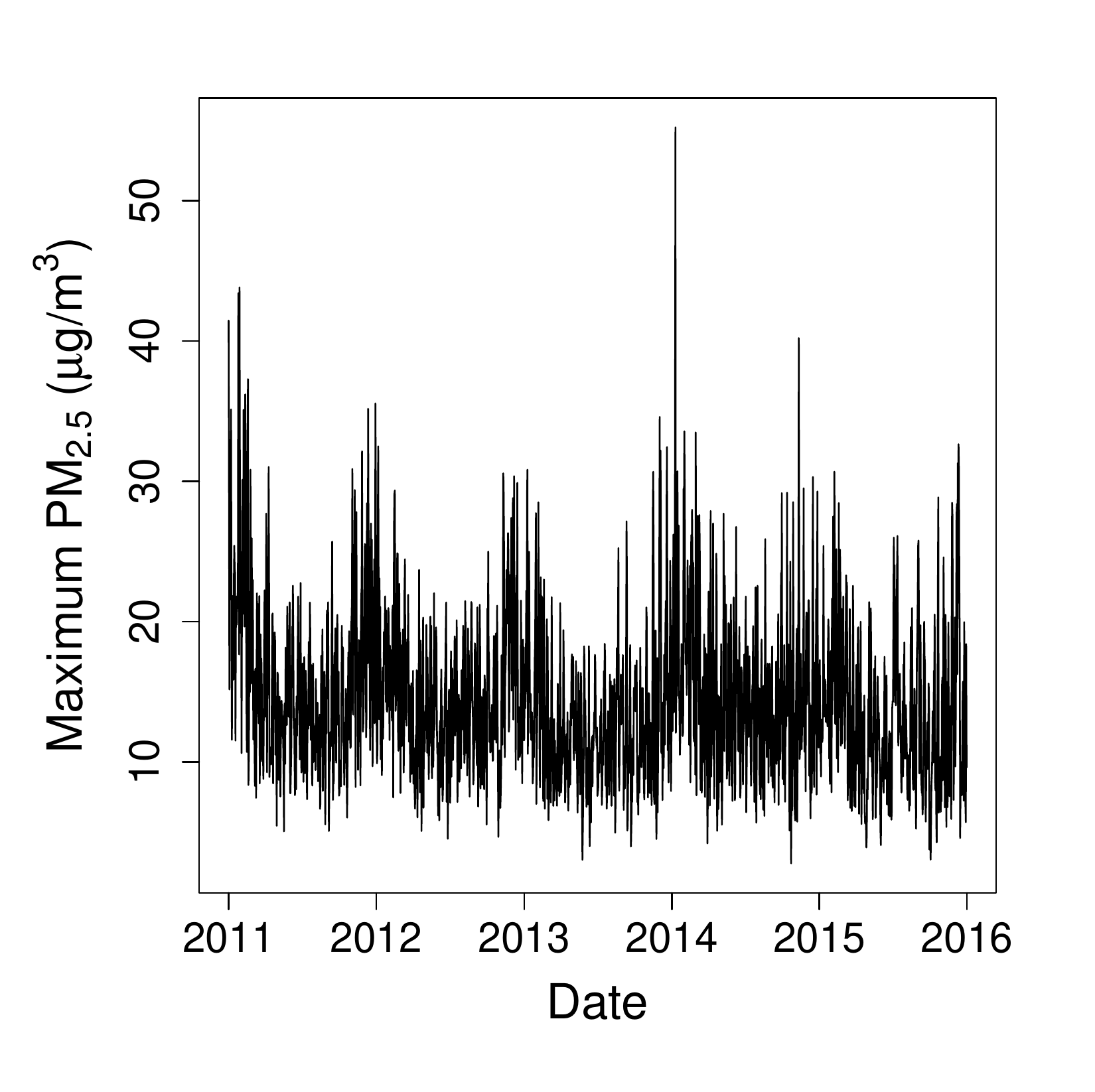}
\includegraphics[scale = 0.32]{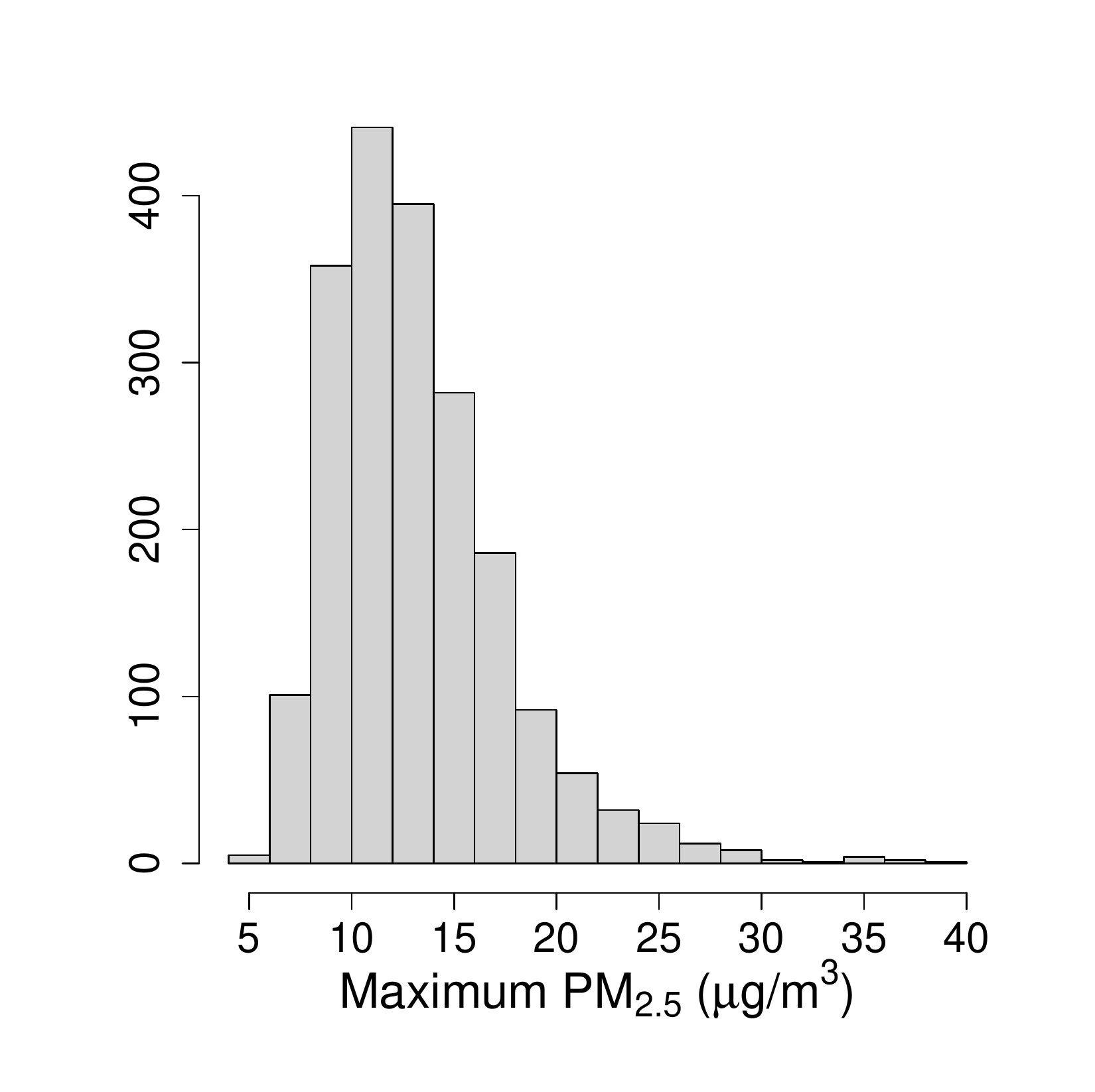}
\caption{Air pollution modeling/prediction study area where gray dots represent locations of estimates from the Community Multiscale Air Quality model and blue dots represent the location of the Air Quality System monitors that were active at any point between 2002-2015 (left panel); posterior mean predicted daily maximum 24-hour average PM$_{2.5}$ from the model in (7-8) (middle panel); histogram of posterior predictive exposure samples on August 1, 2011.}
\end{figure}

\subsection{Stage 1:  PM$_{2.5}$ modeling and prediction}
In the first stage of analysis, we use a hierarchical Bayesian framework for modeling and predicting the daily PM$_{2.5}$ concentrations collected from the AQS using the closest CMAQ estimate as a predictor.  Many of the air pollution monitors in the study region are not active on a given day and the network of monitors are only sparsely located across the study area (see Figure 4).  As a result, we use this model and the spatiotemporal completeness of the CMAQ estimates to predict the AQS data at unobserved locations and days.  We then take the maximum of these predictions within the three NJ counties on each day to estimate the daily maximum of the 24-hour PM$_{2.5}$ concentrations for the study area.  This is used as the primary exposure of interest in the subsequent stillbirth epidemiological analysis.   

The model for the AQS PM$_{2.5}$ data uses the closest CMAQ estimate as a predictor within a flexible spatially- and temporally-varying regression coefficient framework, similar to the original downscaling work of \cite{berrocal2010spatio}, such that \begin{equation}\ln\left\{Z\left(\textbf{s}, t\right) + 0.01\right\} = \eta_{0}\left(\textbf{s}, t\right) + \eta_{1}\left(\textbf{s}, t\right) \ln\left\{\text{C}_{B\left(\textbf{s}\right), t} + 0.01\right\} + \epsilon\left(\textbf{s}, t\right)\end{equation} where $Z\left(\textbf{s},t\right)$ is the AQS PM$_{2.5}$ concentration measured at the monitor located at $\textbf{s}$ on day $t$ (i.e., $t=1$ is January 1, 2002 and $t=5,113$ is December 31, 2015); $\text{C}_{B\left(\textbf{s}\right), t}$ is the corresponding CMAQ estimate at the grid cell centroid located closest to $\textbf{s}$ (i.e., $B\left(\textbf{s}\right)$) on day $t$; and $\epsilon\left(\textbf{s},t\right)|\sigma^2_{\epsilon} \stackrel{\text{iid}}{\sim} \text{N}\left(0, \sigma^2_{\epsilon}\right)$.  We work on the log scale during modeling given that the PM$_{2.5}$ concentrations are $\geq 0$.  

The spatially- and temporally-varying intercept and slope parameters are represented by $\eta_{0}\left(\textbf{s}, t\right)$ and $\eta_{1}\left(\textbf{s}, t\right)$, respectively, and allow the association between the CMAQ estimates and AQS data to flexibly change across space and time if appropriate.  They are modeled as a function of spatial and temporal covariates such that \begin{equation}
    \eta_{j}\left(\textbf{s}, t\right) = \mu_j + \text{lat}\left(\textbf{s}\right)\gamma_{1j} + \text{lon}\left(\textbf{s}\right)\gamma_{2j} + \left\{\text{lat}\left(\textbf{s}\right)\text{lon}\left(\textbf{s}\right)\right\}\gamma_{3j} + \sum_{k=1}^4 \text{x}_k\left(t\right)\delta_{kj},\ j=0,1
\end{equation} where $\text{x}_k\left(t\right)$ corresponds to the $k^{\text{th}}$ column of the B-spline basis matrix for a polynomial spline with four degrees of freedom (df) on day $t$, and $\text{lat}\left(\textbf{s}\right)/\text{lon}\left(\textbf{s}\right)$ are the latitude/longitude at spatial location $\textbf{s}$, respectively.     

We complete the model by specifying non/weakly informative prior distributions for the introduced model parameters.  Specifically, we choose flat prior distributions for all of the regression parameters (i.e., $\mu_j, \gamma_{kj}, \delta_{kj}$), and $\sigma^2_{\epsilon} \sim \text{Inverse Gamma}\left(0.01, 0.01\right)$; resulting in an efficient closed-form Monte Carlo sampling algorithm for obtaining samples from the joint posterior distribution of the model parameters.  We use it to collect $2,000$ independent posterior samples in total.

Next, we use composition sampling \citep{tanner1996tools} to generate independent posterior predictive samples of $Z\left(\textbf{s},t\right)$ at each of the nine CMAQ grid cell locations within the three NJ counties on each day of the study.  On each day and for every joint ppd sample, we calculate the maximum of the predictions across the three counties.  In total, we obtain a $5,113$ (i.e., number of days) by $m=2,000$ matrix of maximum 24-hour average PM$_{2.5}$ ppd samples, denoted by $\textbf{Z}^*$, and use it as the exposure for the stillbirth epidemiological analysis.          

\subsection{Stage 2:  Modeling stillbirth and PM$_{2.5}$}
Given $\textbf{Z}^*$ from Stage 1, we next turn to the stillbirth epidemiological analysis.  We model the total number of stillbirths occurring across the three NJ counties on a specific day as a function of time-varying predictors (i.e., day of week, long-term trend), meteorological variables (i.e., maximum/minimum temperature), and lagged PM$_{2.5}$ exposure.  The model is given as \begin{align}\begin{split} & Y_t|r, p_t \stackrel{\text{ind}}{\sim} \text{Negative Binomial}\left(r, p_t\right),\ t=1,\hdots, n,\\ & \text{logit}\left(p_t\right) = \text{O}_t + \sum_{j=1}^6 1\left\{\text{dow}\left(t\right) = j\right\} \alpha_j + \sum_{j=1}^{35} \text{x}_{1j}\left(t\right) \nu_{j} + \sum_{j=1}^4 \text{x}_{2j}\left(w_{0t}\right) \zeta_{j} + \sum_{j=1}^4 \text{x}_{3j}\left(w_{1t}\right) \kappa_{j} + \text{z}_{t - l} \theta \end{split}\end{align} where $Y_t$ is the number of stillbirths occurring on day $t$ with $t=1$ corresponding to January 1, 2011 and $t=n=1,826$ to December 31, 2015; $p_t$ is the probability parameter that controls the magnitude of counts on day $t$; $r > 0$ represents the dispersion parameter with small values indicating overdispersion in the data; $\text{O}_t$ is the offset variable representing the log of the total number of births occurring on day $t$; $1\left(.\right)$ is an indicator function; $\text{dow}\left(t\right)$ is the day of week that day $t$ occurred on with Saturday (i.e., $\text{dow}\left(t\right) = 7$) serving as the reference category; $\text{x}_{1j}\left(t\right)$, $\text{x}_{2j}\left(w_{0t}\right)$, and $\text{x}_{3j}\left(w_{1t}\right)$ are the $j^{\text{th}}$ columns of the B-spline basis matrices for a natural cubic spline with 35, 4, and 4 df for study day, minimum temperature (i.e., $w_{0t}$), and maximum temperature (i.e., $w_{1t}$), respectively; and $\text{z}_{t-l}$ is the true but unobserved maximum 24-hour average PM$_{2.5}$ exposure $l$ days prior to day $t$.

We choose 35 df for the long-term time trend based on selecting 7 df for each of the 5 study years as in \cite{samet2000fine} while noting that \cite{peng2006model} found reduced bias in effect estimation with more aggressive smoothing in similar time series modeling.  We consider daily lags from two to six days (i.e., $l=2,\hdots,6$), similar to previous acute stillbirth and air pollution modeling work \citep{faiz2013does, sarovar2020case, Enebish2022} and the estimated timing of 48 hours between fetal death and delivery \citep{gardosi1998analysis}.  

Using the model in (9) and $\textbf{Z}^*$, we test several of the existing methods from Section 2 for propagating exposure uncertainty in the health analysis along with the newly developed \textit{UKDE}.  \textit{MKDE} is not considered given its poor performance in simulation and long computing time for the large analysis dataset.  \textit{MI} is not applied due to its lengthy run time (i.e., requires fitting (9) in an MCMC framework $m=2,000$ times) and its overall similarity with \textit{MIA} in the simulation study results.  We separately fit each method and exposure lag ($l=2-6$) and make inference on $\theta$, the parameter that describes the association between maximum PM$_{2.5}$ exposure and stillbirth risk.  

The prior distributions for the parameters in (9) are chosen as $r \sim \text{Discrete Uniform}\left[1,100\right]$ and $\alpha_j, \nu_j, \zeta_j, \kappa_j, \theta \stackrel{\text{iid}}{\sim} \text{N}\left(0, 100^2\right)$.  From all methods, we collect $10,000$ samples from the joint posterior distributions after discarding the first $20,000$ prior to convergence and thinning the remaining $200,000$ by a factor of $20$ to reduce posterior autocorrelation.  We assessed convergence by visually inspecting traceplots of individual parameters and monitoring Geweke's diagnostic \citep{geweke1991evaluating}; neither tool suggested any obvious signs of non-convergence across all model fits.  

\subsection{Results}
In Figure 4, we display daily predictions of the maximum PM$_{2.5}$ exposures from the three NJ counties in 2011-2015 along with a histogram of the ppd samples on a single day (August 1, 2011; other days were similar).  The level of skewness in the ppd samples resembles that from the simulation study due to the log transformation used in (7).  In Figure S6 of the Supplement, we show a scatterplot of these predictions and the observed AQS data (daily maximum of the PM$_{2.5}$ AQS concentrations across all of NJ).  The plot shows that the model is generally predicting well with respect to the observed data.   

In Figure 5, we show results from the stillbirth analyses across all considered methods and lags.  Specifically, we present posterior inference (i.e., posterior means and quantile-based equal-tailed credible intervals) for $\exp\left\{\theta\right\}$ across all analyses, resulting in a relative risk interpretation.  Because we standardize $\textbf{Z}^*$ by subtracting off the median and dividing by the interquartile range (IQR) prior to each analysis, the estimates represent the relative risk of stillbirth for an IQR increase in exposure during a given lag (IQRs ranged from 9.37-9.38 across all lags).  We applied a Bonferroni correction due to the separate lag models being fit and present $99\%$ credible intervals in the figure.  

\begin{figure}[ht!]
\centering
\includegraphics[scale = 0.40]{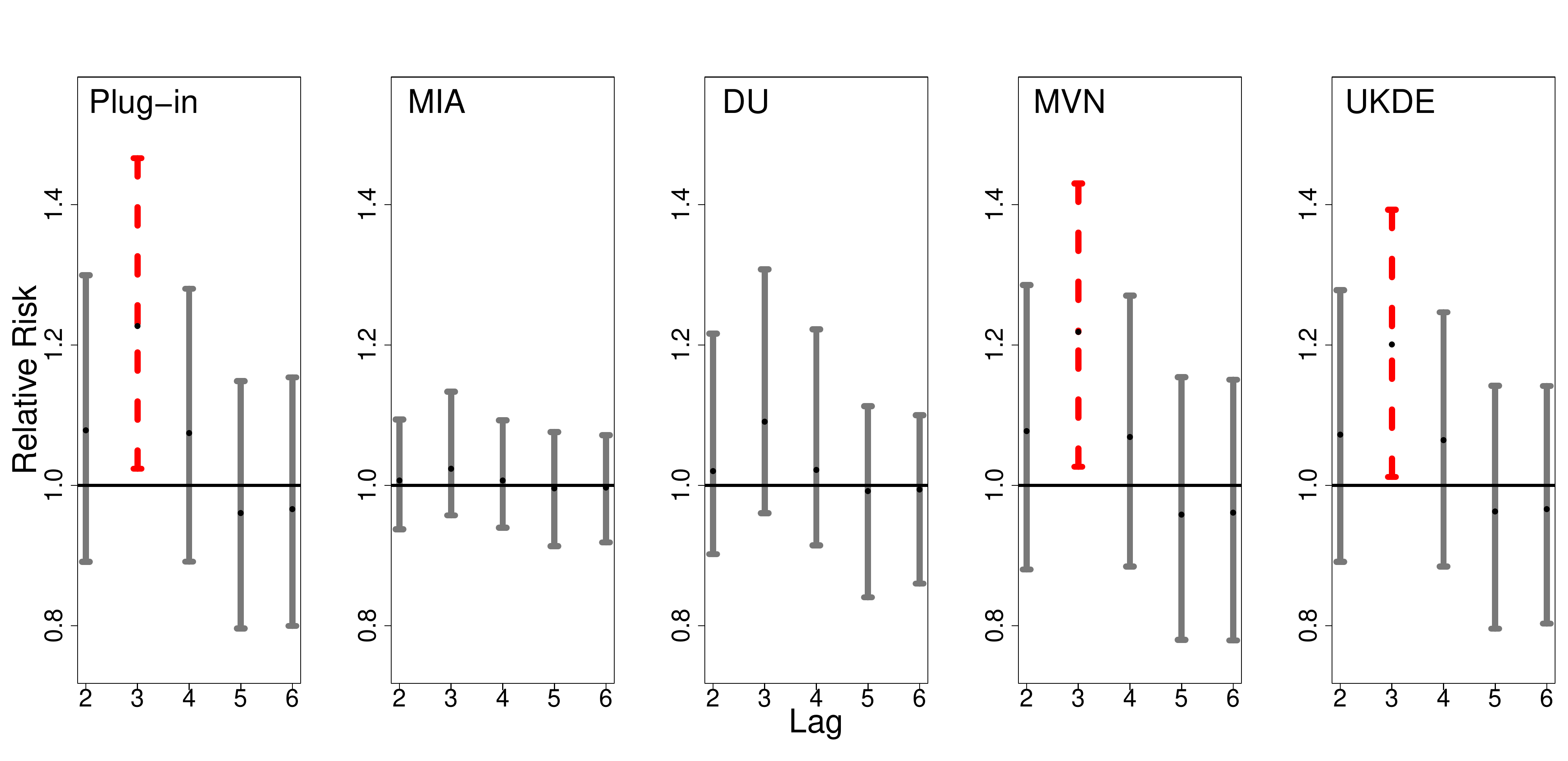}
\caption{Posterior mean and 99\% credible interval plots for $\exp\left\{\theta\right\}$ across the different models and daily lag periods for the New Jersey three county stillbirth and maximum daily 24-hour PM$_{2.5}$ exposure analysis.}
\end{figure}

Generally the findings are in agreement with the simulation study results.  \textit{MIA} and, to a lesser extent, \textit{DU} tend to pull the point estimates towards the null in comparison to the other approaches.  \textit{Plug-in}, \textit{MVN}, and \textit{UKDE} each suggest that elevated ambient levels of maximum 24-hour average PM$_{2.5}$ exposure three days prior to delivery is associated with an increase in stillbirths.  Specifically for \textit{UKDE}, an IQR increase in exposure of 9.4 $\mu$g/m$^3$ three days prior to delivery is associated with a $20.36\%$ increase in stillbirths ($99\%$ credible interval: $2.18-38.59$).  While the differences between these three methods is subtle in this application, \textit{UKDE} has the shortest credible interval followed by \textit{MVN} and \textit{Plug-in}, which may also be in agreement with the improved MSE performance of \textit{UKDE} observed in the simulation study.   

As a sensitivity analysis, we repeated each of the stillbirth analyses while randomly shuffling the order of the rows in $\textbf{Z}^*$.  This mixing breaks the temporal ordering of the exposures from the original analysis and we expect to estimate a null signal in $\theta$ unless there are serious confounding issues that the model fails to capture.  The results shown in Figure S7 of the Supplement show no significant associations across all methods/lags, with point estimates near the null overall.  This finding provides further evidence that population-level PM$_{2.5}$ may play an important role in explaining stillbirth risk.

\section{Discussion}
In this work we developed “UKDE”, a new framework for exposure uncertainty propagation in subsequent health outcome analyses, detailed its connection with existing approaches, derived its closed-form MCMC full conditional distributions, and created an R package for its implementation within several common epidemiological analyses (\textbf{KDExp}; \texttt{https://github.com/warrenjl/KDExp}).  Existing methods for quantifying this uncertainty were detailed within a unified framework, making comparing/contrasting the approaches more accessible.  In a simulation study, we showed that \textit{UKDE} had improved performance overall and particularly when the ppds were skewed.  The multivariate extension of \textit{UKDE}, \textit{MKDE}, was consistently outperformed by the other methods, likely because of the difficulty associated with estimating high dimensional densities using multivariate KDE. 

\cite{Thijssen2020} evaluated several density estimation techniques for performing sequential Bayesian inference, outside of the environmental health setting, using posterior samples collected from a first stage analysis.  However, the focus of these analyses was on making inference on a low-dimensional vector of parameters ($n = 10$ was the maximum considered in the study) that was shared across two or more datasets analyzed sequentially.  The goals and assumptions of this type of analysis differ from those in the environmental health setting.  In environmental health, the parameters shared across both modeling stages (i.e., exposures) are typically high dimensional (e.g., number of participants in a study), limiting the usefulness of some of the presented approaches (e.g., \textit{MKDE}).  Additionally, sequential analysis is primarily concerned with estimation of the set of parameters included in both modeling stages/datasets, whereas in environmental health analyses the emphasis is on correctly characterizing uncertainty in the exposures to improve inference for parameters only included in the second stage health outcome model (i.e., associations between exposure and health). 

In our stillbirth data analysis in NJ, we found that elevated exposure to maximum 24-hour average PM$_{2.5}$ three days prior to delivery was associated with elevated risk of stillbirth (relative risk: 1.20 (1.02, 1.39) for IQR increase).  In a recent study based in Mongolia, \cite{Enebish2022} identified multiple critical daily PM$_{2.5}$ exposure lags with respect to elevated stillbirth risk, including three days prior to delivery (odds ratio (OR): 1.28 (1.00, 1.62) for IQR increase).  Another NJ-based analysis estimated an OR with similar magnitude for an IQR increase in carbon monoxide exposure two days before delivery (OR: 1.20 (1.05, 1.37)) \citep{faiz2013does}.  \cite{sarovar2020case} similarly identified a link between coarse particulate matter exposure two days before delivery and increased odds of stillbirth, while \cite{mendola2017chronic} saw similar links between ozone and stillbirth (with similar magnitudes to our findings) on multiple days before delivery, including day three.  A recent meta-analysis found that stillbirth was positively associated with an increase in ozone of 10 $\mu$g/m$^3$ four days before delivery, but found no significant short-term effects for PM$_{2.5}$ in pooled estimates \citep{zhang2021ambient}.  

Future methods work in this area should focus on developing improved techniques for incorporating correlation between high dimensional ppds.  As methods for investigating the impact of pollution mixtures on health are becoming more common, extensions of the \textit{UKDE} framework for more general application in these settings is needed.  In our work we showed that closed-form MCMC full conditional updates were still available with additive exposure models, but this does not necessarily hold for more complex interactions and hierarchical structures used for the exposures.  Additionally, extensions of this work to accommodate critical exposure window identification/estimation (e.g., \cite{warren2012spatial}) is also needed.  Overall, \textit{UKDE} is shown to be a promising framework for characterizing exposure uncertainty in environmental health studies, representing a more flexible hybrid approach between existing methods, and can be implemented in the R package \textbf{KDExp} for several common regression models.

\section*{Acknowledgments}
This research was supported by the National Institute of Environmental Health Sciences (NIEHS) of the National Institutes of Health (NIH) under Award R01 NIEHS ES028346.

\bibliographystyle{chicago}
\bibliography{References}

\end{document}


\maketitle

\pagebreak

\section{Exposure full conditional distribution derivations}
We consider three likelihood and link function combinations for the health model in (1) from the Main Text and derive the full conditional distribution for the vector of true exposures, $\textbf{z}$, for \textit{MVN}, \textit{UKDE}, and \textit{MKDE}.  For Gaussian distributed data with identity link function we have \begin{equation}Y_i = \textbf{x}_i^{\text{T}} + \text{z}_i \theta + \epsilon_i,\ \epsilon_i|\sigma^2_{\epsilon} \stackrel{\text{iid}}{\sim}\text{N}\left(0, \sigma^2_{\epsilon}\right).\end{equation}  For Bernoulli distributed data with logit link function we have \begin{equation}Y_i | p_i \stackrel{\text{ind}}{\sim} \text{Bernoulli}\left(p_i\right),\ \ln\left(\frac{p_i}{1 - p_i}\right) = \textbf{x}_i^{\text{T}}\boldsymbol{\beta} + \text{z}_i \theta.\end{equation}  For negative binomial distributed data with logit link function we have \begin{equation}Y_i | r, p_i \stackrel{\text{ind}}{\sim} \text{Negative Binomial}\left(r, p_i\right),\ \ln\left(\frac{p_i}{1 - p_i}\right) = \text{O}_i + \textbf{x}_i^{\text{T}}\boldsymbol{\beta} + \text{z}_i \theta.\end{equation}  All terms have been previously described in Section 2 of the Main Text.

While each of these models have different likelihood functions, these functions have a similar general form due to results from \cite{polson2013bayesian}.  Specifically, the authors show that conjugacy is possible for (2) and (3) when observation-level, $\text{P\'olya-Gamma}$ distributed auxiliary variables (i.e., $w_i$) are introduced.  This finding allows for the full conditional distribution of $\textbf{z}$ for each model to be expressed as \begin{align}f\left(\textbf{z} | \boldsymbol{Y}, \boldsymbol{\beta}, \theta, \boldsymbol{\zeta}\right) &\propto f\left(\boldsymbol{Y} | \boldsymbol{\beta}, \theta, \boldsymbol{\zeta}_{-\boldsymbol{w}}, \textbf{z}\right) f\left(\boldsymbol{w}|\boldsymbol{\beta}, \theta, \boldsymbol{\zeta}_{-\boldsymbol{w}}, \textbf{z}\right) f\left(\textbf{z}\right)\\ &\propto \exp\left\{-\frac{1}{2}\left(\textbf{O} + \textbf{X}\boldsymbol{\beta} + \textbf{z}\theta - \widetilde{\boldsymbol{Y}}\right)^{\text{T}} \Omega \left(\textbf{O} + \textbf{X}\boldsymbol{\beta} + \textbf{z}\theta - \widetilde{\boldsymbol{Y}}\right)\right\} f\left(\textbf{z}\right) \end{align} where $\boldsymbol{w}^{\text{T}} = \left(w_1, \hdots, w_n\right)$ is the aforementioned vector of auxiliary variables which is contained in $\boldsymbol{\zeta}$; $\boldsymbol{\zeta}_{-\boldsymbol{w}}$ is the vector of dispersion/variance parameters (i.e., after removing the auxiliary variables); $\textbf{O}^{\text{T}} = \left(\text{O}_1, \hdots, \text{O}_n\right)$ is the vector of offset terms; $\textbf{X}$ is a matrix with $i^{\text{th}}$ row equal to $\textbf{x}_i^{\text{T}}$; and $f\left(\textbf{z}\right)$ is the prior distribution for $\textbf{z}$.  

For (1), $\boldsymbol{w}$ is not introduced in (4).  However, the result in (5) remains the same with $\Omega = \frac{I_n}{\sigma^2_{\epsilon}}$ and $\widetilde{\boldsymbol{Y}}^{\text{T}} = \left(Y_1, \hdots, Y_n\right)$ where $I_n$ is the $n$ by $n$ identity matrix.  For (2), $$w_i|\boldsymbol{\beta}, \theta, \text{z}_i \stackrel{\text{ind}}{\sim} \text{P\'olya-Gamma}\left(1, \textbf{x}_i^{\text{T}}\boldsymbol{\beta} + \text{z}_i \theta\right),$$ $\Omega = \text{diag}\left(w_1, \hdots, w_n\right)$, and $\widetilde{\boldsymbol{Y}}^{\text{T}} = \left(\frac{Y_1 - 0.50}{w_1}, \hdots, \frac{Y_n - 0.50}{w_n}\right)$.  Finally in (3), $$w_i|\boldsymbol{\beta}, \theta, r, \text{z}_i \stackrel{\text{ind}}{\sim} \text{P\'olya-Gamma}\left(r + Y_i, \text{O}_i + \textbf{x}_i^{\text{T}}\boldsymbol{\beta} + \text{z}_i \theta\right),$$ $\Omega = \text{diag}\left(w_1, \hdots, w_n\right)$, and $\widetilde{\boldsymbol{Y}}^{\text{T}} = \left(\frac{0.50\left(Y_1 - r\right)}{w_1}, \hdots, \frac{0.50\left(Y_n - r\right)}{w_n}\right)$.  The general form of the full conditional distribution given in (5) allows for semi-conjugacy during model fitting for the prior distributions specified using \textit{MVN}, \textit{UKDE}, and \textit{MKDE}.  

\subsection{Multivariate normal prior distribution}
Recall for \textit{MVN} that $\textbf{z} \sim \text{MVN}\left(\widehat{\textbf{z}}, \widehat{\Sigma}\right)$ where $\widehat{\textbf{z}}$ and $\widehat{\Sigma}$ are defined in Section 2 of the Main Text.  Therefore, the full conditional distribution in (5) is proportional to $$\exp\left\{-\frac{1}{2}\left(\textbf{O} + \textbf{X}\boldsymbol{\beta} + \textbf{z}\theta - \widetilde{\boldsymbol{Y}}\right)^{\text{T}} \Omega \left(\textbf{O} + \textbf{X}\boldsymbol{\beta} + \textbf{z}\theta - \widetilde{\boldsymbol{Y}}\right)\right\} \exp\left\{-\frac{1}{2}\left(\textbf{z} - \widehat{\textbf{z}}\right)^{\text{T}} \widehat{\Sigma}^{-1} \left(\textbf{z} - \widehat{\textbf{z}}\right)\right\},$$ which can be reorganized to show that $$\textbf{z}|\boldsymbol{Y}, \boldsymbol{\beta}, \theta, \boldsymbol{\zeta} \sim \text{MVN}\left(\boldsymbol{\mu}_{\textbf{z}}, \Sigma_{\textbf{z}}\right)$$ where $\Sigma_{\textbf{z}} = \left(\theta^2 \Omega + \widehat{\Sigma}^{-1}\right)^{-1}$ and $\boldsymbol{\mu}_{\textbf{z}} = \Sigma_{\textbf{z}} \left\{\theta \Omega(\widetilde{\boldsymbol{Y}} - \textbf{O} - \textbf{X}\boldsymbol{\beta}) + \widehat{\Sigma}^{-1}\boldsymbol{\widehat{\textbf{z}}}\right\}$.

\subsection{Univariate kernel density estimation prior distribution}
We update each $\text{z}_i$ separately given the independent prior distributions used for \textit{UKDE}.  Therefore, the full conditional distribution for $\text{z}_i$ based on (5) and the \textit{UKDE} prior distribution is proportional to $$\exp\left\{-\frac{w_i}{2}\left(\text{O}_i + \textbf{x}_i^{\text{T}}\boldsymbol{\beta} + \text{z}_i \theta - \widetilde{Y}_i\right)^2\right\} \sum_{j=1}^m \exp\left\{-\frac{1}{2h_i^2} \left(\text{z}_i - \text{z}_{ij}^*\right)^2\right\},$$ which can also be reorganized to show that $f\left(\text{z}_i | \boldsymbol{Y}, \boldsymbol{\beta}, \theta, \boldsymbol{\zeta}\right)$ is the density function for the mixture of normal distributions shown in (2) and (3) from the Main Text. 

\subsection{Multivariate kernel density estimation prior distribution}
We update the entire $\textbf{z}$ vector given the joint prior distribution specified by $\textit{MKDE}$.  Therefore, the full conditional distribution for $\textbf{z}$ based on (5) and the \textit{MKDE} prior distribution is proportional to $$\exp\left\{-\frac{1}{2}\left(\textbf{O} + \textbf{X}\boldsymbol{\beta} + \textbf{z}\theta - \widetilde{\boldsymbol{Y}}\right)^{\text{T}} \Omega \left(\textbf{O} + \textbf{X}\boldsymbol{\beta} + \textbf{z}\theta - \widetilde{\boldsymbol{Y}}\right)\right\} \sum_{j=1}^m \exp\left\{-\frac{1}{2} \left(\textbf{z} - \textbf{z}_{.j}^*\right)^{\text{T}} \textbf{H}^{-1} \left(\textbf{z} - \textbf{z}_{.j}^*\right)\right\},$$ which can similarly be reorganized to show that $f\left(\textbf{z} | \boldsymbol{Y}, \boldsymbol{\beta}, \theta, \boldsymbol{\zeta}\right)$ is the density function of the mixture of multivariate normal distributions shown in (6) and (7) below.
\clearpage 

\section{Multivariate kernel density estimation prior distribution}
We extend \textit{UKDE} to the multivariate setting (i.e., \textit{MKDE}) to specify a prior distribution for the entire $\textbf{z}$ vector in an attempt to account for any correlation between exposures while maintaining the flexibility allowed by the kernel density estimator.  Specifically, the prior distribution for $\textbf{z}$ is: \begin{equation*}f\left(\textbf{z}\right) = \frac{1}{m} \sum_{j=1}^m \frac{1}{\sqrt{\left(2 \pi\right)^n |\textbf{H}|}} \exp\left\{-\frac{1}{2} \left(\textbf{z} - \textbf{z}_{.j}^*\right)^{\text{T}} \textbf{H}^{-1} \left(\textbf{z} - \textbf{z}_{.j}^*\right)\right\}\end{equation*} where $\textbf{H}$ is the $n$ by $n$ bandwidth matrix variable estimated using standard approaches (e.g., \cite{scott2015multivariate}) based on $\textbf{Z}^*$, and all other terms have been previously described in Sections 2 and 3 of the Main Text.  This prior distribution represents a mixture of $m$ equally weighted multivariate normal distributions with mean vector equal to different columns of $\textbf{Z}$ and variance/covariance matrix $\textbf{H}$.    

The full conditional distribution for $\textbf{z}$ is a mixture of multivariate normal distributions with density given as 
\begin{align}\begin{split}
f\left(\textbf{z}|\boldsymbol{Y}, \boldsymbol{\beta}, \theta, \boldsymbol{\zeta}\right) &= 
\sum_{j=1}^m \left(\frac{d_{j}}{\sum_{k=1}^m d_{k}}\right) \sqrt{|\theta^2 \Omega + \textbf{H}^{-1}|} \times \\ &\phi_n\left(\left(\theta^2 \Omega + \textbf{H}^{-1}\right)^{1/2} \left[\textbf{z} - \left(\theta^2 \Omega + \textbf{H}^{-1}\right)^{-1}\left\{\theta \Omega \left(\widetilde{\boldsymbol{Y}} - \textbf{O} - \textbf{X}\boldsymbol{\beta}\right) + \textbf{H}^{-1}\textbf{z}^*_{.j}\right\}\right]\right)
\end{split}\end{align} 
where $\phi_n\left(.\right)$ is the pdf of the $n$-dimensional standard multivariate normal distribution; $\widetilde{\boldsymbol{Y}}$ is the complete vector of $n$ $\widetilde{Y}_i$ values; $\textbf{O}$ is the complete vector of $n$ offset terms; $\textbf{X}$ is a matrix with $i^{\text{th}}$ row equal to $\textbf{x}_i^{\text{T}}$; the mixture weights are defined by 
\begin{align}\begin{split}
d_j = \exp\left\{-\frac{1}{2} \right. &\left[\textbf{z}^{*\text{T}}_{.j} \textbf{H}^{-1} \textbf{z}^*_{.j} - \left\{\theta \Omega \left(\widetilde{\boldsymbol{Y}} - \textbf{O} - \textbf{X}\boldsymbol{\beta}\right) + \textbf{H}^{-1} \textbf{z}^*_{.j}\right\}^{\text{T}} \right.\\ &\left. \left. \left(\theta^2 \Omega + \textbf{H}^{-1}\right)^{-1} \left\{\theta \Omega \left(\widetilde{\boldsymbol{Y}} - \textbf{O} - \textbf{X}\boldsymbol{\beta}\right) + \textbf{H}^{-1} \textbf{z}^*_{.j}\right\} \right] \right\}; \end{split}\end{align} and all other terms have been previously described in Sections 2 and 3 of the Main Text.

As with \textit{UKDE}, the use of \textit{MKDE} represents a hybrid approach between \textit{DU} and \textit{MVN}.  However, instead of considering one data point's exposure at a time like \textit{UKDE}, \textit{MKDE} uses entire columns of $\textbf{Z}^*$.  Therefore, during a single iteration of the MCMC algorithm it selects a probable exposure vector (using the health data) from the columns of $\textbf{Z}^*$, similar to \textit{DU}, and then samples from the true exposure distribution using a multivariate normal distribution based on this selected vector, similar to \textit{MVN}.  Similar to \textit{UKDE}, it avoids the restrictive assumptions made by \textit{DU} and \textit{MVN} and should be more flexible than \textit{MVN} when the ppds exhibit non-symmetric behavior.  However, given the typically large value of $n$ in most health outcome analyses (i.e., sample size), accurate estimation of $\textbf{H}$ for \textit{MKDE} may be difficult and could drastically impact the quality of the density estimation.  For reference, high dimensionality in the \textit{MKDE} methods development setting has been defined as between 4 and 50 in past work \citep{wang2019nonparametric}.

\clearpage

\section{Supplementary tables}

\begin{landscape}
\begin{table}[h!]
\centering
\small
\caption{Simulation study results for $\theta = 1.00$ and $\tau^2 = 1.00$.  Estimates are presented with the range of standard errors for a group of estimates provided in parentheses.  Bold entries represent the optimal estimate across an entire row (i.e., closest to zero for bias, smallest for mean squared error (MSE), closest to 95 for empirical coverage (EC), and largest for power).  All results are multiplied by 100 for presentation purposes.}
\begin{tabular}{llllrrrrrrrr}
\hline
 & \multicolumn{2}{c}{Settings} & & \multicolumn{8}{c}{Methods} \\
\cline{2-3} \cline{5-12}
Metric & Correlated & Skewed & & True        & Plug-in         & MI              & MIA             & DU              & MVN             & UKDE            & MKDE            \\
\hline
Bias   & No         & No     & &   0.24      & \textbf{0.48}   & -49.72          & -49.75          & -32.77          &  -5.11          &   1.87          & -32.89          \\
       & No         & Yes    & &  -0.17      &  61.60          & -64.34          & -64.36          & -29.95          & \textbf{0.72}   &   5.93          & -32.95          \\
       & Yes        & No     & &   0.05      & \textbf{0.24}   & -46.22          & -46.22          & -21.10          &  -0.64          &   1.75          & -14.94          \\
       & Yes        & Yes    & &  -0.32      &  58.73          & -54.11          & -54.09          &  -4.91          &  -7.60          & \textbf{2.29}   & -32.44          \\
       &            &        & & (0.28-0.49) & (0.47-3.60)     & (0.26-1.06)     & (0.26-1.07)     & (0.29-2.43)     & (0.39-2.16)     & (0.49-1.98)     & (0.30-1.89)     \\
\hline
MSE    & No         & No     & &   0.39      & \textbf{1.14}   &  25.07          &  25.09          &  11.16          &   1.23          &   1.25          &  11.26          \\
       & No         & Yes    & &   0.68      &  92.65          &  44.06          &  44.08          &  15.09          &  14.68          & \textbf{10.92}  &  18.75          \\
       & Yes        & No     & &   0.41      &   1.10          &  21.71          &  21.72          &   5.04          & \textbf{0.77}   &   1.21          &   2.89          \\
       & Yes        & Yes    & &   1.22      &  99.10          &  34.90          &  34.92          &  29.69          &  23.88          & \textbf{19.64}  &  28.36          \\
       &            &        & & (0.03-0.13) & (0.06-11.33)    & (0.25-0.94)     & (0.26-0.94)     & (0.15-3.66)     & (0.05-3.01)     & (0.07-3.02)     & (0.12-2.51)     \\
\hline
EC     & No         & No     & &  95.20      & \textbf{94.80}  &   0.00          &   0.00          &   1.00          &  93.00          &  94.20          &   1.20          \\
       & No         & Yes    & &  94.80      &  49.40          &  19.00          &  19.20          &  27.60          &  70.20          & \textbf{90.80}  &  32.80          \\
       & Yes        & No     & &  95.80      &  95.80          &   0.00          &   0.00          &  27.60          &  94.40          & \textbf{95.00}  &  57.00          \\
       & Yes        & Yes    & &  95.00      &  46.20          &  64.00          &  62.60          &  47.20          &  53.00          & \textbf{70.60}  &  37.20          \\
       &            &        & & (0.90-0.99) & (0.90-2.24)     & (0.00-2.15)     & (0.00-2.16)     & (0.44-2.23)     & (1.03-2.23)     & (0.97-2.04)     & (0.49-2.21)     \\
\hline
Power  & No         & No     & & 100.00      & \textbf{100.00} & \textbf{100.00} & \textbf{100.00} & \textbf{100.00} & \textbf{100.00} & \textbf{100.00} & \textbf{100.00} \\
       & No         & Yes    & & 100.00      &  99.20          &  56.20          &  57.60          & \textbf{99.80}  &  99.40          & \textbf{99.80}  &  99.00          \\
       & Yes        & No     & & 100.00      & \textbf{100.00} & \textbf{100.00} & \textbf{100.00} & \textbf{100.00} & \textbf{100.00} & \textbf{100.00} & \textbf{100.00} \\
       & Yes        & Yes    & &  99.60      &  97.80          &  57.40          &  59.40          &  98.60          &  95.20          & \textbf{99.00}  &  95.40          \\
       &            &        & & (0.00-0.28) & (0.00-0.66)     & (0.00-2.22)     & (0.00-2.21)     & (0.00-0.53)     & (0.00-0.96)     & (0.00-0.44)     & (0.00-0.94)     \\
\hline
\end{tabular}
\end{table}
\end{landscape}
\clearpage

\begin{landscape}
\begin{table}[h!]
\centering
\small
\caption{Simulation study results for $\theta = 0.00$ and $\tau^2 = 0.10$.  Estimates are presented with the range of standard errors for a group of estimates provided in parentheses.  Bold entries represent the optimal estimate across an entire row (i.e., closest to zero for bias, smallest for mean squared error (MSE), closest to 95 for empirical coverage (EC), and closest to 5 for type I error).  All results are multiplied by 100 for presentation purposes.}
\begin{tabular}{llllrrrrrrrr}
\hline
 & \multicolumn{2}{c}{Settings} & & \multicolumn{8}{c}{Methods} \\
\cline{2-3} \cline{5-12}
Metric       & Correlated & Skewed & & True        & Plug-in        & MI             & MIA           & DU          & MVN            & UKDE        & MKDE        \\
\hline
Bias         & No         & No     & &   0.17      &  0.56          &   0.06         & \textbf{0.06} &  0.43       &  0.69          &  0.72       &  0.46       \\
             & No         & Yes    & &  -0.32      &  0.19          &   0.01         & \textbf{0.00} &  0.20       & -0.11          &  0.10       &  0.06       \\
             & Yes        & No     & &  -0.09      & -1.70          & \textbf{-0.19} &  -0.19        & -0.93       & -1.16          & -1.93       & -0.79       \\
             & Yes        & Yes    & &  -0.67      &  0.86          &   0.03         & \textbf{0.03} &  0.04       &  0.13          &  0.07       &  0.14       \\
             &            &        & & (0.29-0.41) & (0.93-2.13)    & (0.08-0.12)    & (0.08-0.12)   & (0.41-0.69) & (0.36-0.83)    & (0.81-1.13) & (0.33-0.41) \\
\hline
MSE          & No         & No     & &   0.41      &  4.29          & \textbf{0.04}  &   0.04        &  0.85       &  3.45          &  5.83       &  0.83       \\
             & No         & Yes    & &   0.54      & 22.57          & \textbf{0.03}  &   0.03        &  0.92       &  2.04          &  3.40       &  0.83       \\
             & Yes        & No     & &   0.48      &  4.88          & \textbf{0.05}  &   0.06        &  0.98       &  1.58          &  6.44       &  0.68       \\
             & Yes        & Yes    & &   0.86      & 22.08          & \textbf{0.07}  &   0.07        &  2.38       &  0.63          &  3.31       &  0.54       \\
             &            &        & & (0.02-0.07) & (0.27-1.46)    & (0.00-0.00)    & (0.00-0.00)   & (0.06-0.25) & (0.06-0.22)    & (0.23-0.38) & (0.05-0.07) \\
\hline
EC           & No         & No     & &  95.20      & \textbf{94.80} & 100.00         & 100.00        & 99.00       & \textbf{95.20} & 92.20       & 98.60       \\
             & No         & Yes    & &  93.80      & \textbf{94.00} & 100.00         & 100.00        & 99.80       & 99.20          & 98.20       & 99.60       \\
             & Yes        & No     & &  94.60      & \textbf{92.80} & 100.00         & 100.00        & 98.20       & 97.80          & 91.20       & 99.40       \\
             & Yes        & Yes    & &  95.40      & \textbf{94.20} & 100.00         & 100.00        & 98.40       & 99.80          & 97.80       & 99.40       \\
             &            &        & & (0.94-1.08) & (0.99-1.16)    & (0.00-0.00)    & (0.00-0.00)   & (0.20-0.59) & (0.20-0.96)    & (0.59-1.27) & (0.28-0.53) \\

\hline
Type I Error & No         & No     & &   4.80      & \textbf{5.20}  &   0.00         &   0.00        &  0.80       & \textbf{4.80}  &  7.80       &  1.40       \\
             & No         & Yes    & &   6.20      & \textbf{6.00}  &   0.00         &   0.00        &  0.20       &  0.80          &  1.80       &  0.40       \\
             & Yes        & No     & &   5.40      & \textbf{7.20}  &   0.00         &   0.00        &  1.80       &  2.20          &  8.80       &  0.60       \\
             & Yes        & Yes    & &   4.60      & \textbf{5.80}  &   0.00         &   0.00        &  1.60       &  0.20          &  2.20       &  0.60       \\
             &            &        & & (0.94-1.08) & (0.99-1.16)    & (0.00-0.00)    & (0.00-0.00)   & (0.20-0.59) & (0.20-0.96)    & (0.59-1.20) & (0.28-0.53) \\
\hline
\end{tabular}
\end{table}
\end{landscape}
\clearpage 

\begin{landscape}
\begin{table}[h!]
\centering
\small
\caption{Simulation study results for $\theta = 0.00$ and $\tau^2 = 1.00$.  Estimates are presented with the range of standard errors for a group of estimates provided in parentheses.  Bold entries represent the optimal estimate across an entire row (i.e., closest to zero for bias, smallest for mean squared error (MSE), closest to 95 for empirical coverage (EC), and closest to 5 for type I error).  All results are multiplied by 100 for presentation purposes.}
\begin{tabular}{llllrrrrrrrr}
\hline
 & \multicolumn{2}{c}{Settings} & & \multicolumn{8}{c}{Methods} \\
\cline{2-3} \cline{5-12}
Metric       & Correlated & Skewed & & True        & Plug-in        & MI             & MIA            & DU             & MVN            & UKDE           & MKDE           \\
\hline
Bias         & No         & No     & &  -0.26      & -0.45          & \textbf{-0.22} &  -0.22         & -0.41          & -0.47          & -0.45          & -0.45          \\
             & No         & Yes    & &   0.94      &  0.78          &   0.24         & \textbf{0.23}  &  0.61          &  0.42          &  0.87          &  0.41          \\
             & Yes        & No     & &  -0.23      & -0.60          &  -0.33         & \textbf{-0.33} & -0.45          & -0.50          & -0.63          & -0.42          \\
             & Yes        & Yes    & &  -0.19      &  1.48          & \textbf{0.53}  &   0.55         &  1.53          &  0.82          &  1.17          &  0.73          \\
             &            &        & & (0.29-0.45) & (0.41-0.96)    & (0.19-0.28)    & (0.19-0.28)    & (0.38-0.75)    & (0.39-0.46)    & (0.41-0.66)    & (0.35-0.40)    \\
\hline
MSE          & No         & No     & &   0.41      &  0.82          & \textbf{0.21}  &   0.21         &  0.79          &  0.83          &  0.84          &  0.79          \\
             & No         & Yes    & &   0.68      &  3.80          & \textbf{0.18}  &   0.19         &  1.48          &  1.04          &  1.66          &  0.71          \\
             & Yes        & No     & &   0.45      &  0.84          & \textbf{0.24}  &   0.24         &  0.72          &  0.75          &  0.87          &  0.62          \\
             & Yes        & Yes    & &   1.00      &  4.60          &   0.40         & \textbf{0.40}  &  2.87          &  0.88          &  2.22          &  0.72          \\
             &            &        & & (0.02-0.08) & (0.05-0.31)    & (0.01-0.03)    & (0.01-0.03)    & (0.05-0.21)    & (0.05-0.08)    & (0.05-0.15)    & (0.04-0.06)    \\
\hline
EC           & No         & No     & &  95.80      & 94.20          & 100.00         & 100.00         & 94.00          & \textbf{94.60} & \textbf{94.60} & \textbf{94.60} \\
             & No         & Yes    & &  95.20      & \textbf{95.60} & 100.00         & 100.00         & 96.00          & 99.60          & 96.80          & 98.20          \\
             & Yes        & No     & &  94.20      & 93.80          &  99.80         &  99.80         & 94.40          & 94.20          & 94.00          & \textbf{95.00} \\
             & Yes        & Yes    & &  94.80      & 93.40          & 100.00         & 100.00         & \textbf{95.00} & 99.40          & 95.40          & 99.40          \\
             &            &        & & (0.90-1.05) & (0.92-1.11)    & (0.00-0.20)    & (0.00-0.20)    & (0.88-1.06)    & (0.28-1.05)    & (0.79-1.06)    & (0.35-1.01)    \\
\hline
Type I Error & No         & No     & &   4.20      &  5.80          &   0.00         &   0.00         &  6.00          & \textbf{5.40}  & \textbf{5.40}  &  \textbf{5.40} \\
             & No         & Yes    & &   4.80      &  \textbf{4.40} &   0.00         &   0.00         &  4.00          &  0.40          &  3.20          &  1.80          \\
             & Yes        & No     & &   5.80      &  6.20          &   0.20         &   0.20         &  5.60          &  5.80          &  6.00          &  \textbf{5.00} \\
             & Yes        & Yes    & &   5.20      &  6.60          &   0.00         &   0.00         & \textbf{5.00}  &  0.60          &  4.60          &  0.60          \\
             &            &        & & (0.90-1.05) & (0.92-1.11)    & (0.00-0.20)    & (0.00-0.20)    & (0.88-1.06)    & (0.28-1.05)    & (0.79-1.06)    & (0.35-1.01)    \\
\hline
\end{tabular}
\end{table}
\end{landscape}
\clearpage

\section{Supplemental figures}

\begin{landscape}
\begin{figure}[ht]
\centering
\includegraphics[scale = 0.75]{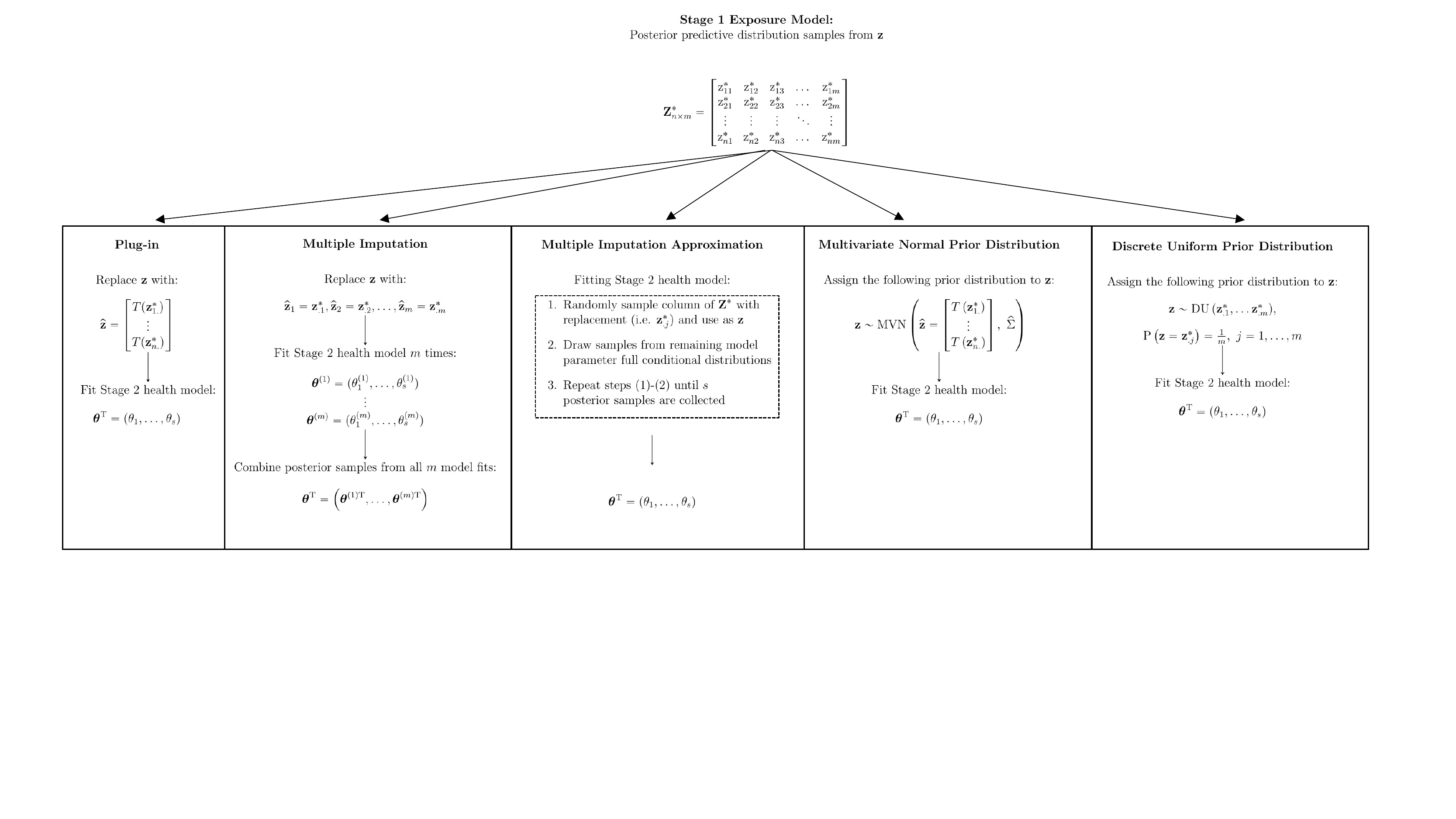}
\caption{Overview of selected existing approaches for incorporating exposure uncertainty into health analyses.}
\end{figure}
\end{landscape}
\clearpage 

\begin{figure}[ht]
\centering
\includegraphics[scale = 0.24]{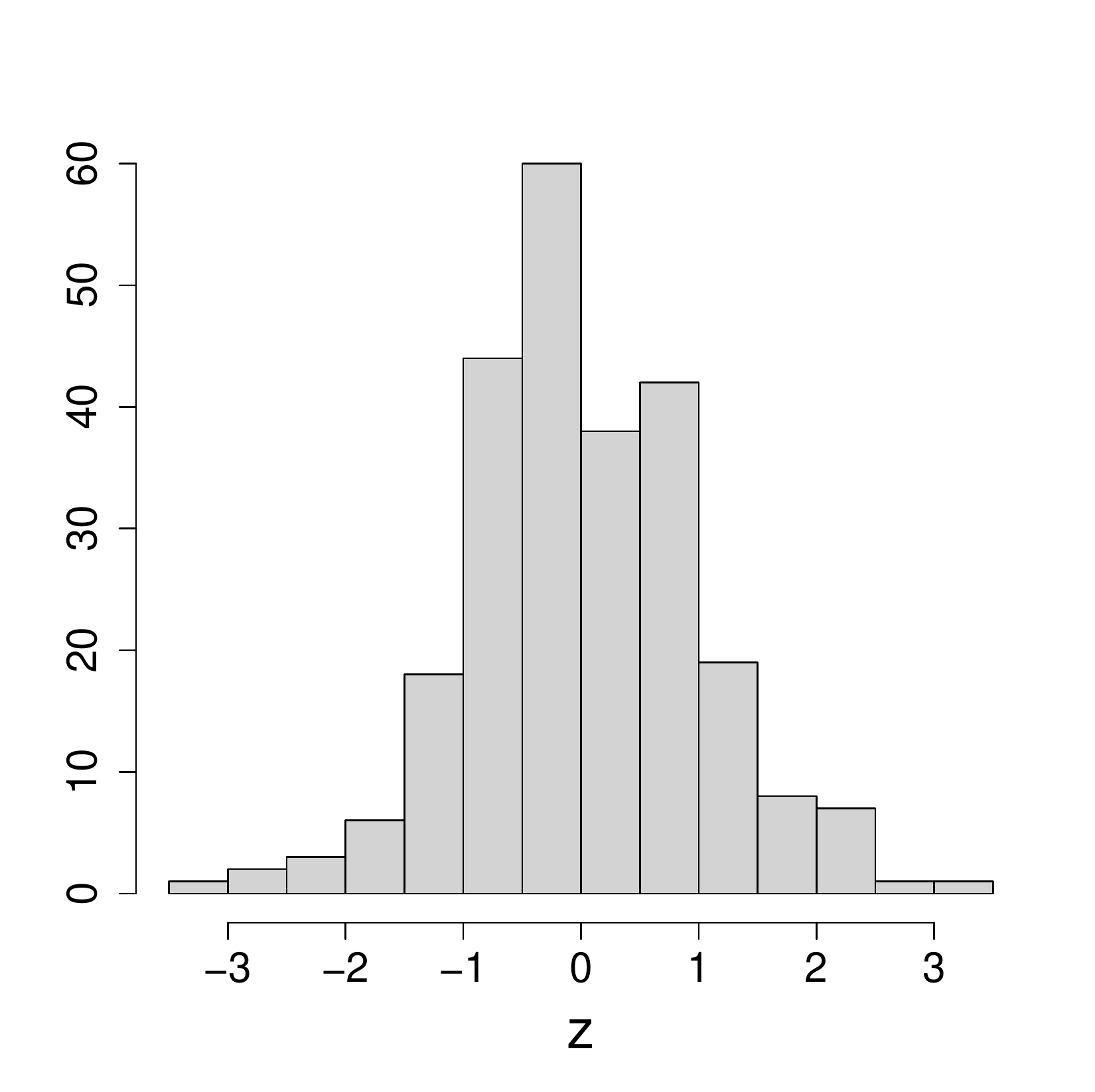}
\includegraphics[scale = 0.24]{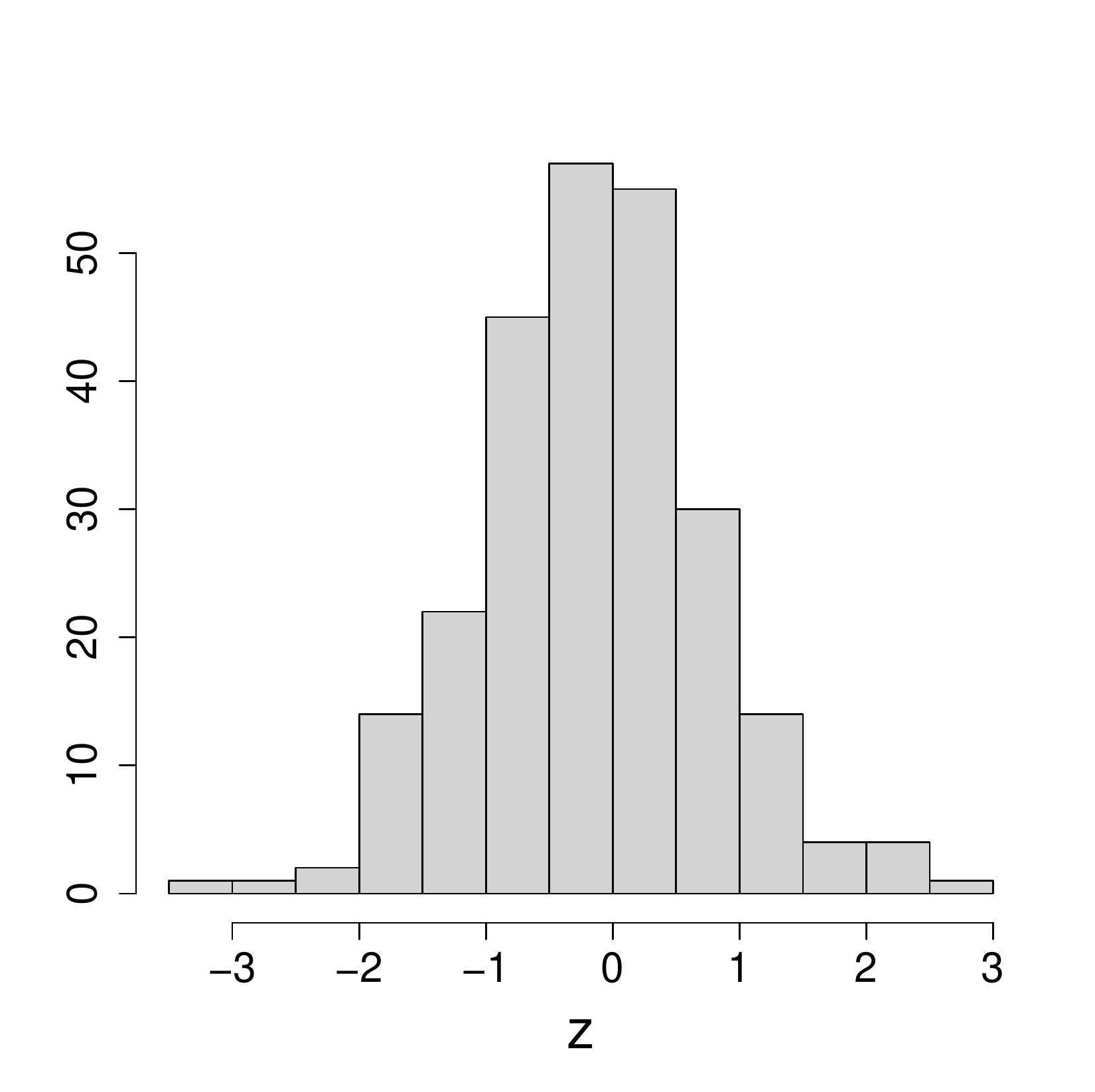}
\includegraphics[scale = 0.24]{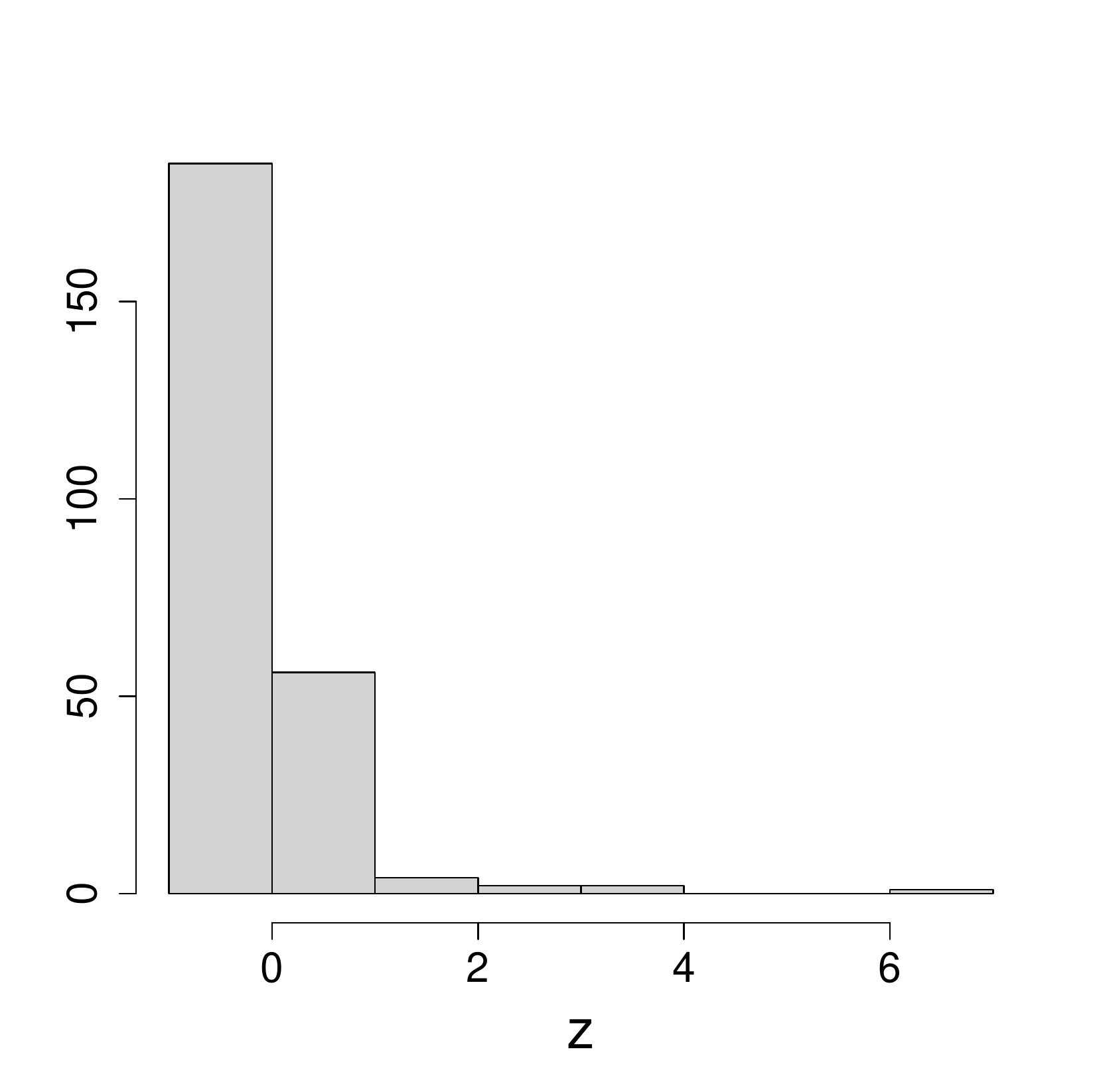}
\includegraphics[scale = 0.24]{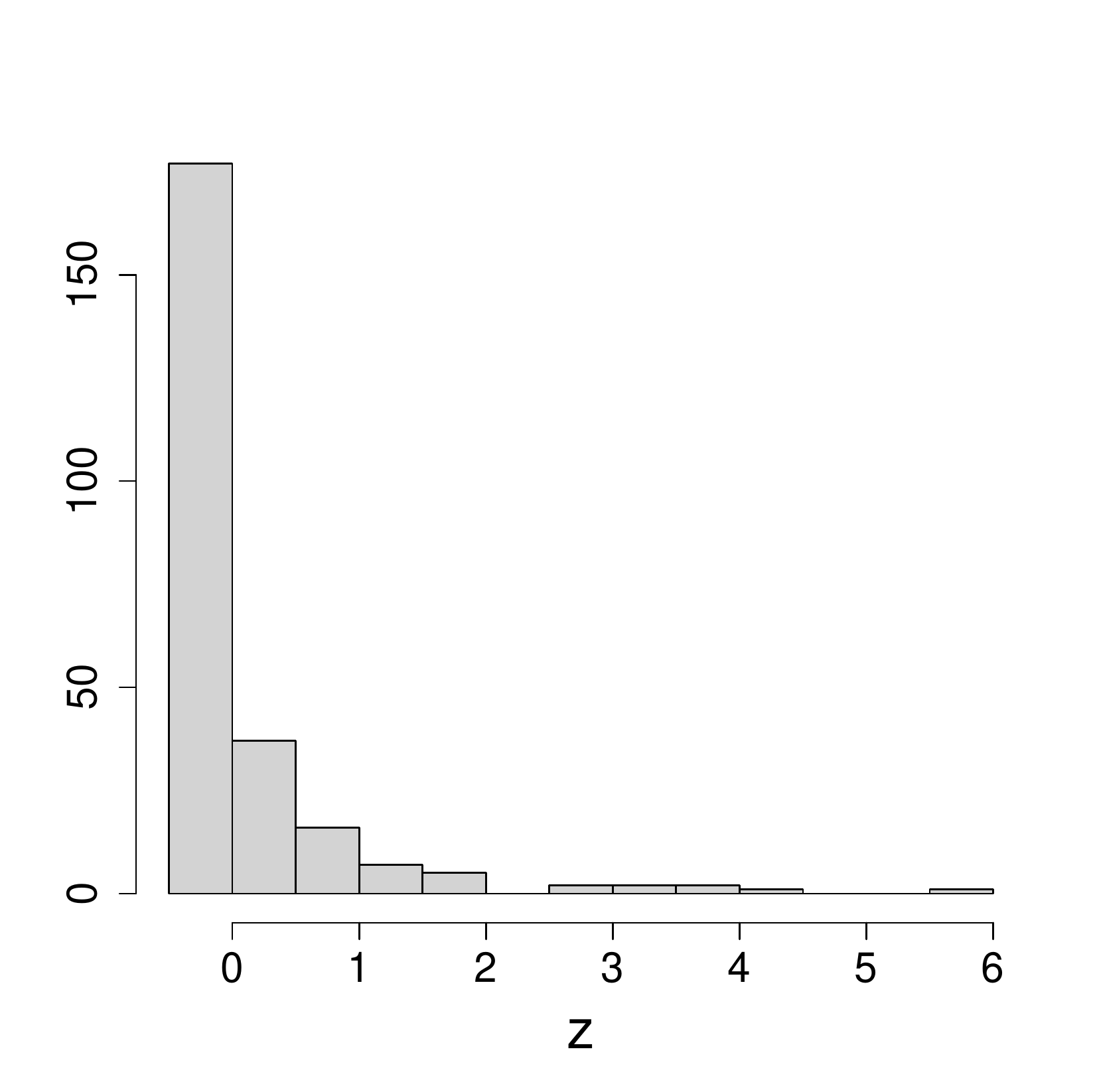}
\includegraphics[scale = 0.18]{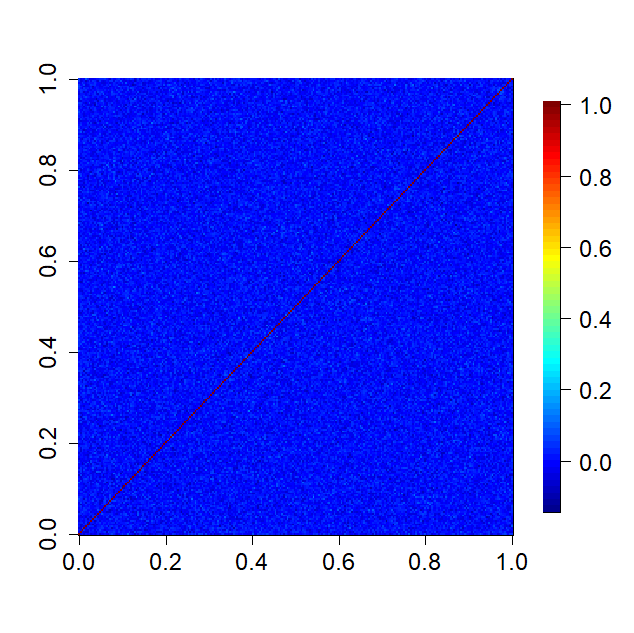}
\includegraphics[scale = 0.18]{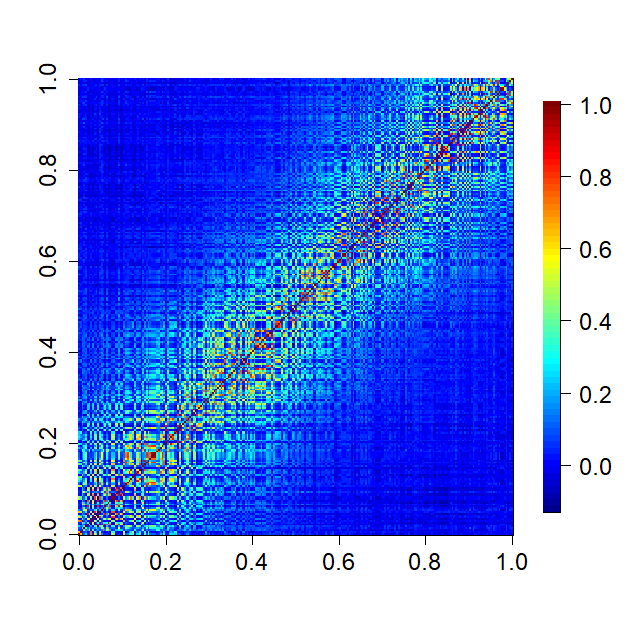}
\includegraphics[scale = 0.18]{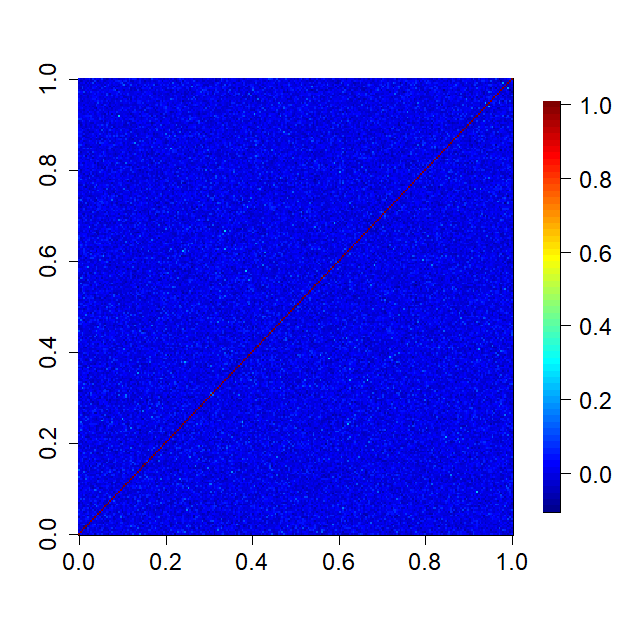}
\includegraphics[scale = 0.18]{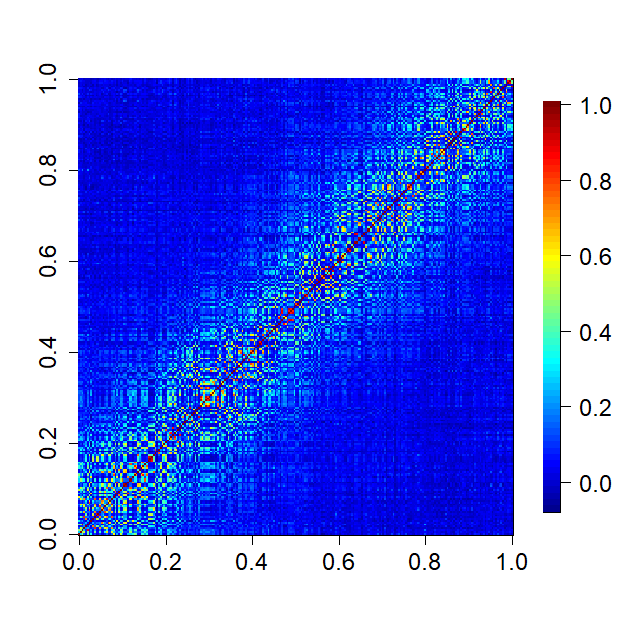}
\includegraphics[scale = 0.24]{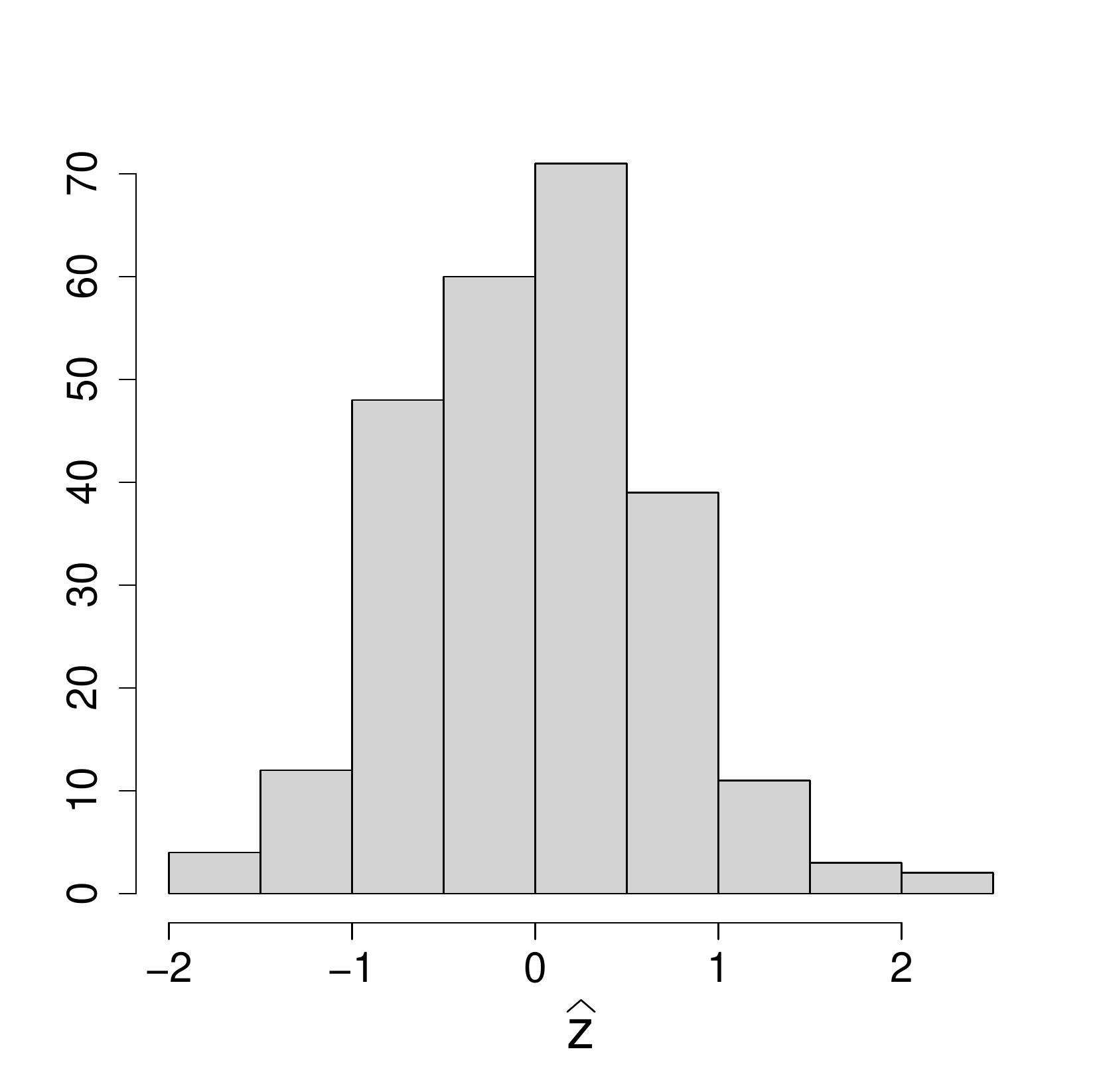}
\includegraphics[scale = 0.24]{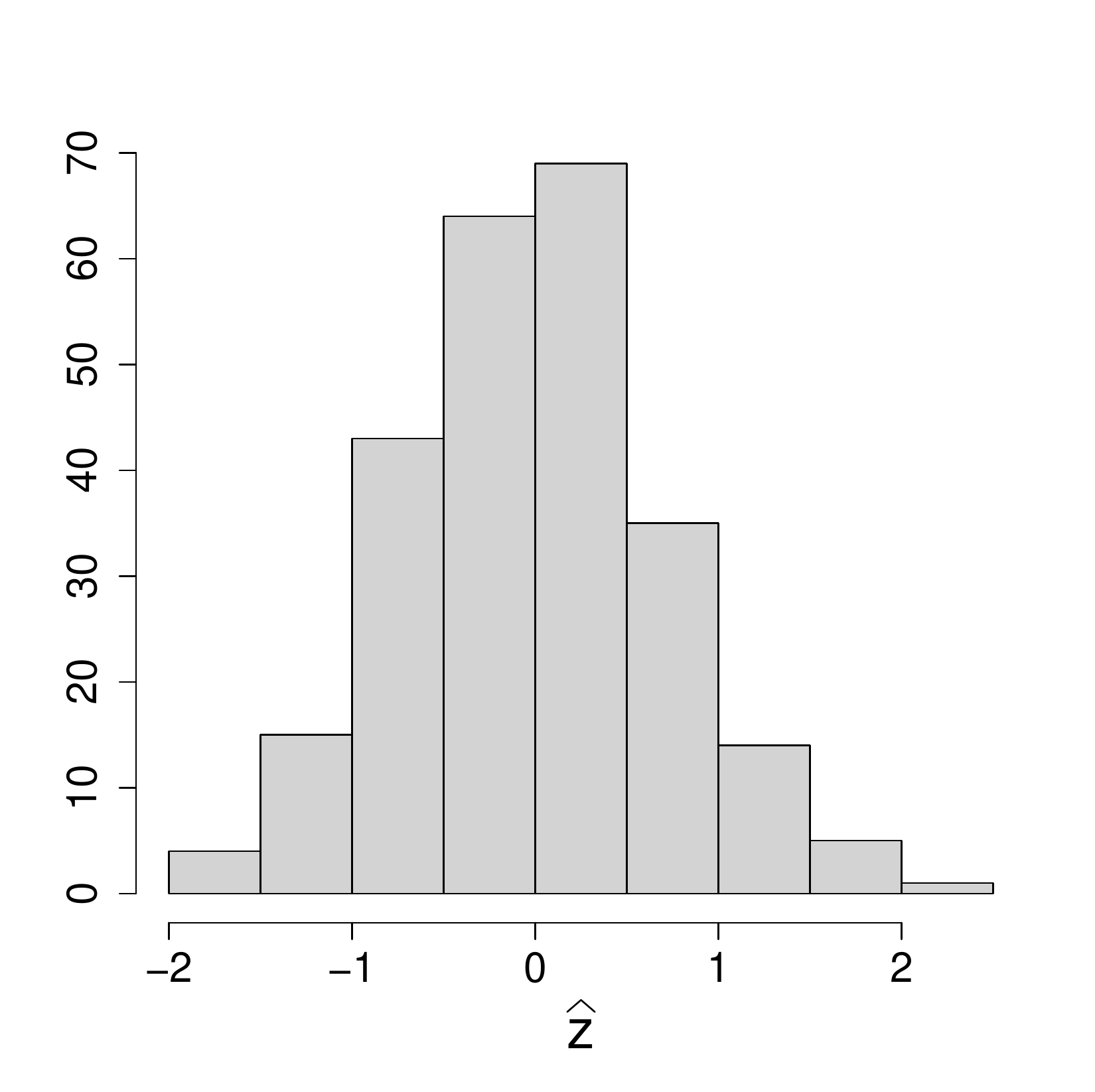}
\includegraphics[scale = 0.24]{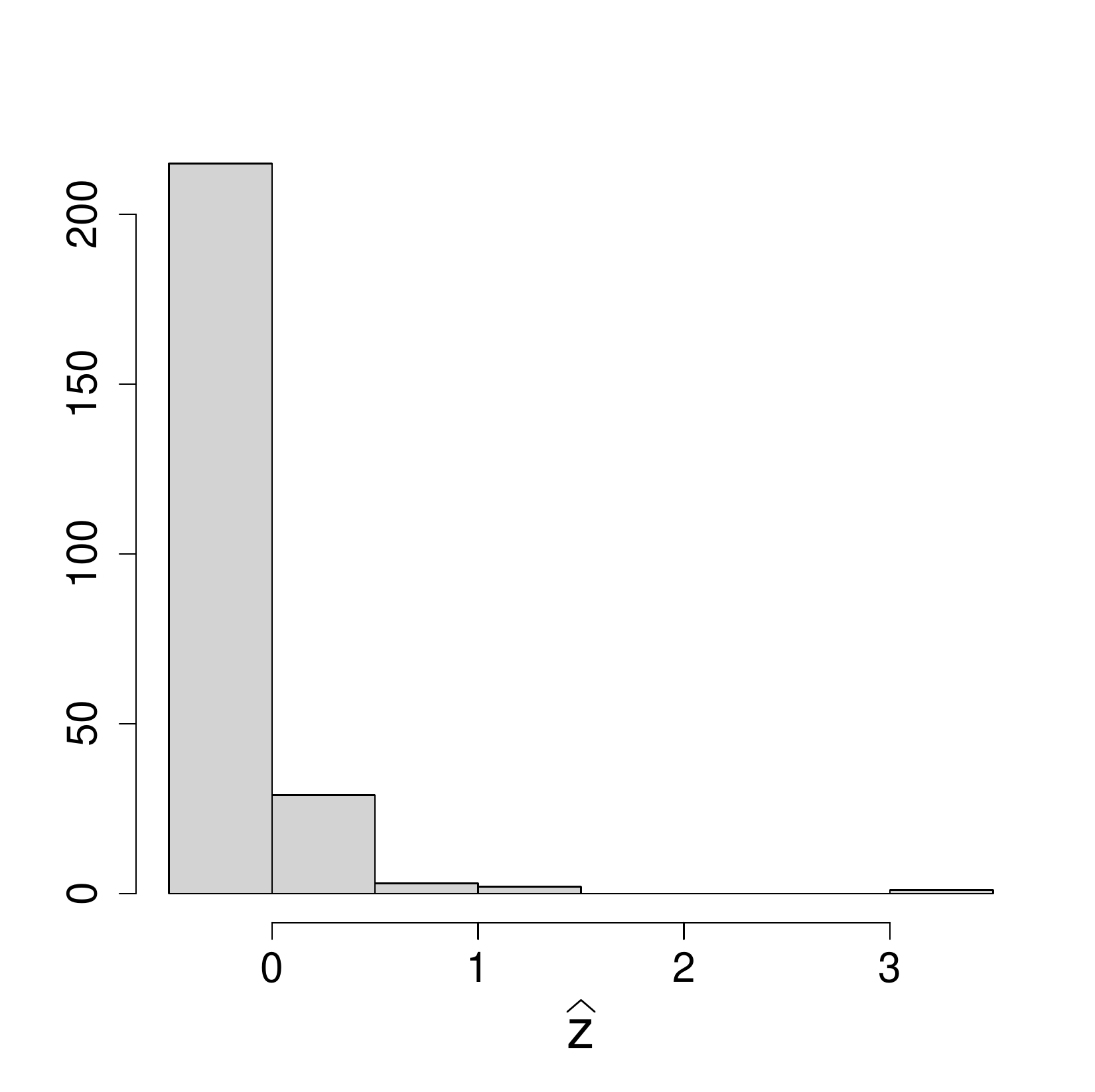}
\includegraphics[scale = 0.24]{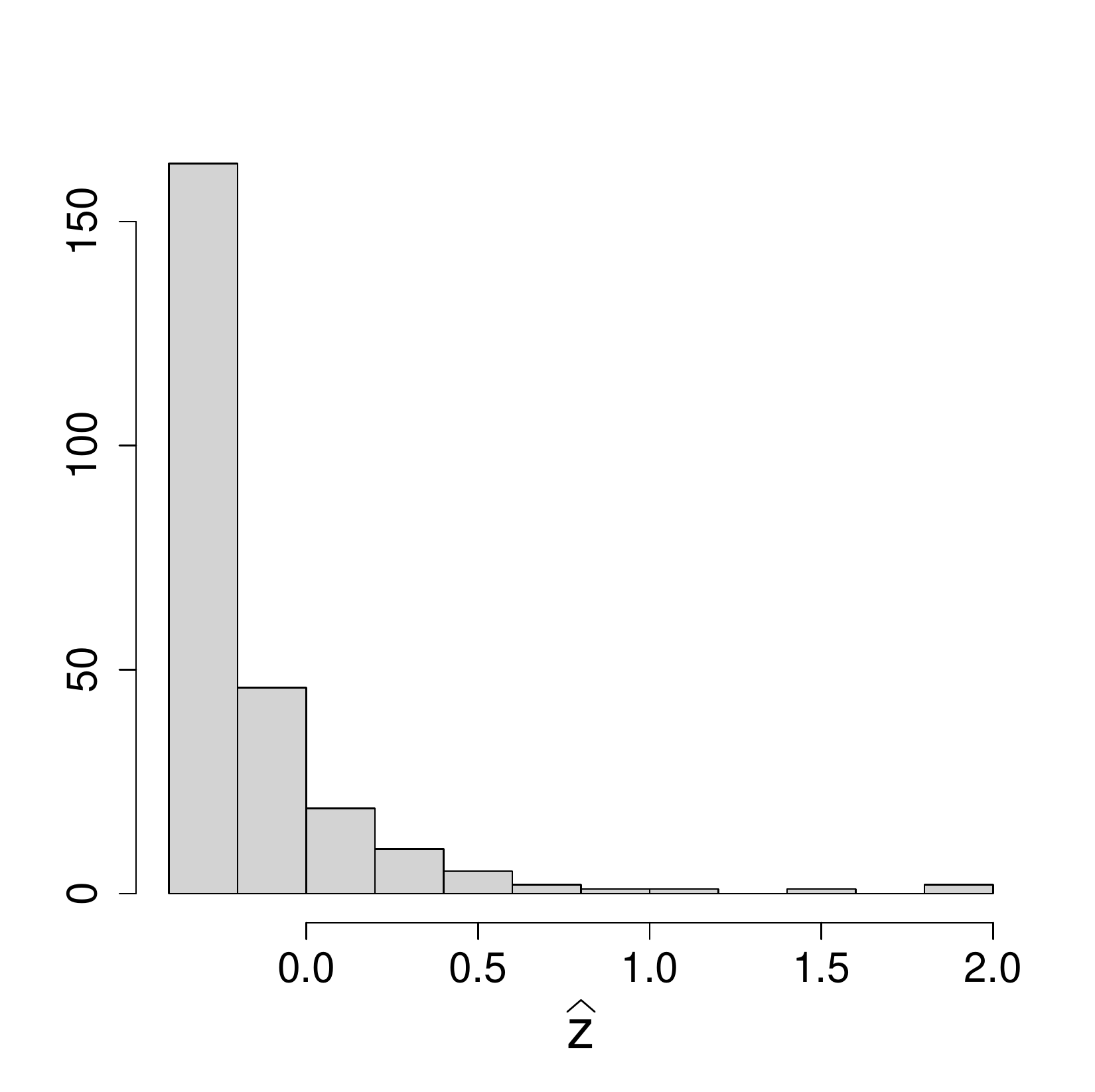}
\caption{Simulated datasets from each simulation setting in Table 1 where $\tau^2 = 1.00$ (first column: uncorrelated, not skewed; second column: correlated, not skewed; third column: uncorrelated, skewed; fourth column: correlated, skewed).  The first row shows the histogram of true exposures (i.e., $\textbf{z}$), the second row shows the sample covariance matrix of the ppd samples (i.e., $\textbf{Z}^*$), and the third row shows the histogram of the median of each column of $\textbf{Z}^*$ (i.e., $\widehat{\text{z}}_i$).}
\end{figure}
\clearpage 

\begin{figure}[ht]
\centering
\includegraphics[trim={0.5cm 0cm 1.5cm 0cm}, clip, scale = 0.27]{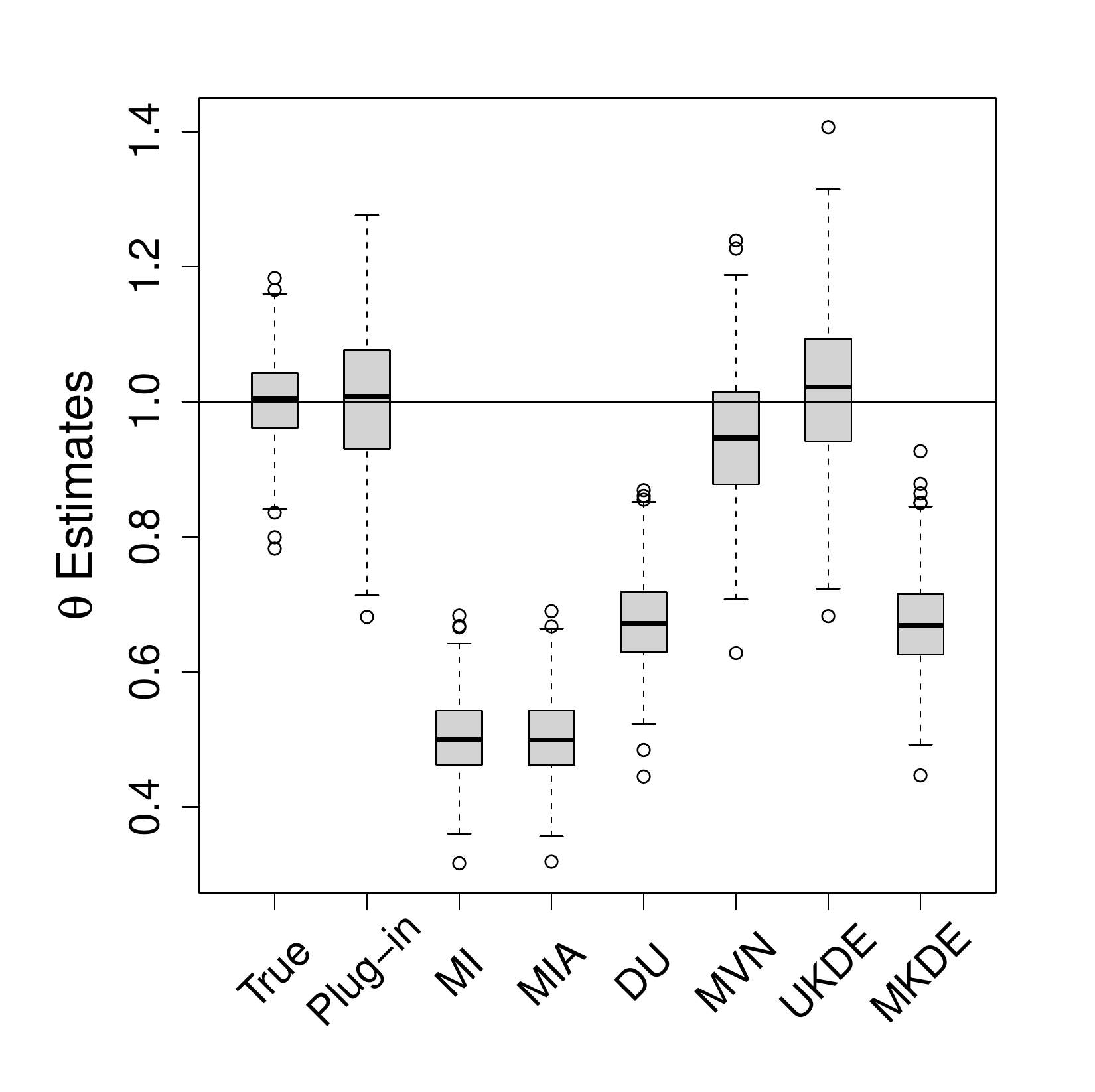}
\includegraphics[trim={0.5cm 0cm 1.5cm 0cm}, clip, scale = 0.27]{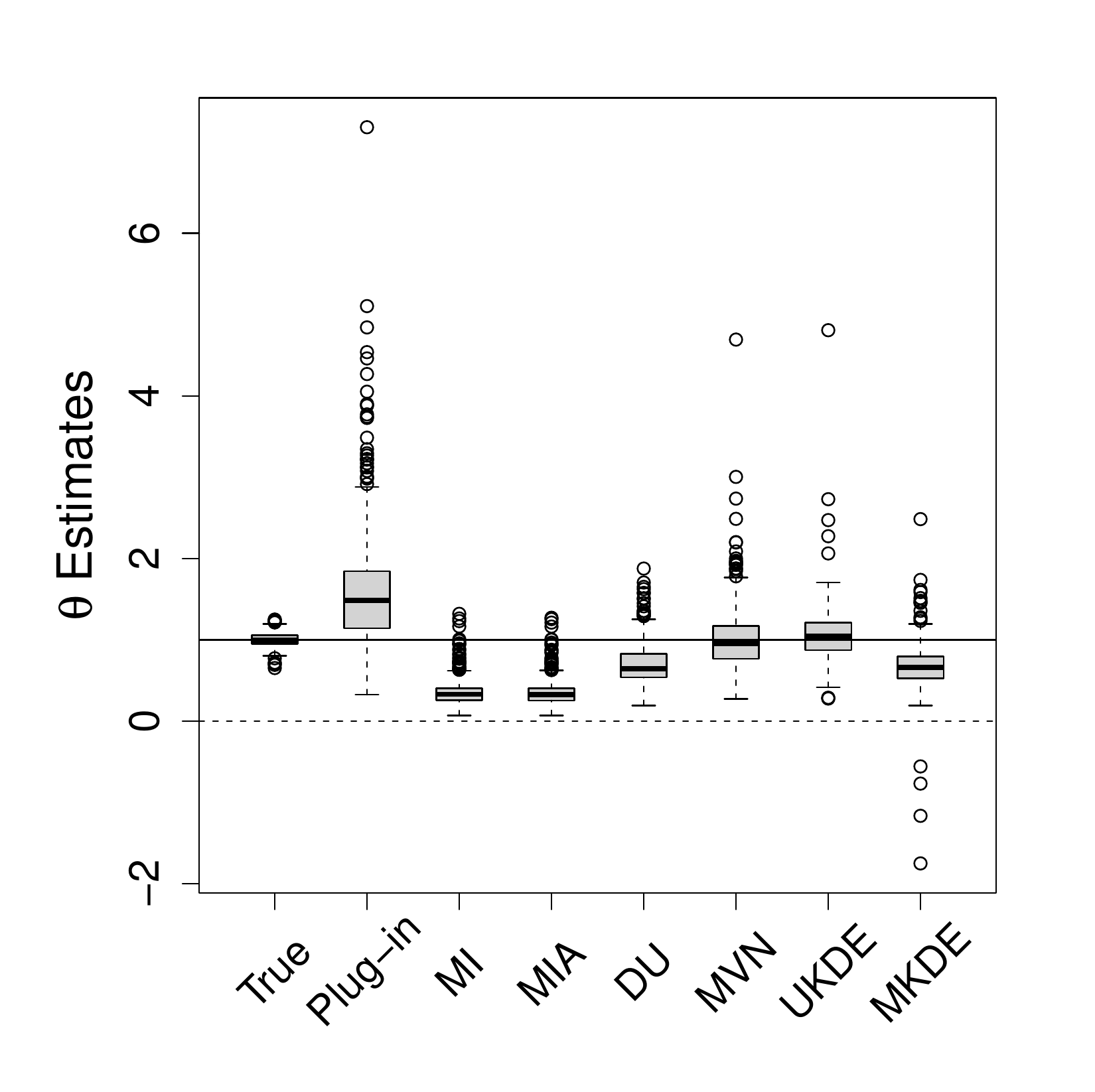}
\includegraphics[trim={0.5cm 0cm 1.5cm 0cm}, clip, scale = 0.27]{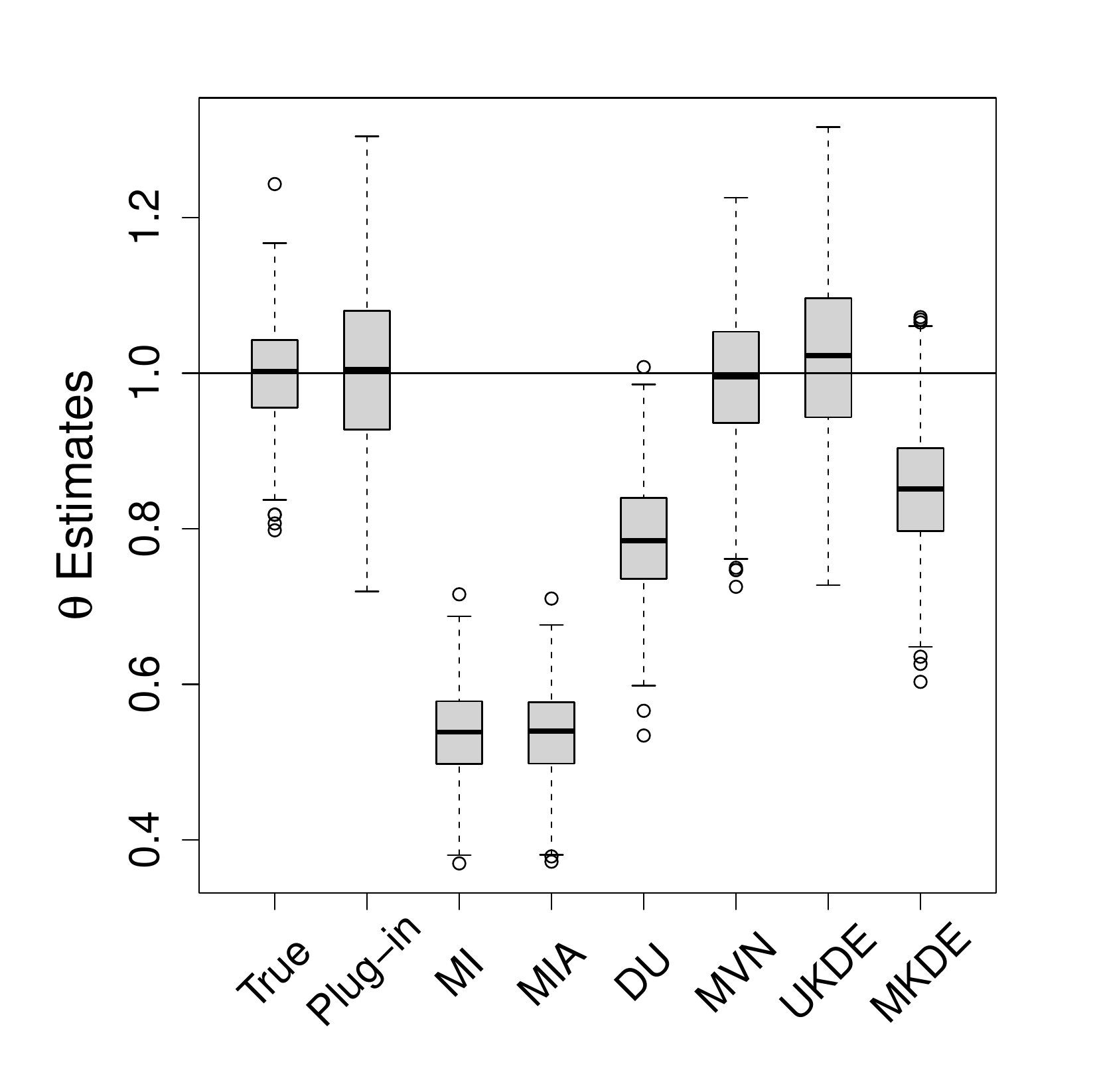}
\includegraphics[trim={0.5cm 0cm 1.5cm 0cm}, clip, scale = 0.27]{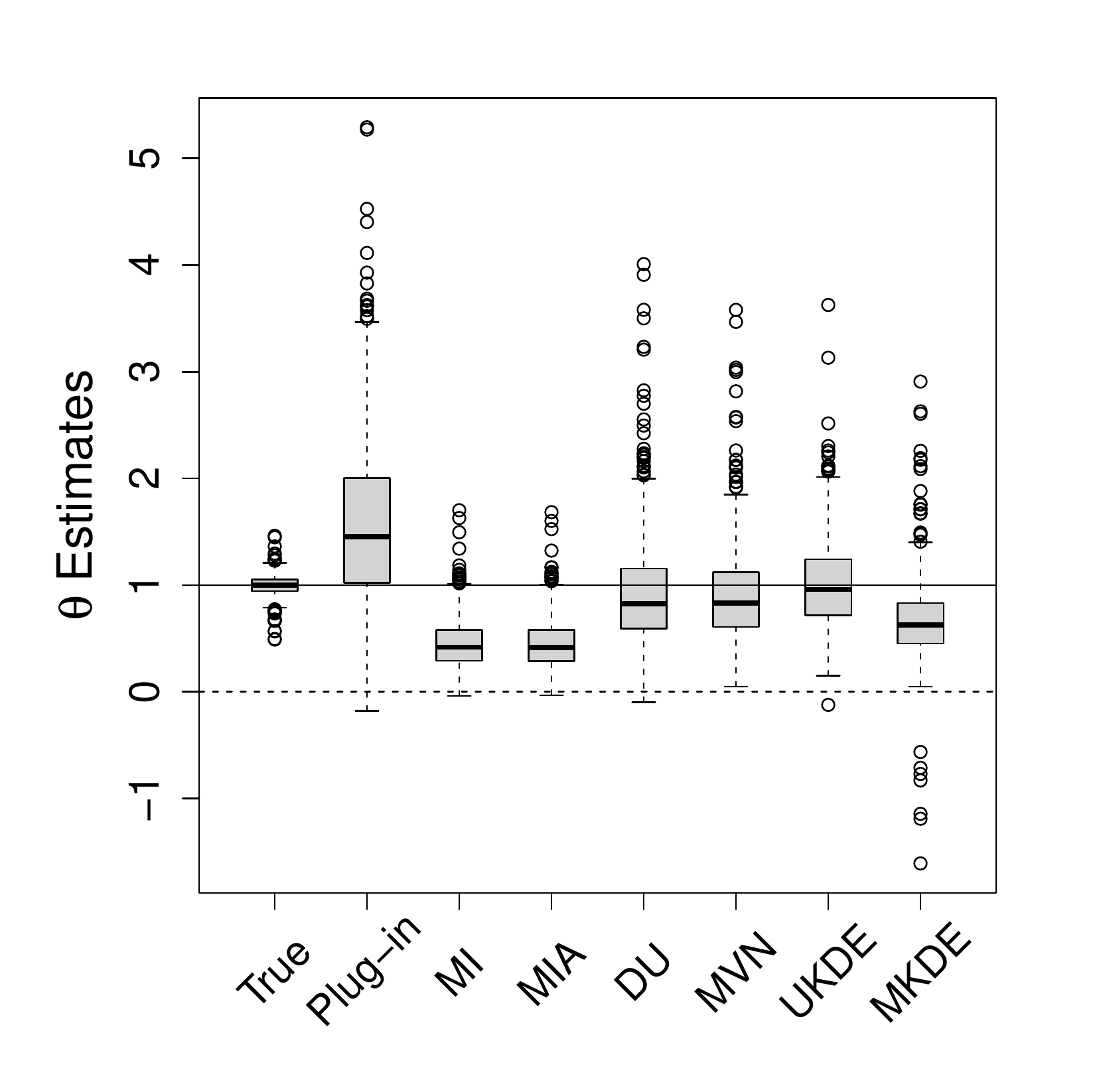}
\caption{Posterior mean estimates of $\theta$ for each method and correlation/skewness settings across all 500 analyses for the $\theta=1.00$ and $\tau^2 = 1.00$ scenario (first panel: uncorrelated, not skewed; second panel: uncorrelated, skewed; third panel: correlated, not skewed; fourth panel: correlated, skewed).  The solid horizontal line represents the true value of $\theta$.}
\end{figure}
\clearpage 

\begin{figure}[ht]
\centering
\includegraphics[trim={0.5cm 0cm 1.5cm 0cm}, clip, scale = 0.27]{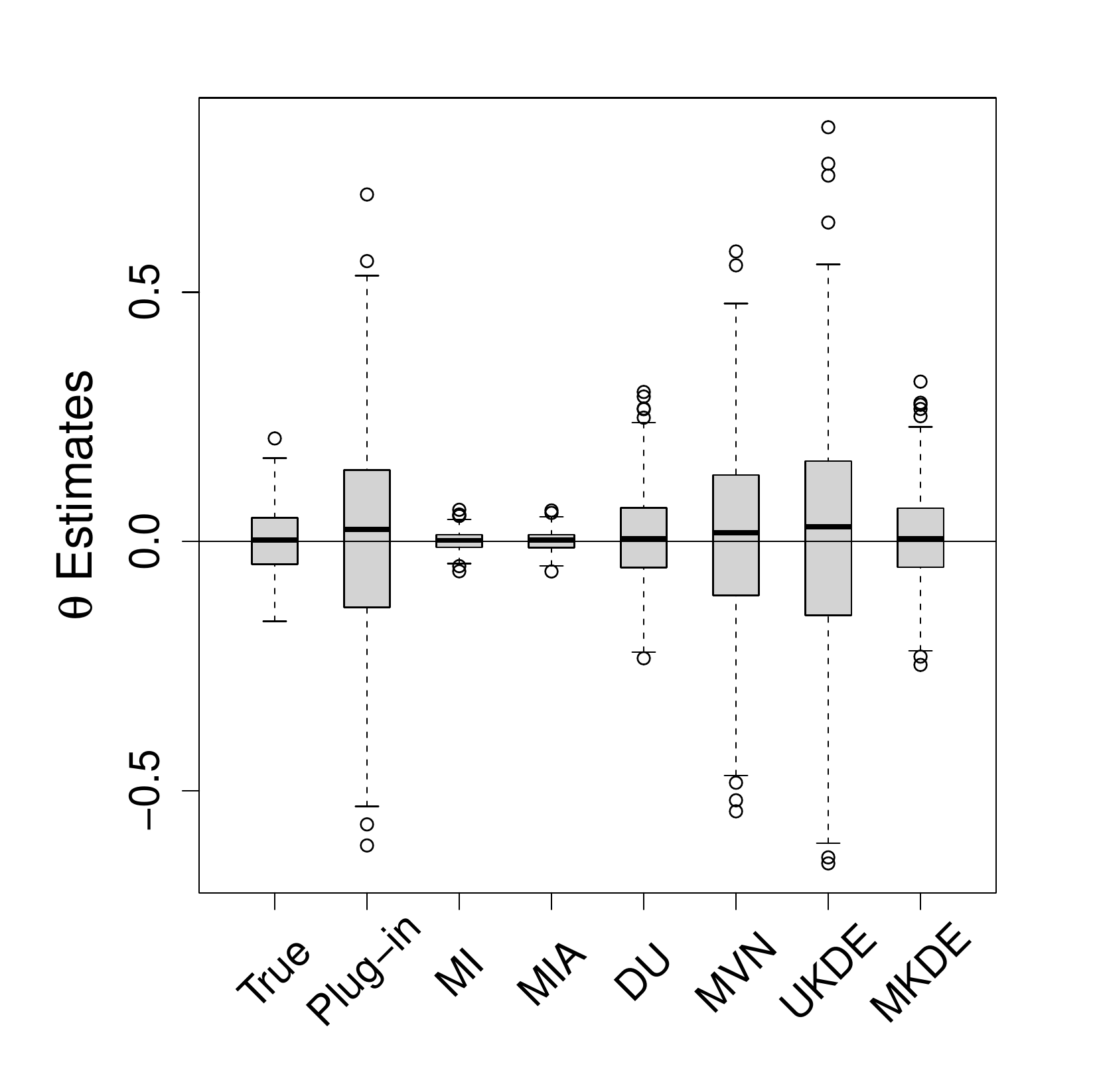}
\includegraphics[trim={0.5cm 0cm 1.5cm 0cm}, clip, scale = 0.27]{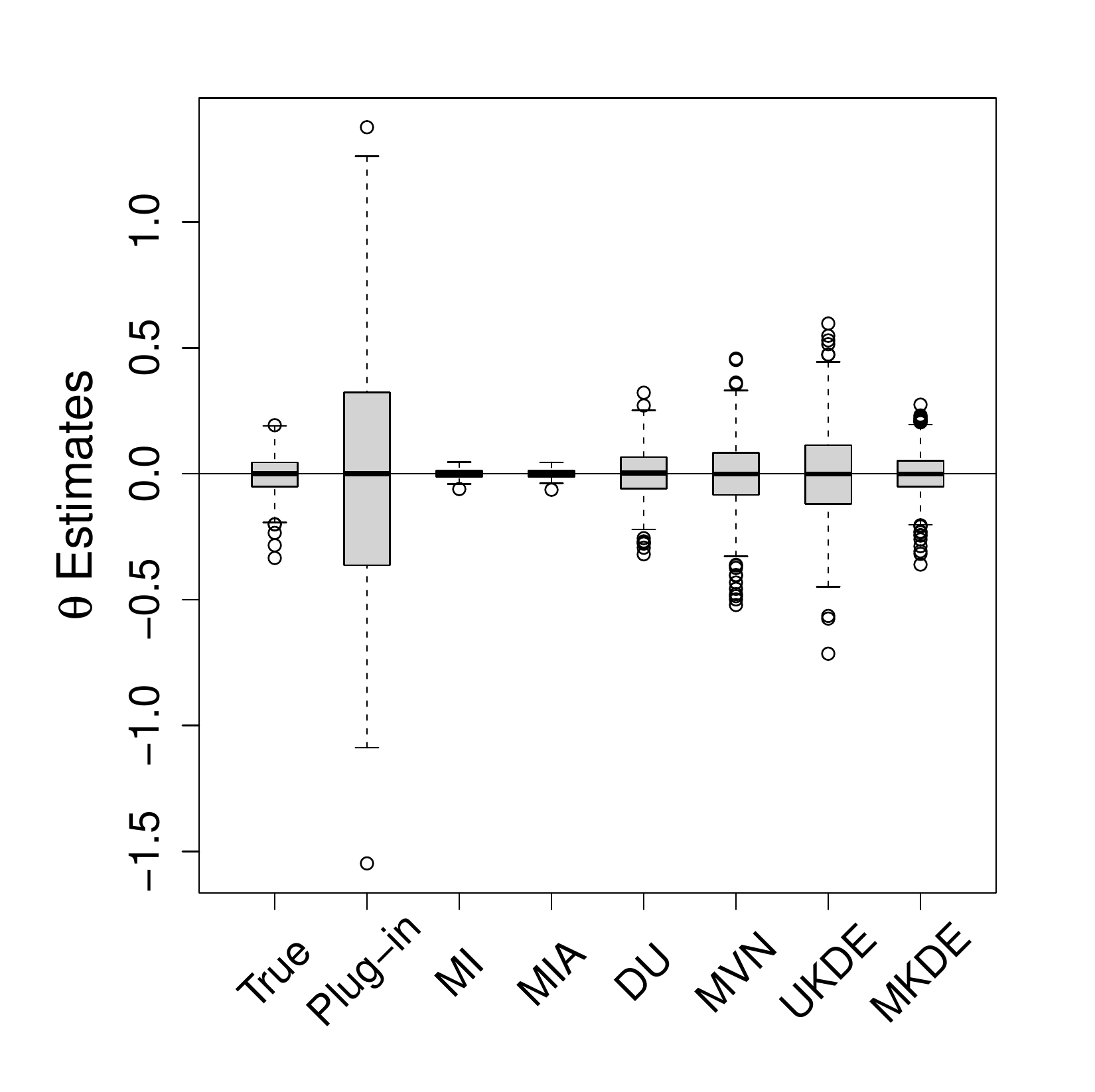}
\includegraphics[trim={0.5cm 0cm 1.5cm 0cm}, clip, scale = 0.27]{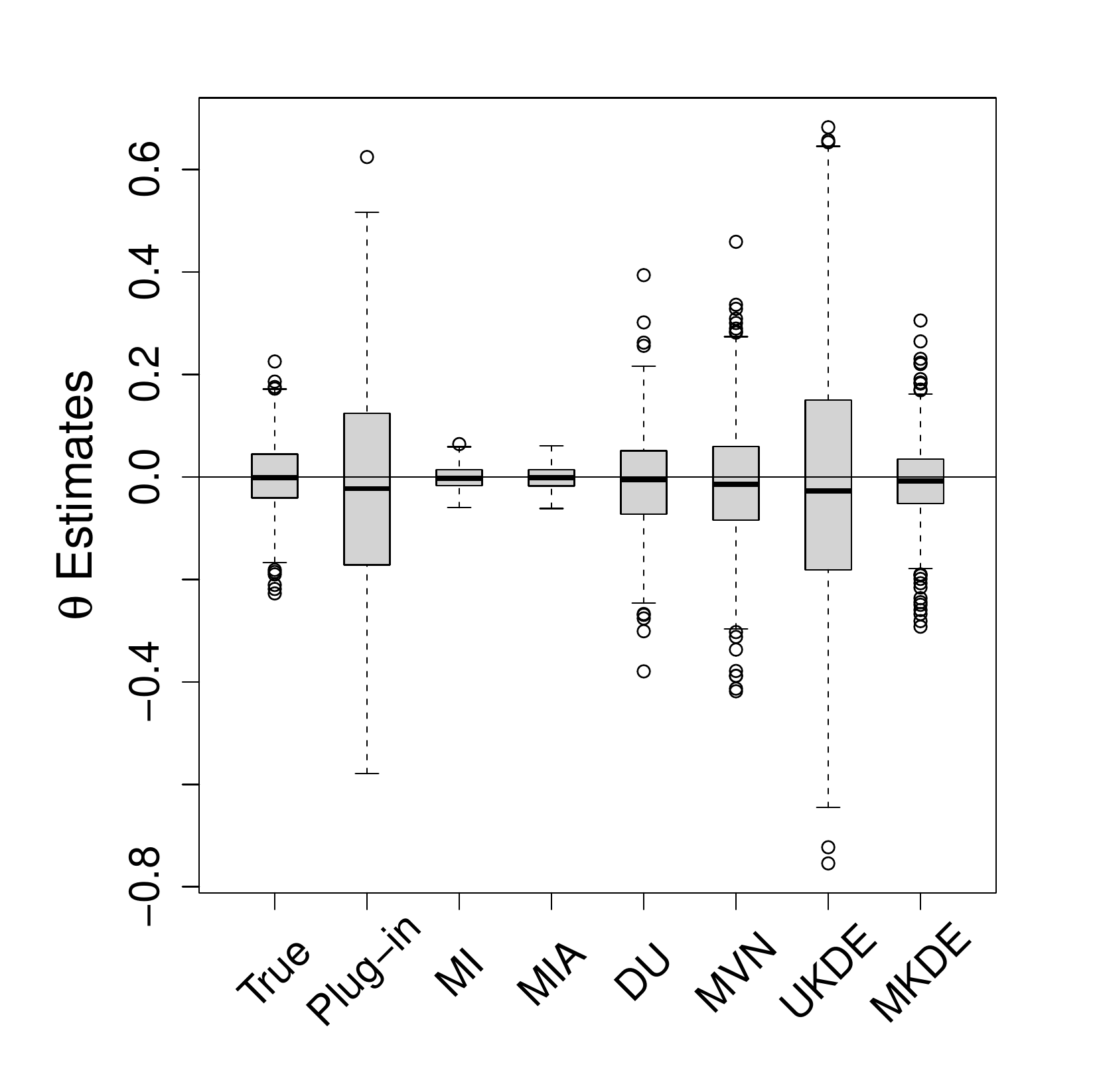}
\includegraphics[trim={0.5cm 0cm 1.5cm 0cm}, clip, scale = 0.27]{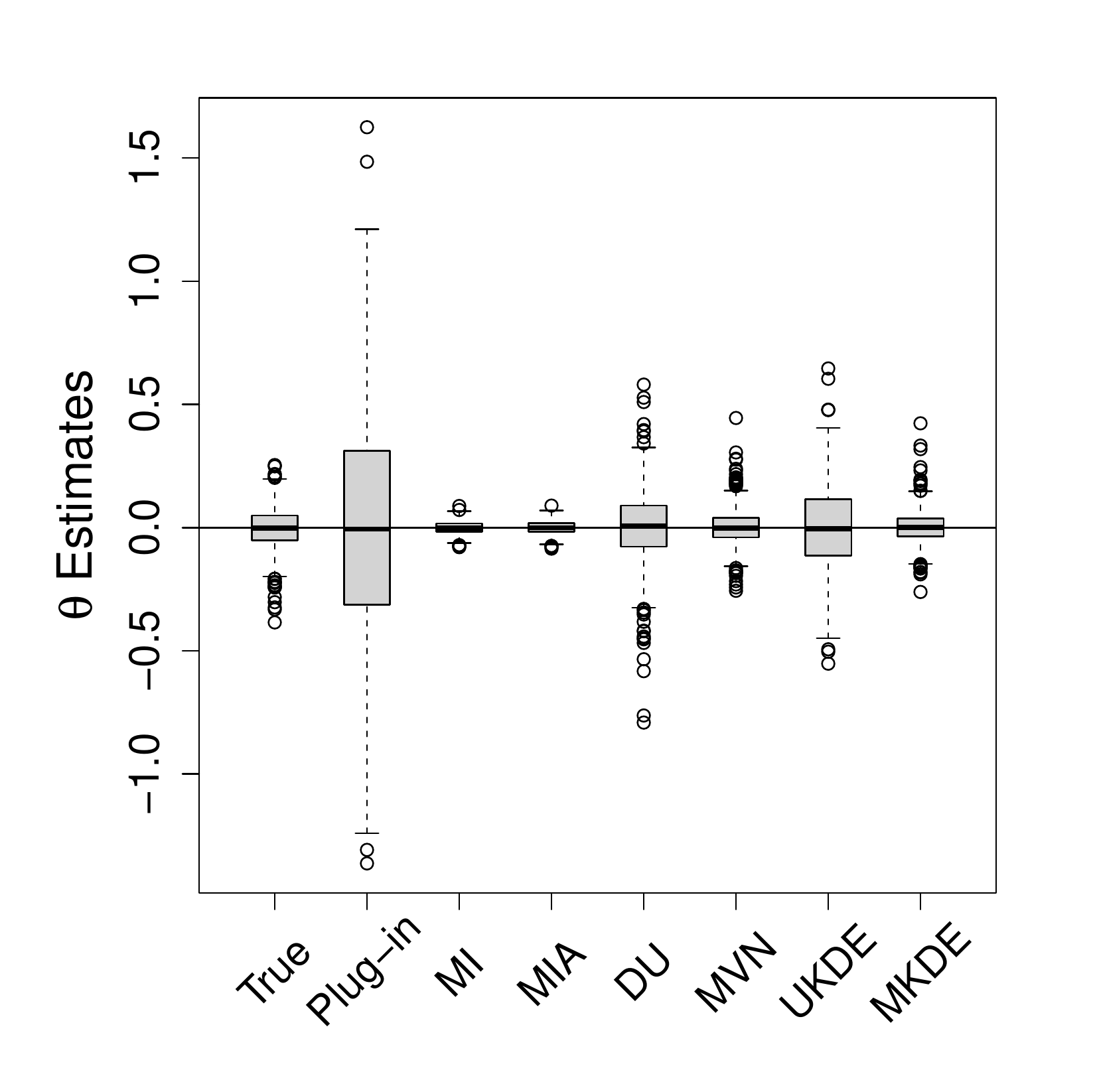}
\caption{Posterior mean estimates of $\theta$ for each method and correlation/skewness settings across all 500 analyses for the $\theta=0.00$ and $\tau^2 = 0.01$ scenario (first panel: uncorrelated, not skewed; second panel: uncorrelated, skewed; third panel: correlated, not skewed; fourth panel: correlated, skewed).  The solid horizontal line represents the true value of $\theta$.}
\end{figure}
\clearpage 

\begin{figure}[ht]
\centering
\includegraphics[trim={0.5cm 0cm 1.5cm 0cm}, clip, scale = 0.27]{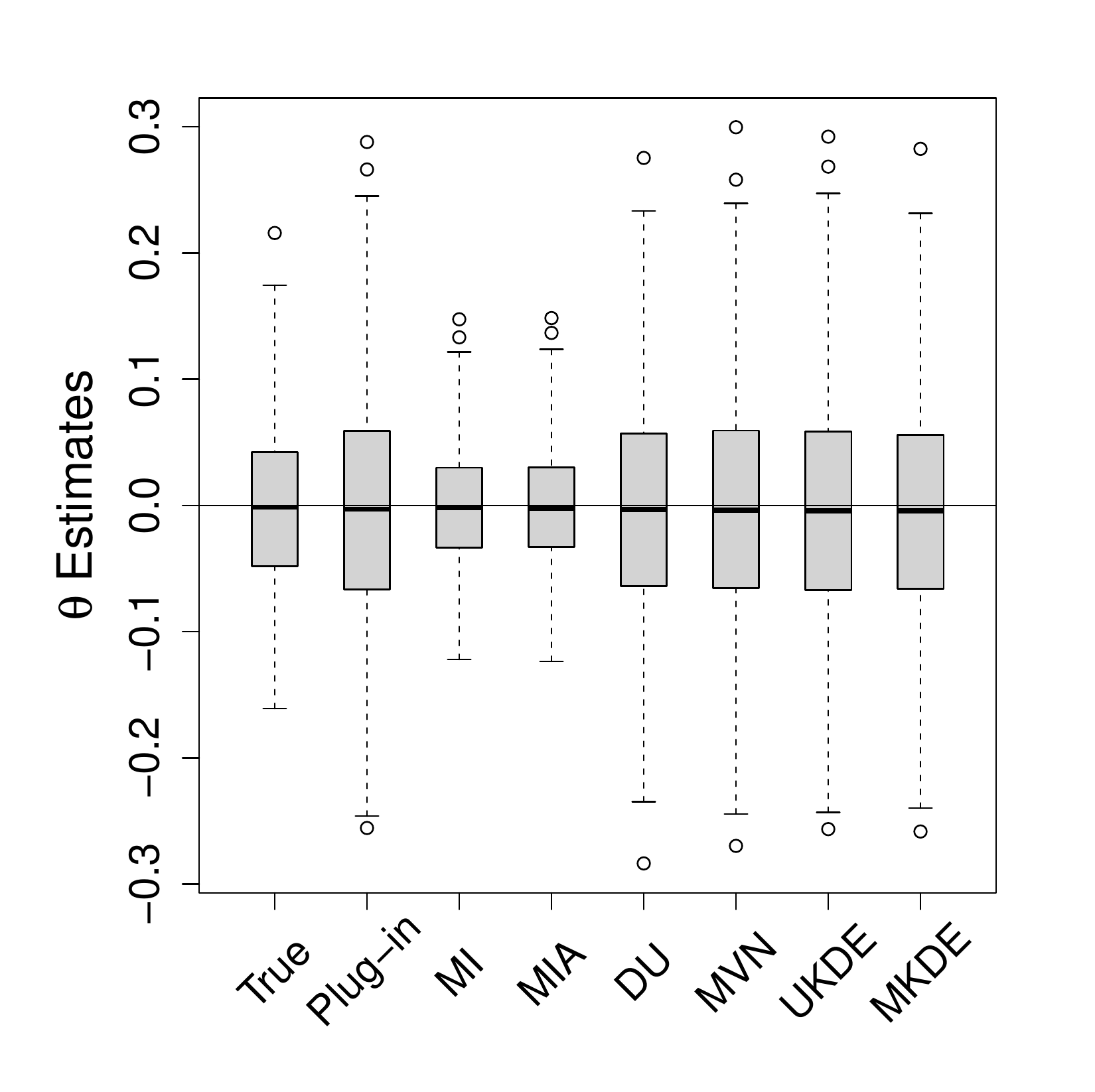}
\includegraphics[trim={0.5cm 0cm 1.5cm 0cm}, clip, scale = 0.27]{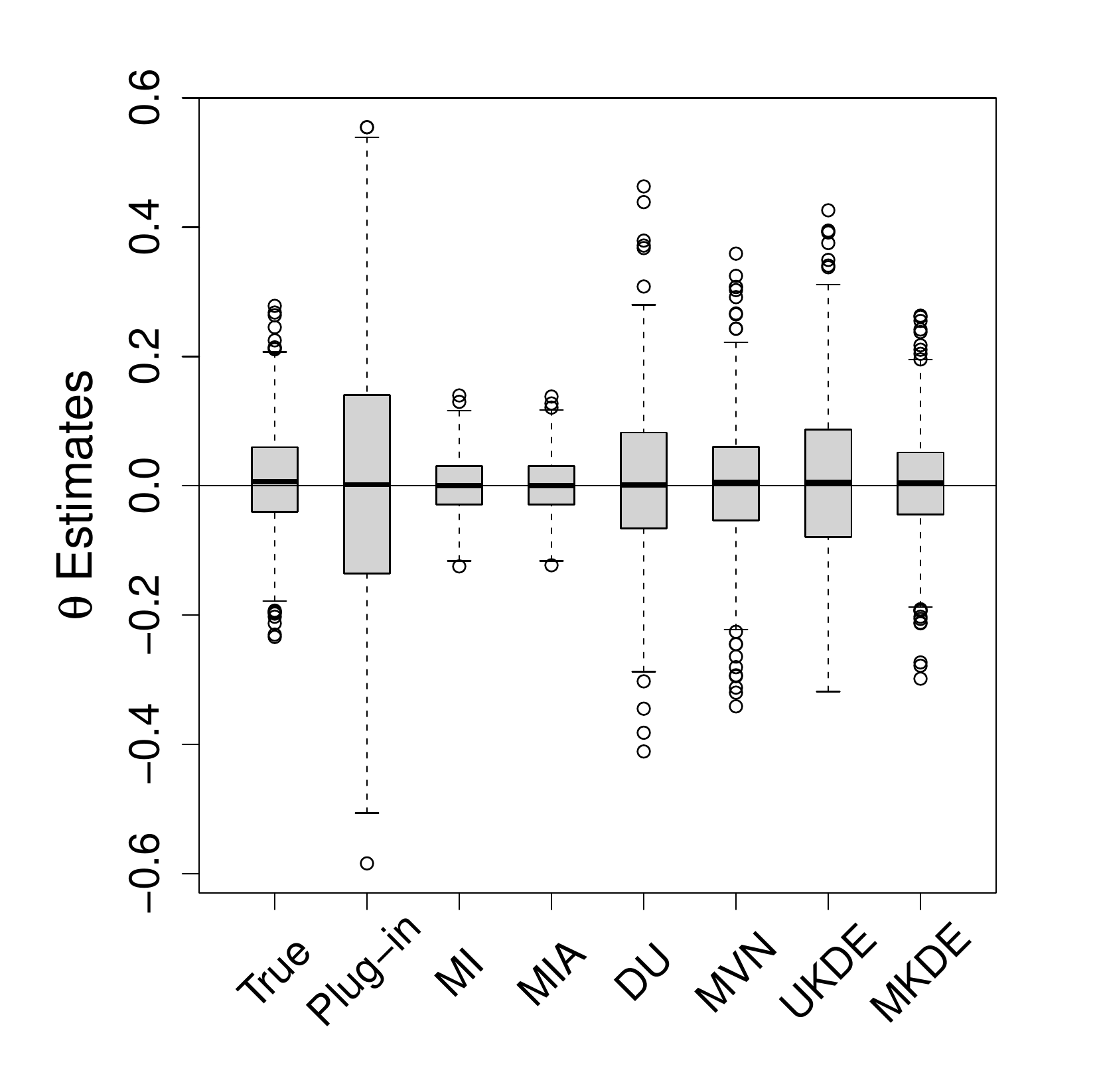}
\includegraphics[trim={0.5cm 0cm 1.5cm 0cm}, clip, scale = 0.27]{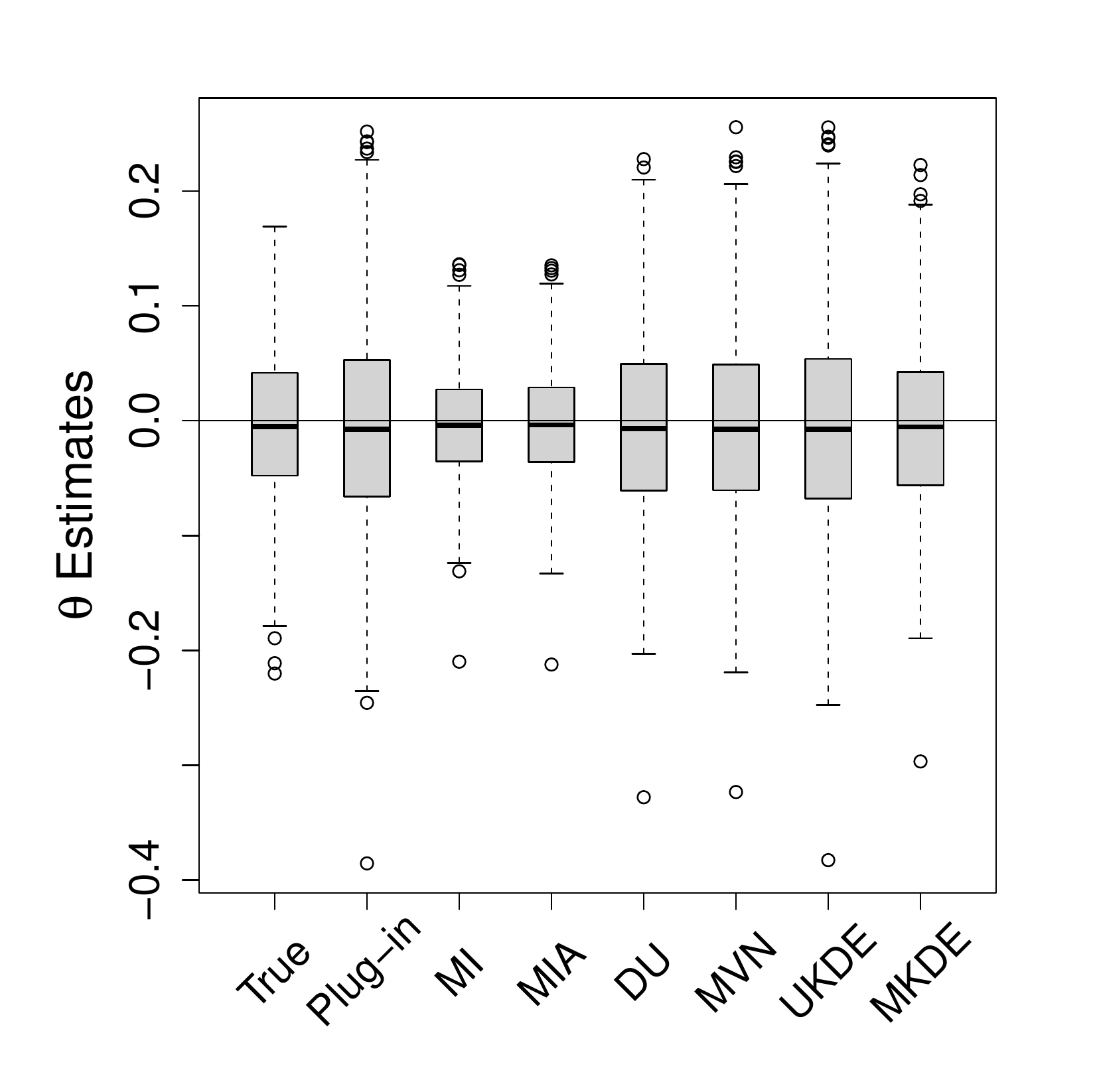}
\includegraphics[trim={0.5cm 0cm 1.5cm 0cm}, clip, scale = 0.27]{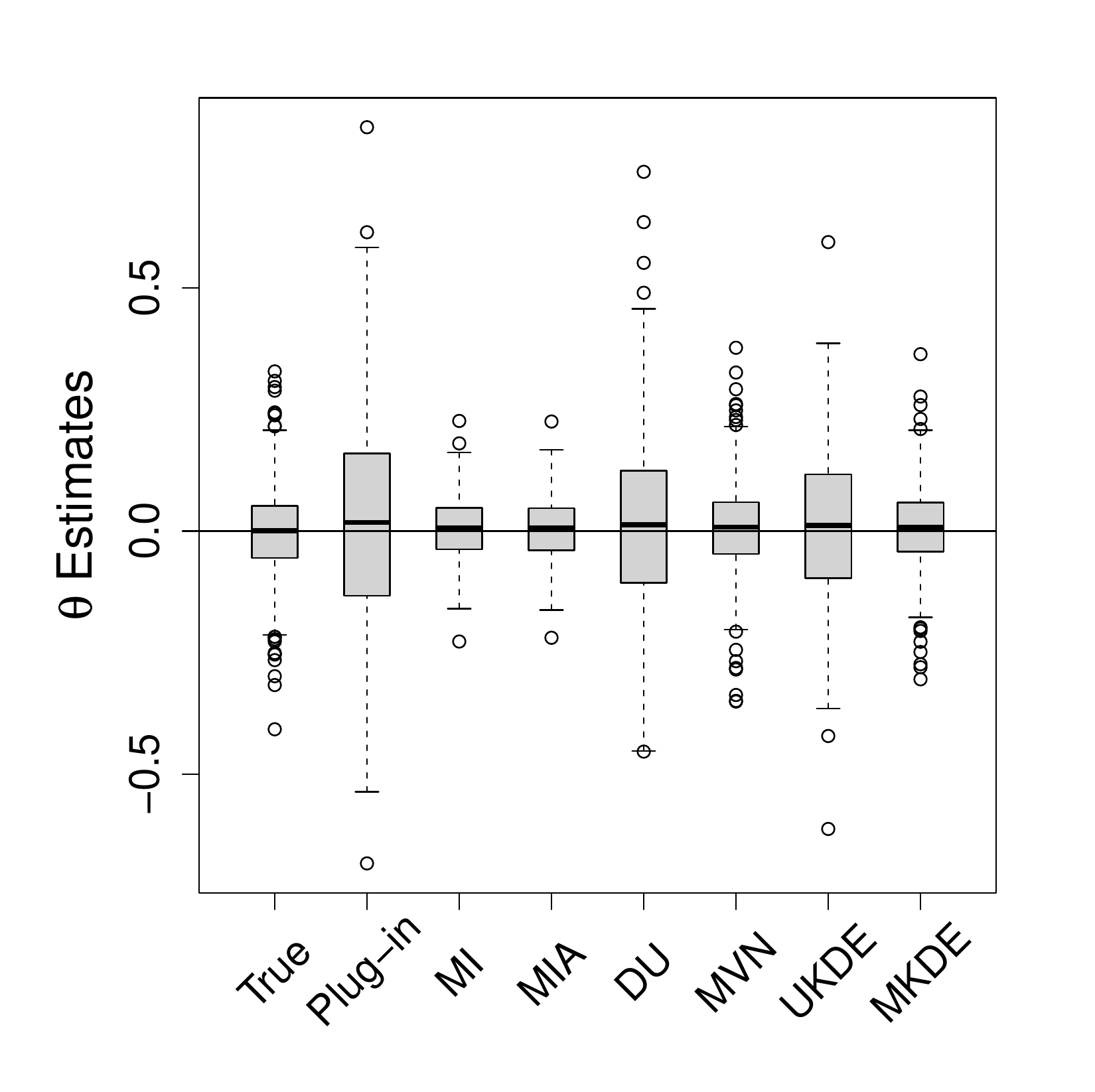}
\caption{Posterior mean estimates of $\theta$ for each method and correlation/skewness settings across all 500 analyses for the $\theta=0.00$ and $\tau^2 = 1.00$ scenario (first panel: uncorrelated, not skewed; second panel: uncorrelated, skewed; third panel: correlated, not skewed; fourth panel: correlated, skewed).  The solid horizontal line represents the true value of $\theta$.}
\end{figure}
\clearpage 

\begin{figure}[ht]
\centering
\includegraphics[scale = 1.00]{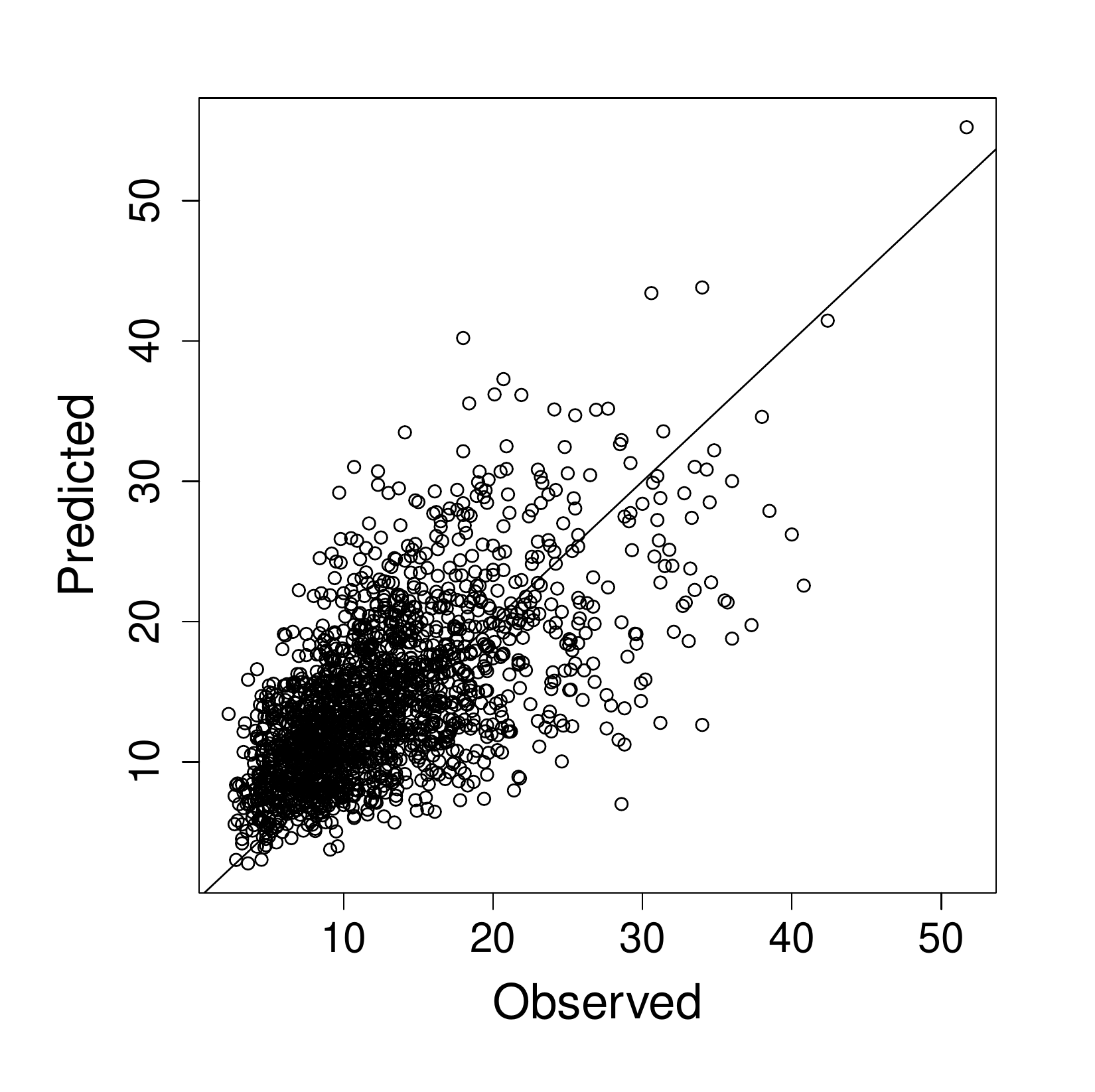}
\caption{Scatterplot of the model-based predictions (posterior medians) of daily maximum PM$_{2.5}$ ($\mu$g/m$^3$) in the three New Jersey counties during the study period and the observed daily maximum of the PM$_{2.5}$ AQS concentrations ($\mu$g/m$^3$) across all of New Jersey on the same days.}
\end{figure}
\clearpage 

\begin{figure}[ht]
\centering
\includegraphics[scale = 0.40]{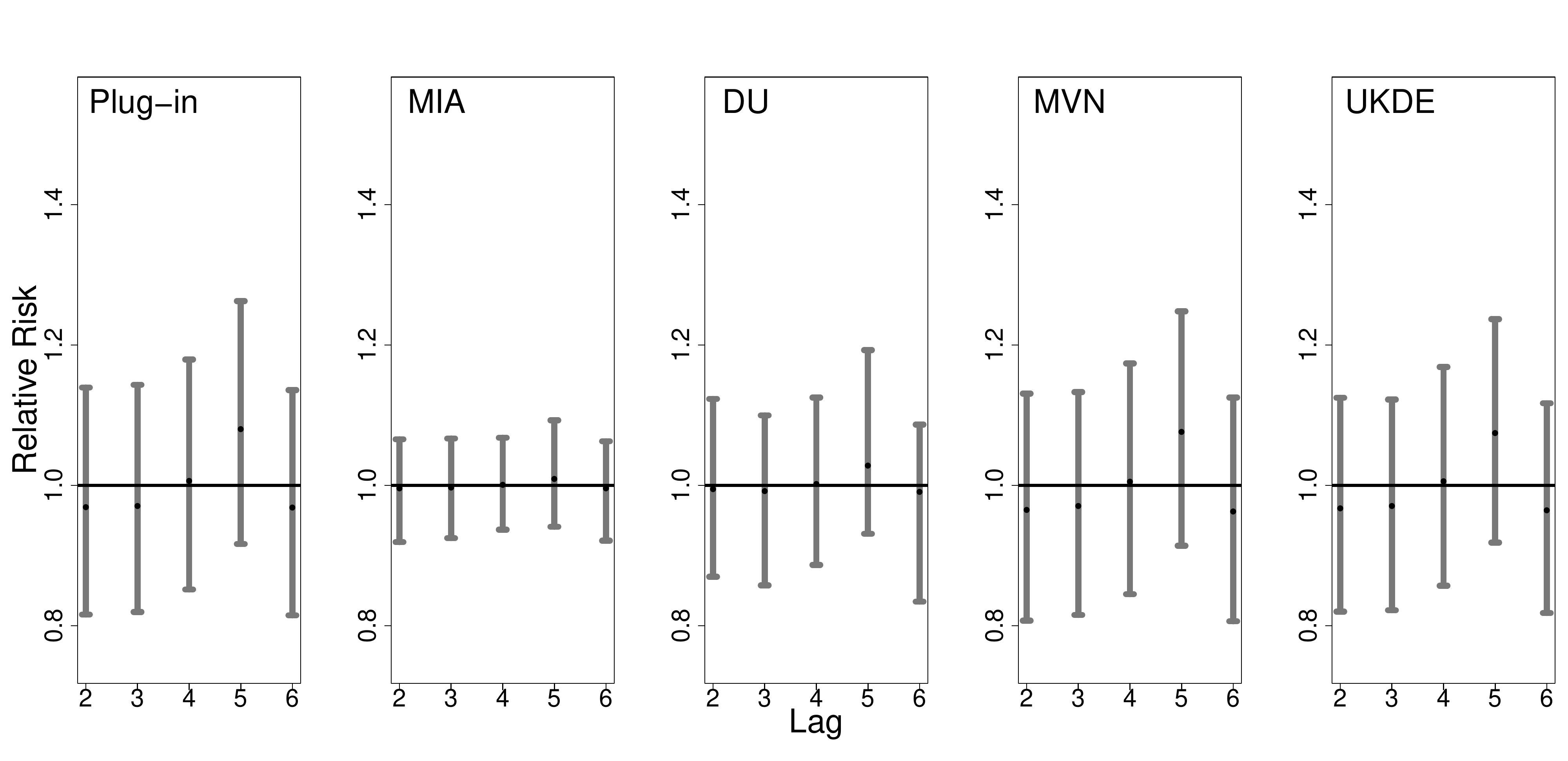}
\caption{Posterior mean and 99\% credible interval plots for $\exp\left\{\theta\right\}$ across the different models and daily lag periods for the New Jersey three county stillbirth and maximum daily 24-hour PM$_{2.5}$ exposure sensitivity analysis where the temporal ordering of exposure has been shuffled.}
\end{figure}
\clearpage 

\bibliographystyle{chicago}
\bibliography{References}